\documentclass[sigconf]{acmart}

\usepackage[disable]{todonotes}

\usepackage{amsmath}

\usepackage{graphicx}
\usepackage{balance}
\usepackage{bm}
\usepackage{color}
\usepackage{listings}
\usepackage{colortbl}
\usepackage{verbatim}
\usepackage{textcomp}
\usepackage{afterpage}
\usepackage{subcaption}

\usepackage{amscd}
\usepackage{amsthm}
\usepackage{mathtools}

\usepackage{algorithm}
\usepackage{algorithmicx}
\usepackage[noend]{algpseudocode}

\usepackage{tikz}

\usepackage{array}
\usepackage{tabularx}
\usepackage{multirow}
\usepackage{xspace}

\usetikzlibrary{shapes,backgrounds,arrows}
\usepackage{listings}
\usepackage{caption}

\usepackage{supertabular}
\usepackage{paralist}
\usepackage{cleveref}

\usepackage{enumitem}

\newcommand{\partitle}[1]{\smallskip\noindent\textbf{#1}.}

\renewcommand*\Call[2]{\textproc{\textcolor{darkdarkpurple}{#1}}(#2)}
\newcommand{\card}[1]{\left|{#1}\right|}

\newtheorem{theo}{Theorem}
\newtheorem{lem}[theo]{Lemma}

\newtheorem{exam}{Example}
\newtheorem{defi}{Definition}

\newcommand{\proofpar}[1]{\smallskip\noindent\underline{{#1}}}

\definecolor{black}{rgb}{0,0,0}
\definecolor{grey}{rgb}{0.8,0.8,0.8}
\definecolor{red}{rgb}{1,0,0}
\definecolor{green}{rgb}{0,1,0}
\definecolor{darkgreen}{rgb}{0,0.5,0}
\definecolor{darkpurple}{rgb}{0.5,0,0.5}
\definecolor{darkdarkpurple}{rgb}{0.3,0,0.3}
\definecolor{blue}{rgb}{0,0,1}
\definecolor{shadegreen}{rgb}{0.95,1,0.95}
\definecolor{shadeblue}{rgb}{0.95,0.95,1}
\definecolor{shadered}{rgb}{1,0.85,0.85}
\definecolor{shadegrey}{rgb}{0.85,0.85,0.85}
\definecolor{oddRowGrey}{rgb}{0.80,0.80,0.80}
\definecolor{evenRowGrey}{rgb}{0.85,0.85,0.85}

\newcommand{\defas}{:=}

\newcommand{\projection}{\Pi}
\newcommand{\selection}{\sigma}

\newcommand{\union}{\cup}

\newcommand{\join}{\bowtie}

\newcommand{\schema}[1]{\textsc{Sch}(#1)}

\newcommand{\T}{\mathbf{true}}
\newcommand{\F}{\mathbf{false}}
\newcommand{\isnull}{\,\mathbf{isnull}}

\newcommand{\sqlCase}[3]{\mathbf{if}\thickspace #1 \thickspace
  \mathbf{then} \thickspace #2 \thickspace \mathbf{else} \thickspace #3}

\newcommand{\update}[2]{\mathcal{U}_{#1,#2}}
\newcommand{\aupdate}{\update{\pset}{\cond}}
\newcommand{\delete}[1]{\mathcal{D}_{#1}}
\newcommand{\adelete}{\delete{\cond}}
\newcommand{\ins}[1]{\mathcal{I}_{#1}}
\newcommand{\ainsert}{\ins{t}}
\newcommand{\aqinsert}{\ins{\query}}
\newcommand{\up}{u}
\newcommand{\upPos}{pos}

\newcommand{\pos}{\mathsf{Pos}}

\newcommand{\db}{D}
\newcommand{\tup}{t}
\newcommand{\rel}{R}

\newcommand{\dataDomain}{\mathbb{D}}

\newcommand{\dbver}[1]{\db_{#1}}

\newcommand{\ractSymbol}{\mathcal{R}}
\newcommand{\ract}[1]{\ractSymbol_{#1}}
\newcommand{\subract}[2]{\ractSymbol^{#1}_{#2}}

\newcommand{\mul}{\cdot}

\newcommand{\query}{Q}

\newcommand{\hwhatif}{\mathcal{H}}
\newcommand{\ahwhatif}{(\history,\db,\deltaHist)}
\newcommand{\history}{H}
\newcommand{\histpre}[1]{\history_{#1}}

\newcommand{\histslice}[2]{\history_{#1,#2}}
\newcommand{\hsliceOf}[3]{{#1}_{#2,#3}}
\newcommand{\idxs}{\mathcal{I}}
\newcommand{\idx}{i}
\newcommand{\hislice}[1]{\history_{#1}}
\newcommand{\hisliceOf}[2]{{#1}_{#2}}

\newcommand{\modi}{m}
\newcommand{\mdel}[1]{\ensuremath{\mathbf{del}(#1)}\xspace}
\newcommand{\minsert}[2]{\ensuremath{\mathbf{ins}_{#2}(#1)}\xspace}

\newcommand{\deltaHist}{\mathcal{M}}
\newcommand{\hmod}[2]{#1[#2]}
\newcommand{\ahmod}{\hmod{\history}{\deltaHist}}

\newcommand{\diffsym}{\Delta}
\newcommand{\iDiff}[2]{\diffsym(#1,#2)}

\newcommand{\pset}{\textit{Set}}
\newcommand{\psetOf}[1]{\pset_{#1}}
\newcommand{\attr}[1]{A_{#1}}

\newcommand{\cond}{\theta}
\newcommand{\condOf}[1]{\theta_{#1}}
\newcommand{\expr}{e}
\newcommand{\subst}[3]{{#1}[#2 \gets #3]}
\newcommand{\cons}{c}
\newcommand{\condbf}{\boldsymbol{\phi}}
\newcommand{\exprbf}{\bf{e}}

\newcommand{\varbf}{\bf{v}}

\newcommand{\hslice}[2]{{#1}_{#2}}

\newcommand{\exclusion}[3]{\zeta\left({#1}, {#2}, {#3}\right)}
\newcommand{\slicetest}[3]{\zeta({#1},{#2},{#3})}
\newcommand{\aslicetest}{\slicetest{\hwhatif}{\idxs}{\adbconstr}}

\newcommand{\qDSh}{\query_{\history}^{DS}}
\newcommand{\qDSm}{\query_{\history[\deltaHist]}^{DS}}
\newcommand{\condDS}[2]{\cond_{#1}^{DS}(#2)}
\newcommand{\condDSh}[1]{\condDS{\history}{#1}}
\newcommand{\condDSm}[1]{\condDS{\history[\deltaHist]}{#1}}

\newcommand{\pushCond}[2]{{#1}\downarrow^{#2}}
\newcommand{\qpushCond}[2]{(#1)\downarrow^{#2}}

\newcommand{\mcondDS}[3]{\cond_{#1}^{DS}[#3](#2)}
\newcommand{\mcondDSh}[2]{\mcondDS{\history}{#1}{#2}}

\newcommand{\mpushCond}[3]{{#1}[#3]\downarrow^{#2}}
\newcommand{\mqpushCond}[3]{(#1)[#3]\downarrow^{#2}}

\newcommand{\world}{\db}
\newcommand{\worlds}{\mathcal{D}}

\newcommand{\varDom}{\Sigma}
\newcommand{\varAssign}{\lambda}
\newcommand{\allVarAssigns}{\Lambda}
\newcommand{\vcdb}{\mathbf{D}}
\newcommand{\vcrel}{\mathbf{R}}
\newcommand{\vct}{\mathbf{t}}
\newcommand{\singtupH}[1]{\vct_{#1}}
\newcommand{\vcOf}[1]{\mathbf{#1}}
\newcommand{\vctn}{\vcOf{t_{new}}}
\newcommand{\vcdbini}{\vcOf{\db_0}}

\newcommand{\lcond}{\phi}
\newcommand{\gcond}{\Phi}
\newcommand{\dbconstr}[1]{\gcond_{#1}}
\newcommand{\adbconstr}{\dbconstr{\db}}
\newcommand{\worldsOf}[1]{Mod({#1})}

\newcommand{\upBound}{M}
\newcommand{\var}{v}
\newcommand{\bvar}{b}

\newcommand{\thead}[1]{{\cellcolor{black}{\textcolor{white}{\textbf{#1}}}}}

    \makeatletter
     \newbox\sf@box
       {\def\sf@one{#1}%
        \def\sf@two{#2}%
        \setbox\sf@box\hbox
          \bgroup}%
       {  \egroup
        \ifx\@empty\sf@two\@empty\relax
          \def\sf@two{\@empty}
        \fi
        \ifx\@empty\sf@one\@empty\relax
          \subfloat[\sf@two]{\box\sf@box}%
        \else
          \subfloat[\sf@one][\sf@two]{\box\sf@box}%
        \fi}
     \makeatother

\newcommand{\eat}[1]{}
{%
\noindent \textsc{Proof Sketch.}%
}%
{\qedsymbol}%

\newcommand{\abbrHW}{HWQ\xspace}
\newcommand{\abbrHWs}{HWQs\xspace}

\newbool{Techreport}
\newrobustcmd{\ifnottechreport}[1]{\ifbool{Techreport}{}{#1}}
\newrobustcmd{\iftechreport}[1]{\ifbool{Techreport}{#1}{}}

\booltrue{Techreport}

\graphicspath{{imgs/}}

\begin{document}

\title{Reenactment for Predictive Analytics or Whatif Queries}
\title{Answering Historical What-if Queries with Provenance, Reenactment, and Symbolic Execution}
\title{SAHIF: A System for Answering Historical What-if Queries}
\title{MAHIF: A Middleware for Answering Historical What-if Queries}
\title{Efficient Answering of Historical What-if Queries}

\author{Felix S. Campbell}
\affiliation{%
  \institution{Illinois Institute of Technology}
  \country{USA}
  \city{Chicago}
}
\email{fcampbell@hawk.iit.edu}

\author{Bahareh Sadat Arab}
\affiliation{%
  \institution{Illinois Institute of Technology}
  \country{USA}
  \city{Chicago}
}
\email{barab@hawk.iit.edu}

\author{Boris Glavic}
\affiliation{%
  \institution{Illinois Institute of Technology}
  \country{USA}
  \city{Chicago}
}
\email{bglavic@iit.edu}

\renewcommand{\shortauthors}{Campbell, et al.}



\definecolor{lstpurple}{rgb}{0.5,0,0.5}
\definecolor{lstred}{rgb}{1,0,0}
\definecolor{lstreddark}{rgb}{0.7,0,0}
\definecolor{lstredl}{rgb}{0.64,0.08,0.08}
\definecolor{lstmildblue}{rgb}{0.66,0.72,0.78}
\definecolor{lstblue}{rgb}{0,0,1}
\definecolor{lstmildgreen}{rgb}{0.42,0.53,0.39}
\definecolor{lstgreen}{rgb}{0,0.5,0}
\definecolor{lstorangedark}{rgb}{0.6,0.3,0}	
\definecolor{lstorange}{rgb}{0.75,0.52,0.005}
\definecolor{lstorangelight}{rgb}{0.89,0.81,0.67}
\definecolor{lstbeige}{rgb}{0.90,0.86,0.45}

\DeclareFontShape{OT1}{cmtt}{bx}{n}{<5><6><7><8><9><10><10.95><12><14.4><17.28><20.74><24.88>cmttb10}{}

\lstdefinestyle{psql}
{
tabsize=2,
basicstyle=\small\upshape\ttfamily,
language=SQL,
morekeywords={PROVENANCE,BASERELATION,INFLUENCE,COPY,ON,TRANSPROV,TRANSSQL,TRANSXML,CONTRIBUTION,COMPLETE,TRANSITIVE,NONTRANSITIVE,EXPLAIN,SQLTEXT,GRAPH,IS,ANNOT,THIS,XSLT,MAPPROV,cxpath,OF,TRANSACTION,SERIALIZABLE,COMMITTED,INSERT,INTO,WITH,SCN,UPDATED},
extendedchars=false,
keywordstyle=\bfseries,
mathescape=true,
escapechar=@,
sensitive=true
}

\lstdefinestyle{psqlcolor}
{
tabsize=2,
basicstyle=\small\upshape\ttfamily,
language=SQL,
morekeywords={PROVENANCE,BASERELATION,INFLUENCE,COPY,ON,TRANSPROV,TRANSSQL,TRANSXML,CONTRIBUTION,COMPLETE,TRANSITIVE,NONTRANSITIVE,EXPLAIN,SQLTEXT,GRAPH,IS,ANNOT,THIS,XSLT,MAPPROV,cxpath,OF,TRANSACTION,SERIALIZABLE,COMMITTED,INSERT,INTO,WITH,SCN,UPDATED},
extendedchars=false,
keywordstyle=\bfseries\color{lstpurple},
deletekeywords={count,min,max,avg,sum},
keywords=[2]{count,min,max,avg,sum},
keywordstyle=[2]\color{lstblue},
stringstyle=\color{lstreddark},
commentstyle=\color{lstgreen},
mathescape=true,
escapechar=@,
sensitive=true
}

\lstdefinestyle{datalog}
{
basicstyle=\footnotesize\upshape\ttfamily,
language=prolog
}

\lstdefinestyle{pseudocode}
{
  tabsize=3,
  basicstyle=\small,
  language=c,
  morekeywords={if,else,foreach,case,return,in,or},
  extendedchars=true,
  mathescape=true,
  literate={:=}{{$\gets$}}1 {<=}{{$\leq$}}1 {!=}{{$\neq$}}1 {append}{{$\listconcat$}}1 {calP}{{$\cal P$}}{2},
  keywordstyle=\color{lstpurple},
  escapechar=&,
  numbers=left,
  numberstyle=\color{lstgreen}\small\bfseries, 
  stepnumber=1, 
  numbersep=5pt,
}

\lstdefinestyle{xmlstyle}
{
  tabsize=3,
  basicstyle=\small\upshape\ttfamily,
  language=xml,
  extendedchars=true,
  mathescape=true,
  escapechar=£,
  tagstyle=\bfseries,
  usekeywordsintag=true,
  morekeywords={alias,name,id},
  keywordstyle=\color{lstred}
}

\lstdefinestyle{xmlstyle-color}
{
  tabsize=3,
  basicstyle=\small\upshape\ttfamily,
  language=xml,
  extendedchars=true,
  mathescape=true,
  escapechar=£,
  tagstyle=\color{keywordpurple},
  usekeywordsintag=true,
  morekeywords={alias,name,id},
  keywordstyle=\color{lstred}
}

\lstset{style=psqlcolor}

\definecolor{revgreen}{rgb}{0,0.5,0}
\newrobustcmd{\reva}[1]{{#1}}
\newrobustcmd{\revb}[1]{{#1}}
\newrobustcmd{\revc}[1]{{#1}}
\newrobustcmd{\revm}[1]{{#1}}
\newrobustcmd{\revdel}[1]{}

\begin{abstract}
  We introduce \emph{historical what-if queries}, a novel type of what-if analysis that determines the effect of a hypothetical change to the transactional history of a database. For example, \emph{``how would revenue be affected if we would have charged an additional \$6 for shipping?''} %
\iftechreport{Such queries may lead to more actionable insights than traditional what-if queries as their results can be used to inform future actions, e.g., increasing shipping fees.}
We develop efficient techniques for answering historical what-if queries, i.e., determining how  a modified history affects the current database state. Our techniques are based on \emph{reenactment}, a %
 replay technique for transactional histories.
We optimize this process using
 program and data slicing techniques  that determine which updates and what data can be excluded from reenactment without affecting the result. Using an implementation of our techniques in \emph{Mahif} (a Middleware for Answering Historical what-IF queries) we demonstrate their effectiveness experimentally. %

\end{abstract}

\maketitle

\section{Introduction}
\label{sec:introduction}
What-if analysis~\cite{hung17,deutch13} determines how a hypothetical update to a database instance affects the result of a query.
Consider the following what-if query: \textit{``How would a 10\% increase in sales  affect our company’s revenue this year?''}
While the result of this query can help an analyst to understand how revenue is affected by sales,
its practical utility is limited because it does not provide any insights about how this increase in sales could have been achieved in the first place.
We argue that this problem is not specific to this example, but rather is a fundamental issue with classical what-if analysis since the hypothetical update to the database is part of the input.
We propose \emph{historical what-if queries} (\emph{\abbrHW}), a novel type of what-if queries where the user postulates a hypothetical change to the transactional history of the database.

\begin{figure}[t]
  \centering

    \resizebox{1\columnwidth}{!}{
      \begin{minipage}{1.5\columnwidth}
        \centering
        {\bf\large Order}\\[1mm]
    \begin{tabular}{|c|c|c|c|c|l}
      \thead{ID} & \thead{Customer} & \thead{Country} & \thead{Price} & \thead{ShippingFee} & \\ \cline{1-5}
                                                          11 & Susan  & UK &  20 &  5 & $o_1$ \\
      12 & Alex  & UK &  50 & 5 & $o_2$ \\
                                                          13 & Jack  & US &  60 & 3 & $o_3$ \\
                                                          14 & Mark  & US &  30 & 4 & $o_4$ \\ \cline{1-5}
    \end{tabular}
  \end{minipage}
} \vspace{-3mm}
  \caption{Running example database instance.}
  \label{fig:running-example-instance}
  \vspace{-4mm}
\end{figure}
\begin{figure}[t]
  \resizebox{1\columnwidth}{!}{
    \begin{minipage}{1.5\linewidth}
      \centering
      \begin{tabular}{|l|l|c|}
        \hline
        \thead{U}                          & \multicolumn{1}{c|}{\thead{SQL}}                                                              \\ \hline
        \rowcolor{shadeblue}     $u_1$     & \lstinline! UPDATE Order SET ShippingFee=0  WHERE Price>=50;!                                \\  [1mm]
        \rowcolor{pink}         ${u_1}'$   & \lstinline! UPDATE Order SET ShippingFee=0  WHERE Price>=60;!                                \\ [1mm]
        $u_2$                              & \lstinline!  UPDATE Order SET ShippingFee=ShippingFee+5 WHERE Country='UK' AND Price <=100;!  \\[1mm]
        \rowcolor{shadeblue}        $u_3$  & \lstinline! UPDATE Order SET ShippingFee=ShippingFee-2 WHERE Price <=30 AND ShippingFee>=10;! \\ [1mm]
        \hline
      \end{tabular}
    \end{minipage}
  }                                                                                                                                       \\[-3mm]
  \caption{History $\history$ implementing the shipping fee policy and a hypothetical change of the policy (update ${u_1}'$ replaces $u_1$ to raise the price for waiving shipping fees to \$60).}
  \label{fig:Transitive-Transactions-Example}
\end{figure}
\begin{figure}[t]
  $\,$                                                                                                                                                   \\[-5mm]
  \centering
  \begin{minipage}{1\linewidth}
    \centering
    \resizebox{1\columnwidth}{!}{
      \begin{minipage}{1.5\columnwidth}
        \centering {\large \bf Order}                                                                                                                      \\[2mm]
        \begin{tabular}{|c|c|c|c|c|l}
          \thead{ID}                                                                                                                                                                                                                         & \thead{Customer} & \thead{Country} & \thead{Price} & \thead{ShippingFee} & \\ \cline{1-5}
                                                                  11                                                                                                                                                                         & Susan            & UK              & 20            & 8                   & $o_5$                                                        \\
                                                                                                                          12                                                                                                                 & Alex             & UK              & 50            & 5                   & $o_6$ \\
                                                                                                                                                                                  13                                                         & Jack             & US              & 60            & 0                   & $o_7$                                                          \\
                                                                                                                                                                                                                                          14 & Mark             & US              & 30            & 4                   & $o_8$                                                           \\ \cline{1-5}
        \end{tabular}
      \end{minipage}
    }                                                                                                                                                     \\[-3mm]
    \caption{Result of  executing the original history $\history$.}
    \label{fig:updated-example-instance}
  \end{minipage}
\end{figure}
\begin{figure}[t]
  $\,$                                                                                                                                                   \\[-5mm]
  \centering
  \begin{minipage}{1\linewidth}
    \centering
    \resizebox{1\columnwidth}{!}{
      \begin{minipage}{1.5\columnwidth}
        \centering {\large \bf Order}                                                                                                                      \\[2mm]
        \begin{tabular}{|c|c|c|c|c|l}
          \thead{ID}                                                                                                                                                                                                                         & \thead{Customer} & \thead{Country} & \thead{Price} & \thead{ShippingFee}    & \\ \cline{1-5}
                                                                  11                                                                                                                                                                         & Susan            & UK              & 20            & 8                      & $o_5$ \\
                                                                                                                          12                                                                                                                 & Alex             & UK              & 50            & \cellcolor{shadered}10 & $o_6'$ \\
                                                                                                                                                                                  13                                                         & Jack             & US              & 60            & 0                      & $o_7$                                                          \\
                                                                                                                                                                                                                                          14 & Mark             & US              & 30            & 4                      & $o_8$                                                           \\ \cline{1-5}
        \end{tabular}
      \end{minipage}
    }                                                                                                                                                     \\[-3mm]
    \caption{Result of executing the hypothetical history $\ahmod$.}
    \label{fig:whatif-example-instance}
  \end{minipage}
\end{figure}

\begin{exam}
  \label{ex:running-example}

  Consider an online retailer that has developed a new shipping fees policy. An example database instance is shown in \Cref{fig:running-example-instance}.
\iftechreport{
  The new policy was implemented by updating the shipping fees for existing orders as follows: %
the fee for orders with price equal or greater than \$50 was set to \$0, orders of less than or equal to \$100 with a destination in the UK were charged an additional \$5 shipping fee, and orders with a  price equal or less than \$30 and shipping fee equal or more than \$10 received a \$2 discount for their shipping fee.} \Cref{fig:Transitive-Transactions-Example} shows a transactional history with three updates $\up_1$, $\up_2$ and $\up_3$ that implement this policy which %
resulted in the database state shown in \Cref{fig:updated-example-instance}. %
For example, $\up_1$ waives shipping fees for orders of at least \$40.
Bob, an analyst, wants to understand how a larger order price threshold for waiving shipping fees, say \$60, would have affected revenue.
Bob's request can be expressed as a \emph{historical what-if query} which replaces the update $\up_1$ with update ${\up_1}'$ (highlighted in red in \Cref{fig:Transitive-Transactions-Example}). \Cref{fig:whatif-example-instance} shows the new state of the database after executing the modified transactional history over the database from \Cref{fig:running-example-instance}. %
The hypothetical change results in an increase of the shipping fee for the record with ID 12 (highlighted in red).
By evaluating the effect of changing a past action (an update) instead of changing the current state of the database as in classical what-if analysis,
the answer to a historical what-if query can inform future actions. For example, if revenue is increased significantly by using a \$60 cutoff for waiving shipping fees, then we may apply this higher threshold in the future.
\end{exam}

In this paper, we study how to efficiently answer
historical what-if queries (\abbrHWs) such as the one from \Cref{ex:running-example}.
A \abbrHW $\hwhatif$ is a triple $(\history, \db, \deltaHist)$ where $\history$ is a transactional history (a sequence of insert/update/delete statements), $\db$ is the state of the database before the execution of the transactional history $\history$, and $\deltaHist$ is a set of modifications to the history, i.e., it replaces some updates from $\history$ with hypothetical updates (or inserts new / deletes existing update statements). We use $\ahmod$ to denote the history that is the result of applying $\deltaHist$ to $\history$. The result of $\hwhatif$ is the symmetric difference ($\Delta$) of the database instances produced by evaluating $\history$ ($\history[\deltaHist$]) over database $\db$, i.e., the set of tuples in the result of the history that are affected by the modification. For our running example, the symmetric difference  would contain the two versions of the tuple with ID 12 produced by the original and modified history. We focus on deterministic updates (given the same input, multiple executions of an update are guaranteed to return the same result).
The existence of an update in a transactional history is often dependent on the existence of other updates in the history and/or on external events (e.g., user interactions) which are not observed by the DBMS. For instance, if we delete a statement that inserted a customer, then this customer could have never submitted any orders. Consequently, all insert statements corresponding to orders by this customer should be removed. While dealing with such causal relationships is important for helping users to formulate realistic hypothetical scenarios, it is orthogonal to the problem we study in this work: how to efficiently answer \abbrHWs. Learning such causal relationships between the updates of a history %
  and then using them to augment a user-provided \abbrHW is an interesting and challenging problem that we leave to future work.

\begin{figure}[t]
  \begin{minipage}{1.0\linewidth}
    \centering
    \includegraphics[width=0.9\linewidth]{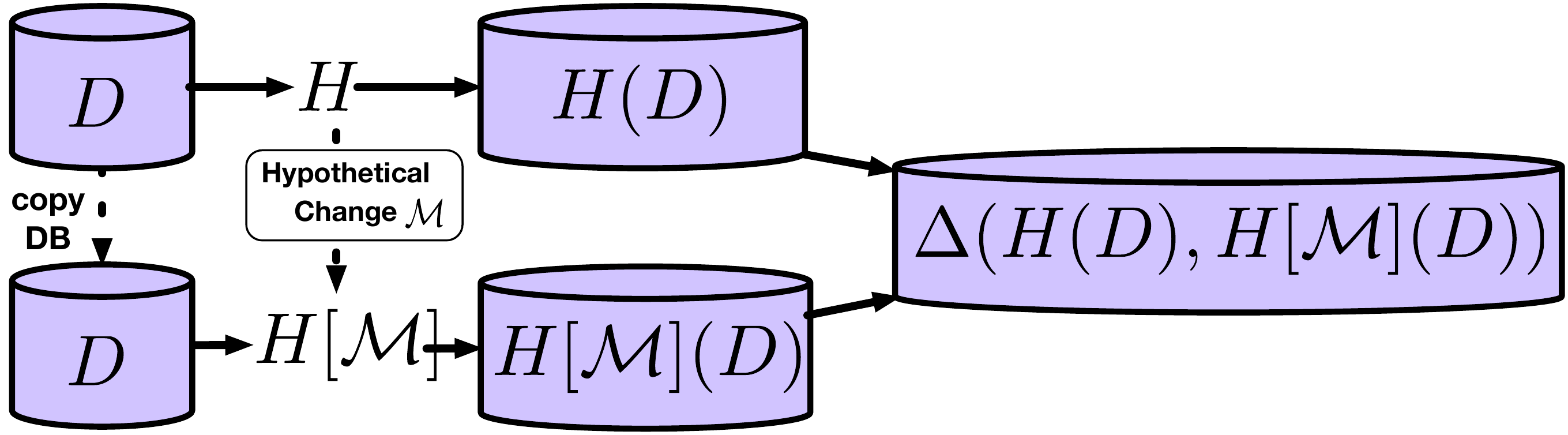}
    \vspace{-4.5mm}
    \caption{The naïve method requires evaluating the modified history over a copy of the original database.}
    \label{fig:Naive-Method}
  \end{minipage}
  \begin{minipage}{1.0\linewidth}
    \centering
    \includegraphics[width=0.9\linewidth]{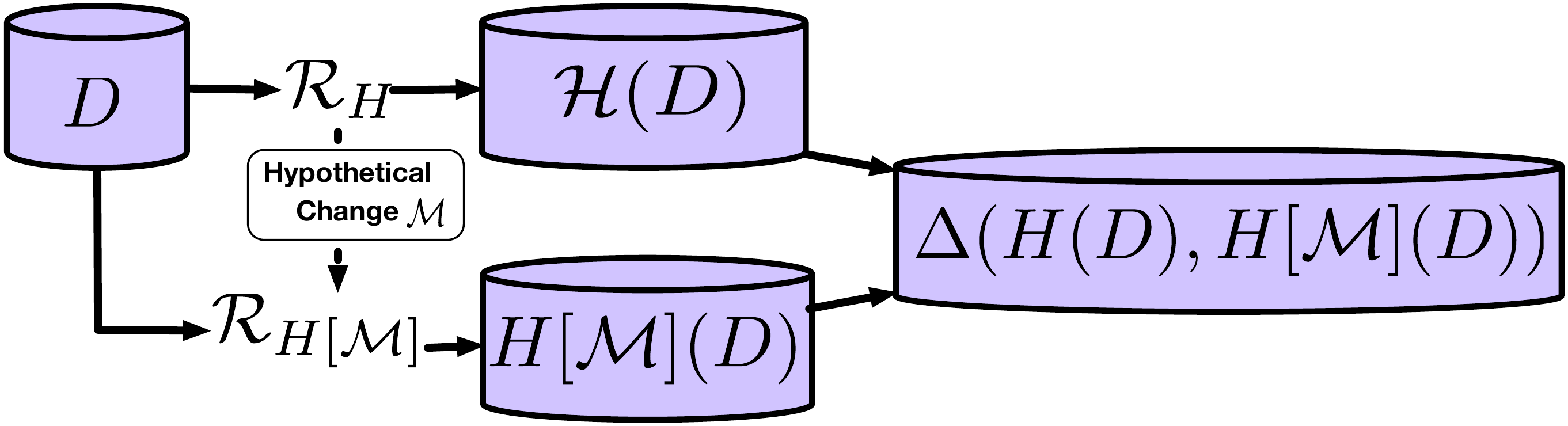}
    \vspace{-4.5mm}
    \caption{Reenactment-based method implemented in Mahif}
    \label{fig:Prop-Method}
  \end{minipage}
\end{figure}

A \textbf{naïve approach} for answering a \abbrHW is shown in \Cref{fig:Naive-Method}. This method creates a copy of the database as it was before the execution of the first update that has been modified by $\deltaHist$, and then executes the modified history on this copy. %
It then computes the symmetric difference between the current database state (which is the result of evaluating the original transactional history $\history$ over $\db$) and the database state that is the result of evaluating the modified history $\ahmod$ over the copy of database $\db$.  %
Note that this requires access to a past database state $\db$ before the execution of the first update of the history, e.g., we can use a DBMS with support for \textit{time travel} to access $\db$ (e.g., Oracle, SQLsever, DB2).
The naïve method requires additional storage to store the copy of $\db$ and the evaluation of the modified history results in a large amount of write I/O. %
However, an even larger concern is that the modifications  $\deltaHist$ may only affect a small fraction of the data and many updates in the history may be irrelevant for computing the symmetric difference. %

Our \textbf{proposed method} is shown in \Cref{fig:Prop-Method}. In order to overcome the limitations of the naïve method, we propose Mahif as a system that answers \abbrHWs using reenactment~\cite{AG14,AG17,AG18}%
, a declarative technique for replaying transactional histories using queries.
Our approach also uses time travel to access $\db$, the state of the database just before the time the first modified update was executed. In contrast to the naïve method, the database does not need to be copied. %
Instead, the modified history is reenacted over $\db$ by running a query $\ract{\ahmod}$.
Thus, reenactment has the advantage of not incurring write I/O.
The result of query $\ract{\ahmod}$ is equal to the result of executing $\ahmod$ over $\db$.  %
We then compute the symmetric difference between the result of the modified history (returned by $\ract{\ahmod}$) and the current database state ($\history(\db)$) computed by reenacting $\history$ over $\db$. Reenacting $\history$, while seemingly redundant, allows us to develop novel optimizations which %
exclude irrelevant updates from the history and irrelevant data from reenactment. %

\partitle{Program Slicing}
To be able to identify updates that can safely be excluded from the evaluation of an \abbrHW, we introduce the notion of a \emph{slice}. A slice for a \abbrHW $\hwhatif$ is a subset of the updates from $\history$ and $\ahmod$ that is sufficient for computing the result of $\hwhatif$. We identify a property called tuple-independence which holds for a large class of updates (corresponding to SQL update and delete statements without joins and subqueries, and \lstinline!INSERT ... VALUES ...! statements). Tuple independence ensures that we can determine whether a subset of updates is a slice by testing for each individual tuple from the database whether the subset produces the same result for $\hwhatif$ than for the full histories. To improve the efficiency of slicing, we compress $\db$ into a set of constraints that compactly over-approximate the database. Inspired by program slicing and symbolic execution techniques~\cite{bucur14,luckow14}, and ideas from incomplete databases~\cite{AG85,IL84a}, we develop a technique that evaluates updates from a history over a single tuple symbolic instance (a tuple with variables as attribute values) subject to the constraints from the compressed database. The result of symbolic evaluation is a single tuple symbolic instance that encodes all possible tuples in the result of the history for any input tuple fulfilling the compressed database constraints. We then use a constraint solver %
  to determine whether a candidate slice produces the same result for $\hwhatif$ as the full histories for every possible input tuple. If that is the case, then it is safe to use the slice instead of $\history$ and $\ahmod$ to answer $\hwhatif$.
The cost of program slicing only depends on the number of updates in the history and the size of the constraints encoding the data distribution of the database. %

\partitle{Data Slicing}
We also propose \emph{data slicing} to prune data that we can prove is irrelevant for computing the answer to a \abbrHW. Based on the observation that any tuple in the symmetric difference has to be affected by at least one statement that was modified by $\deltaHist$, we filter the input of reenactment to remove tuples which are guaranteed to not be affected by any update modified by $\deltaHist$. In addition to the class of queries supported by program slicing, data slicing is also applicable to insert statements with queries (\lstinline!INSERT ... SELECT! in SQL).
The main contributions of this paper are:
\begin{itemize}[noitemsep,topsep=0pt,parsep=0pt,partopsep=0pt,leftmargin=*]
\item
We formalize historical what-if queries and present a novel method for answering such queries based on reenactment.

\item We present two optimization techniques, \emph{program slicing} and \emph{data slicing}, which determine which updates and what data can be safely excluded when answering a \abbrHW.

\item We demonstrate experimentally that our approach outperforms the naïve approach and that our optimizations result in significant additional performance improvements. %

\end{itemize}

\section{Background and Notation}
\label{sec:background}
\begin{figure}[t]
\centering
\begin{minipage}{1.0\linewidth}
\begin{align*}
\exprbf                                & := \varbf | \cons | \exprbf \lbrace +,-,\times,\div\rbrace \exprbf | \sqlCase{\condbf}{\exprbf}{\exprbf} \\
\condbf                                & := \exprbf \lbrace =,\neq,<,\leq,>,\geq \rbrace \exprbf | \condbf \lbrace \wedge ,\vee \rbrace \condbf | \exprbf \isnull | \neg \condbf | \T | \F
\end{align*}
\end{minipage}\\[-5mm]
\caption{Syntax of expressions $\exprbf$ and conditions $\condbf$}
  \label{fig:expr-grammar}
\end{figure}
\iftechreport{
\begin{figure}[t]
\centering
\begin{align*}
\exprbf + \exprbf'                     & = \exprbf' + \exprbf & \exprbf \times \exprbf' & = \exprbf' \times \exprbf \tag{commutativity}
\end{align*}
\begin{align*}
\begin{split}
\exprbf + (\exprbf' + \exprbf'')       & = (\exprbf + \exprbf') + \exprbf''\\
\exprbf \mul (\exprbf' \mul \exprbf'') & = (\exprbf \mul \exprbf') \mul \exprbf''
  \end{split}  \tag{associtivity}
\end{align*}\\[-5mm]
  \caption{Equivalence rules for expressions \exprbf}
  \label{fig:equ-for-expr}
\end{figure}
}

Given a universal value domain $\dataDomain$, a relation $\rel$ (instance) of %
arity $n$ is a subset of $\dataDomain^{n}$. A database instance (or database for short) $\db$ %
is a set of relations $\rel_1$ to $\rel_n$. We use $\schema{\rel}$ to denote the schema of relation $\rel$. %
We consider three type of update operations: updates, inserts, and deletes. In the following, we will use the term \textit{update statement}, or statement for short, as an umbrella term for updates, deletes, and inserts. We view statements as functions that take a relation $\rel$ (or database in the case of inserts with a query) as input and return an updated version of $\rel$. We use $\up$ to denote any such statement and use $\up(\rel)$ (and sometimes abusing notation also $\up(\db)$) to denote the result of applying statement $\up$ to relation $\rel$. An insert $\ainsert(\rel)$ inserts tuple $t$ with the same arity as $\rel$ into relation $\rel$. An insert $\aqinsert(\rel)$ inserts the result of the query $\query$ evaluated over database $\db$ into $\rel$. A delete $\adelete(\rel)$ removes all tuples from $\rel$ that do not fulfill condition $\cond$. Finally, an update $\aupdate(\rel)$ updates the values of each tuple $t$ that fulfills condition $\cond$ based on a list of expressions $\pset$ and returns all other input tuples unmodified. $\pset$ is a list of expressions $(\expr_1,...,\expr_n)$ with the same arity as $\rel$. Each such expression is over the schema of $\rel$. We will sometimes use $(\attr{i_1}\leftarrow\expr_1,...,\attr{i_m}\leftarrow\expr_m)$ as a notional shortcut assuming that the expression for each attribute that is not explicitly mentioned is the identity. For instance, $\pset = (B \leftarrow B + 3)$ over schema $(A,B,C)$ denotes $(A, B + 3, C)$. For an update or delete $\up$ we use $\condOf{\up}$ to denote the update's (delete's) condition. Similarly, $\psetOf{\up}$ for an update $\up$ denotes the update's list of $\pset$ expressions.

A condition $\cond$ (as used in updates and deletions) is a Boolean expression over %
comparisons between scalar expressions containing variables and constants. The grammar defining the syntax of $\pset$ and $\cond$ expressions is shown in \Cref{fig:expr-grammar}. For any expression $\expr$, $\expr'$, and $\expr''$ we use $\subst{\expr}{\expr'}{\expr''}$ to denote the result of substituting each occurrence of $\expr'$ in $\expr$ with $\expr''$. We write $\pset(t)$ to denote the tuple produced by evaluating the expressions from $\pset$ over input tuple $t$ (required to be of the same arty as $\pset$). For example, for a relation $R(A, B, C)$, tuple $t = (1, 1, 1)$, and  $\pset = (A, A + B, 20)$ we get $\pset(t) = (1, 2, 20)$. %
Sometimes, we will us $\up(t)$ to denote the tuple that is the result of applying a statement $\up$ to a single tuple $t$.
We formally define the semantics of evaluating statements over a database $\db$ below. Note that the update statements we define here correspond to SQL update and delete statements without nested subqueries and joins and to \lstinline!INSERT INTO ... VALUES ...! and \lstinline!INSERT INTO ... SELECT ...!.
\begin{align}
  \aupdate(R) &= \{ \pset(t) \mid t \in R \wedge \cond(t) \} \cup \{ t \mid t \in R \wedge \neg \cond(t) \} \label{eq:update-sem}\\
  \adelete(R) &= \{ t \mid t \in R \wedge \neg \cond(t) \} \label{eq:delete-sem}\\
  \ainsert(R) &= R \cup \{ t \} \label{eq:const-insert-sem}\\
  \aqinsert(R) &= R \cup \query(\db) \label{eq:query-insert-sem}
\end{align}

A \textbf{history} $\history = \up_1, \ldots, \up_n$ %
over a database $\db$ is a sequence of updates over $\db$. %
Given a history $\history = \up_1, \ldots,\up_n$, we use $\histslice{i}{j}$ for $i \leq j \in [1,n]$ to denote $u_i, u_{i+1}, \ldots, u_j$. Similarly, $\histpre{i}$, called a prefix of $\history$, denotes $\history_{1,i}$. Furthermore, for a set of indices $\idxs = \{ \idx_1, \ldots, \idx_m \}$ such that $\idx_j < \idx_k$ if $j < k$ and $\idx_j, \idx_k \in [1,n]$, we use $\hislice{\idxs}$ to denote $(\up_{\idx_1}, \ldots, \up_{\idx_m})$.
We use $\history(\db)$ to denote the result of evaluating the history $\history$ over a database instance $\db$ (recursively defined below using the fact that $\histpre{n} = \history$) and will use $\dbver{\idx}$ to denote $\histpre{\idx}(\db)$.
\begin{align*}
  \db_1 &= \up_1(\db) &\db_i &= \up_i(\db_{i-1}) \tag{for $1 < i \leq n$}
\end{align*}

Our program slicing technique relies on a property we call \emph{tuple independence}. Intuitively, statements that fulfill this property process each input tuple individually.

\begin{defi}[Tuple independence]\label{def:tuple-independence}
 A statement $\up$ is \emph{tuple independent} if for every database $\db$, we have $\up(\db) = \bigcup_{t \in \db} \up(\{t\})$
\end{defi}
In SQL, all updates and deletes without nested subqueries or joins and inserts without queries are tuple independent. Thus, all of our statements with the exception of $\aqinsert$ are tuple independent.

\begin{lem}[Tuple independent statements]\label{lem:tuple-independent-operator}
All updates $\aupdate$, deletes $\adelete$, and inserts $\ainsert$ are tuple independent.
\end{lem}
\iftechreport{  \begin{proof}
    WLOG consider a database $\db$ containing tuples $\{s_1, \ldots, s_m\}$ and let $\{t_1, \ldots, t_n\}$ be the instance of the relation $R$ to which a statement is applied to. Note that for any set comprehension $\{ e \mid e \in S \land \psi \}$ where $S = \{e_1, \ldots, e_n\}$ is a set and $\psi$ is a condition over $e$, the following equivalence holds if $\psi$ does not reference $S$:

    \begin{align}
      \label{eq:factor-comprehensions}
      \{ e \mid e \in S \land \psi \} = \bigcup_{e \in S} \{ e \mid \psi \}
    \end{align}

    For deletes, updates, and insert of constant tuples ($\ainsert$), their result only depends on $\rel$ and no other relation in $\db$. Thus, they return $\emptyset$ for any single tuple instance $\{ s_i\}$ unless tuple $s_i$ belongs to $R$ and we trivially have for any statement $\up$ where $\up$ is either an update $\aupdate$, delete $\adelete$, or insert of a constant tuple $\ainsert$:

    \begin{align}
      \label{eq:rel-update-same-as-database-update}
      \bigcup_{t \in R}  \up(\{t\}) = \bigcup_{t \in \db} \up(\{t\})
    \end{align}

    \proofpar{Updates}:
    Consider an update $\aupdate$.
    \begin{align*}
      \aupdate(R) &= \{ \pset(t) \mid t \in R \wedge \cond(t) \} \cup \{ t \mid t \in R \wedge \neg \cond(t) \}\\
                  &= \bigcup_{t \in R} \{ \pset(t) \mid \cond(t) \} \cup
                    \bigcup_{t \in R} \{ t \mid \neg \cond(t) \} \Cref{eq:factor-comprehensions}\\
                  &= \bigcup_{t \in R} \{ \pset(t) \mid \cond(t) \} \cup \{ t \mid \neg \cond(t) \}\\
                  &= \bigcup_{t \in R}  \aupdate(\{t\}) \tag{\Cref{eq:update-sem}}\\
                  &= \bigcup_{t \in \db}  \aupdate(\{t\}) \tag{\Cref{eq:rel-update-same-as-database-update}}
    \end{align*}
    \proofpar{Deletes}
    Consider a delete $\adelete$.
    \begin{align*}
      \adelete(R) &= \{ t \mid t \in R \wedge \neg \cond(t) \} \\
                  &= \bigcup_{t \in R} \{ t \mid \cond(t) \} \tag{\Cref{eq:factor-comprehensions}}\\
                  &= \bigcup_{t \in R}  \adelete(\{t\}) \tag{\Cref{eq:delete-sem}}\\
                  &= \bigcup_{t \in \db}  \adelete(\{t\}) \tag{\Cref{eq:rel-update-same-as-database-update}}
    \end{align*}
    \proofpar{Inserts}
    Consider an insert $\ainsert(R)$.
    \begin{align*}
      \ainsert(R) &= \rel \union \{ t\}\\
                  &= \bigcup_{s \in R} \{ s \} \union \{ t\}\\
                  &= \bigcup_{s \in R} (\{ s \} \union \{ t\})\\
                  &= \bigcup_{s \in R} \ainsert(\{s\}) \tag{\Cref{eq:const-insert-sem}}\\
                  &= \bigcup_{s \in \db}  \ainsert(\{s\}) \tag{\Cref{eq:rel-update-same-as-database-update}}
    \end{align*}
    \proofpar{Inserts with queries}
    Inserts $\aqinsert(R)$ are not tuple independent. As a counterexample, consider $\up = \ins{\projection_{B,B}(R \join_{B=C} S)}(R)$ over $\rel(A,B)$ and $S(C)$ and database instance $R = \{(1,2)\}$ and $S = \{(2)\}$:
    \[
      \up(\db) = \{ (1,2), (2,2) \}
    \]
    while
    \begin{align*}
      \bigcup_{t \in \db} \up(\{t\}) &= \up(R=\{(1,2)\},S=\emptyset) \cup \up(R=\emptyset,S=\{(2)\}) = \{(1,2)\} \cup \emptyset\\
                                       &= \{(1,2)\}
    \end{align*}
    \end{proof}

}

\section{Historical What-if Queries}
\label{sec:whif-def}
We now formally define historical what-if queries. %
Let $\history$ be a history containing an update $\up$.
Historical what-if queries are based on \textbf{modifications} $\modi = \up \gets \up'$ that replace the statement $\up$ in $\history$ with another statement $\up'$, delete the statement $\up$ at position $i$ ($\modi = \mdel{i}$), or insert a new statement $\up$ at position $i$ ($\modi = \minsert{\up}{i}$).
We use $\deltaHist$ to denote a sequence of modifications and $\ahmod$ to denote the result of applying the modifications
$\deltaHist$ to the history $\history$. For example, for a history $\history = u_1, u_2, u_3$ and $\deltaHist = (u_1 \gets u_1', \mdel{3})$ we get $\ahmod = u_1', u_2$. %
Replacing a statement $\up$ with a statement $\up'$ of a different type, e.g., replacing an update with a delete, can be achieved by deleting $\up$ and then inserting $\up'$.

To answer a historical what-if query, we need to compute the difference between the current state of the database, i.e., $\history(\db)$ and the database produced by evaluating the modified history, i.e., $\ahmod(\db)$.
For that we introduce the notion of a database delta.
A \emph{database delta} $\iDiff{\db}{\db'}$ contains all tuples that only occur in $\db$ or in $\db'$. Tuples that exclusively are in $\db'$ are annotated with a $+$ and tuples that exclusively appear in $\db$ are annotated with $-$.  %

  \begin{align*}
    \iDiff{\db}{\db'} &=
\{ +t \mid t \not\in \db \wedge t \in \db' \} \cup \{ -t \mid t \in \db \wedge t \not\in \db' \}
  \end{align*}

  We define a historical what-if query and an answer to such query based on the delta of $\history(\db)$ and $\ahmod(\db)$.

\begin{defi}[Historical What-If Queries]
A \textbf{historical what-if query} $\hwhatif$ is a tuple $(\history, \db, \deltaHist)$ where $\history$ is a  history executed over database instance $\db$, and $\deltaHist$ denotes a sequence of modifications to $\history$ as introduced above. %
The answer to $\hwhatif$ is defined as:
\begin{align*}
  \iDiff{\history(\db)}{\ahmod(\db)}
  \end{align*}
\end{defi}

\begin{exam}
\label{ex:change-prob-example}
Let $\db$ and $\history$ be the database shown in \Cref{fig:running-example-instance} and history shown in \Cref{fig:updated-example-instance}, respectively. Consider the modification $\deltaHist_1 = (\up_1 \gets {\up_1}')$ where $\up_1$ and ${\up_1}'$ are the updates shown in \Cref{fig:Transitive-Transactions-Example}. $\deltaHist_1$ increases the minimum price for waving shipping fees. %
Bob's historical what-if query from this example can be written as $\hwhatif_{Bob} = (\history, \db, \deltaHist_1)$ in our framework.
Evaluating $\history[\deltaHist_1]$ results in the modified database instance shown in \Cref{fig:whatif-example-instance}. For convenience, we have highlighted modified tuple values. The answer of the \abbrHW $\hwhatif_{Bob}$ is
\[
  \iDiff{\history(\db)}{\history[\deltaHist_1](\db)} = \{ -o_6, +o_6'\}
\]
That is, the shipping fee for Alex's order is increased by \$5 because it is no longer eligible for free shipping under the new policy (${\up_1}'$).
\end{exam}

\begin{algorithm}[t]
  \caption{Naïve \abbrHW Algorithm}
  \label{alg:naive-sol}
  \begin{algorithmic}[1]
    \Procedure{Naive-WhatIf}{$\history$, $\db$, $\db_{cur}$, $\deltaHist$}
    \State $\db' \gets \Call{Copy}{\db}$ \label{alg-line:naive-dbzero}
    \State $\db_{mod} \gets \history[\deltaHist](\db')$ \label{alg-line:naive-mod-history}
    \State \Return $\iDiff{\db_{cur}}{\db_{mod}}$ \label{alg-line:naive-sym-diff}
    \EndProcedure
  \end{algorithmic}
\end{algorithm}

\section{Naïve Algorithm}\label{sec:naive-solution}

Before giving an overview of our approach, we briefly revisit the naïve algorithm (\Cref{alg:naive-sol}) in more detail. WLOG assume that $\deltaHist$ modifies the first update in the history (and possibly others). If this is not the case, then we can simply ignore the prefix of the history before the first modified statement and use the state of the database before that statement instead of the database before first statement in the history. The input to the algorithm is the history $\history$, the database state before the first statement of $\history$ was executed ($\db)$, the current state of the database $\db_{current}$ which is assumed to be equal to $\history(\db)$, and the modifications $\deltaHist$ of the historical what-if  query $\hwhatif$. We assume that $\db$ can be accessed using time travel.
The algorithm first creates a copy of $\db'$ of $\db$.  %
Note that we only need to copy relations that are accessed by the history. The state of any relation not accessed by $\history$ will be the same in $\history(\db)$ and $\ahmod(\db')$.  We rename the relations in $\db'$ to avoid name clashes.
We then execute $\ahmod$ over the copy $\db'$  resulting in $\db_{modified} = \ahmod(\db')$ (\Cref{alg-line:naive-mod-history}).
In the last step (\Cref{alg-line:naive-mod-history}), the delta of $\db_{current}$ and $\db_{modified}$ is computed. The delta computation is implemented as a single query for each relation of $\db$ accessed by $\history$. For instance, a relational algebra query computing the delta for a  relation $\rel$ with schema $\schema{\rel} = (A,B)$ is shown below. Note that $+$ and $-$ are constants, i.e., the projections add an additional column storing the annotation of a tuple.

\[
  \projection_{A,B,-}(\rel_{cur} - \rel_{mod}) \union   \projection_{A,B,+}(\rel_{mod} - \rel_{cur})
\]

\begin{algorithm}[t]
  \caption{Optimized, Reenactment-based \abbrHW Algorithm}
  \label{alg:whatif-algo}
  \begin{algorithmic}[1]
    \Procedure{WhatIf}{$\history$, $\db$, $\deltaHist$}
    \State $\idxs \gets \Call{ProgramSlicing}{\history,\ahmod}$ \Comment{Compute Slice $\idxs$} \label{alg-line:opt-ps}
    \State $\ract{\hslice{\history}{\idxs}} \gets \Call{GenReenactmentQuery}{\hslice{\history}{\idxs}}$ \label{alg-line:opt-h-renact}
    \State $\ract{\hslice{\history}{\idxs}}^{DS} \gets \Call{DataSlicing}{\history,\deltaHist,\ract{\hslice{\history}{\idxs}}}$ \label{alg-line:opt-h-ds}
    \State $\ract{\hslice{\ahmod}{\idxs}} \gets \Call{GenReenactmentQuery}{\hslice{\ahmod}{\idxs}}$ \label{alg-line:opt-m-renact}
    \State $\ract{\hslice{\ahmod}{\idxs}}^{DS} \gets \Call{DataSlicing}{\history,\deltaHist,\hslice{\ahmod}{\idxs}}$ \label{alg-line:opt-m-ds}
    \State \Return $\iDiff{\ract{\hslice{\history}{\idxs}}^{DS}}{\ract{\hslice{\ahmod}{\idxs}}^{DS}}$ \label{alg-line:opt-diff}
    \EndProcedure
  \end{algorithmic}
\end{algorithm}

\section{Overview of Our Approach}
\label{sec:overview}

We now give a high-level overview of our \Cref{alg:whatif-algo} for answering a \abbrHW $\hwhatif = (\history, \db, \deltaHist)$.  %
To answer a historical what-if query, we need to compute $\history(\db)$ and $\ahmod(\db)$, and compute the delta of $\history(\db)$ and $\ahmod(\db)$. As mentioned earlier, we utilize a technique called reenactment for this purpose. In the following we first give an overview of reenactment and then discuss how it is applied by our approach. %

\subsection{Reenactment}
\label{sec:reenactment}

Reenactment~\cite{AG18,AG14} is a technique
for simulating a transactional history through queries. For simplicity we limit the discussion to a history $\history$ over a single relation $\rel$ even though our approach supports histories over multiple relations. Using reenactment, we
can construct a query $\ract{\history}$ such that
$\history(\rel) = \ract{\history}(\rel)$\iftechreport{\footnote{\cite{AG18} did prove a
  stronger result, demonstrating equivalence for annotated relations which
  implies equivalence for set and bag semantics as a special case.}}. Reenactment
was originally developed for capturing provenance for transactional workloads
under multiversioning concurrency control protocols. For our purpose, we only
need reenactment for set semantics and introduce a simplified translation for this case.
We use $\ract{\up}$ ($\ract{\history}$) to denote the reenactment query for a single statement $\up$ (history $\history$).

\begin{defi}[Reenactment Queries]\label{def:reenactment-queries}
  Let be a statement $\up$ (update $\aupdate$, delete $\adelete$, insert $\ainsert$, or insert $\aqinsert$) over a relation $\rel$ with schema $(A_1, \ldots, A_n)$ and let $\pset = (\expr_1, \ldots, \expr_n)$. The reenactment query  $\ract{\up}$ for $\up$ is defined as shown below:
  \begin{align*}
    \ract{\aupdate} & \defas \projection_{\sqlCase{\cond}{e_1}{A_1}, \ldots, \sqlCase{\cond}{e_n}{A_n}}(\rel)
  \end{align*}\\[-9mm]
  \begin{align*}
        \ract{\adelete} & \defas \selection_{\neg\, \cond}(\rel) &
    \ract{\ainsert} & \defas \rel \union \{ t \} &
    \ract{\aqinsert} & \defas \rel \union \query
  \end{align*}

Let $\history = (\up_1, \ldots, \up_n)$ be a history. The reenactment query $\ract{\history}$ for $\history$ is constructed from the reenactment queries for $\up_i$ for $i \in \{1,\ldots,n\}$ by substituting the reference to relation $\rel$ in $\ract{\up_i}$ with $\ract{u_{i-1}}$.
\end{defi}

An insert is reenacted as the union between the current state of relation $R$ and the inserted tuple ($\ainsert$) or the result of query $\query$ (for $\aqinsert$). For a delete $\adelete$, we have to remove all tuples fulfilling the condition of the delete. This is achieved by using a selection to only retain tuples that do not fulfill this condition, i.e., we filter based on $\neg \cond$. To reenact an update, we have to update the attribute values of all tuples fulfilling the condition $\cond$ using the expressions $\pset$. All other tuples are just copied from the input. For that, we project on conditional expressions that for each attribute $A_i$ return $\expr_i$ if the tuple fulfills $\cond$ and $A_i$ otherwise.
For a history $\history$ which accesses multiple relations, a separate query,  $\subract{\rel}{\history}$, is constructed for each relation $\rel$ based on all statements from  history $\history$ that access $\rel$.

\begin{exam}\label{ex:reenact-example}
  Consider \Cref{ex:running-example} and let $I,U,C,P$, and $F$ denote attributes \emph{ID}, \emph{Customer}, \emph{Country}, \emph{Price}, and \emph{ShippingFee} of relation \emph{Order} (abbreviated as \emph{O}). The reenactment query $\subract{O}{\history}$ for the history $\history$ from \Cref{fig:Transitive-Transactions-Example} is:

\vspace{-1mm}
\resizebox{1\linewidth}{!}{
\begin{minipage}{1.0\linewidth}
\begin{align*}
  \subract{O}{\history} & = \projection_{I,U,C,P,\sqlCase{P \leq 30 \land F \geq 10}{F-2}{F}}(                                                                                                                                                                                              \\
                        &\hspace{4mm}\projection_{I,U,C,P,\sqlCase{U = UK \land P \leq 30}{F+5}{F}}(                                                                                                                                                                                                   \\
                        &\hspace{4mm} \projection_{I,U,C,P,\sqlCase{P \geq 50}{0}{F}}(O)))\\[-3mm]
\end{align*}
\end{minipage}
}

Recall that $\history[\deltaHist]$ differs from $\history$ in that ${\up_1}'$ replaces $\up_1$ and that the condition of ${\up_1}'$ is $P \geq 60$. Thus,
$\subract{O}{\history[\deltaHist]}$ differs from $\subract{O}{\history}$ in that condition $P \geq 50$ in the first selection is replaced with $P \geq 60$.
\end{exam}

\subsection{Reenacting Historical What-if Queries}
\label{sec:answ-hist-what}

As shown above, we use reenactment to simulate the evaluation of histories. Given the reenactment queries for $\history$ and $\ahmod$, what remains to be done is to compute their delta.
Continuing with our example from above, the result $\iDiff{\subract{O}{\history}(\db)}{\subract{O}{\ahmod}(\db)}$ of $\hwhatif$ is computed as shown below.

\begin{align*}
  \iDiff{\subract{O}{\history}(\db)}{\subract{O}{\history[\deltaHist]}(\db)}&=  \projection_{I,U,C,P,F,-}(\subract{O}{\history}(\db)- \subract{O}{\history[\deltaHist]}(\db))\\
                                                                            &\hspace{4mm}  \union \projection_{I,U,C,P,F,+}(\subract{O}{\history[\deltaHist]}(\db)-\subract{O}{\history}(\db))
\end{align*}

We use \Cref{alg:whatif-algo} to answer historical what-if queries. This algorithm applies two novel optimizations that significantly improve  performance. Program slicing (\Cref{alg-line:opt-ps}, discussed in \Cref{sec:dep-ana})  determines subsets of histories (encoded as a set of positions $\idxs$ called a \emph{slice}) which are sufficient for computing the answer to the what-if query $\hwhatif$. We then generate reenactment queries (\Cref{alg-line:opt-h-renact,alg-line:opt-m-renact})  for the slices of $\history$ and $\ahmod$ according to $\idxs$. Recall that $\hslice{\history}{\idxs}$ denotes the history generated from $\history$ by removing all statements not in $\idxs$. Afterwards (\Cref{alg-line:opt-h-ds,alg-line:opt-m-ds}), we apply our second optimization, data slicing (discussed in \Cref{sec:filter}). Data slicing injects selection conditions into the reenactment query that filter out data that is irrelevant for computing the result of the \abbrHW. The result of data and program slicing is an optimized version of a reenactment query that has to process significantly less data and avoids reenacting updates that are irrelevant for $\hwhatif$. We then calculate the delta of these two queries and return it as the answer for $\hwhatif$ (\Cref{alg-line:opt-diff}).

\section{Data Slicing}
\label{sec:filter}

In this section, we present \textit{data slicing}, a technique which excludes data from reenactment for a \abbrHW $\hwhatif$ without affecting the result. Our technique is based on the observation that any difference between $\history(\db)$ and $\history[\deltaHist](\db)$ has to be caused  by a difference between $\history$ and $\history[\deltaHist]$. Thus, any tuple that is in the result of $\hwhatif$ has to be derived from a tuple that was affected (e.g., fulfills the condition of an update) by a statement affected by $\deltaHist$ in either the original history, the modified history, or both (but in different ways).

For example, in our running example from \Cref{fig:Transitive-Transactions-Example} the original update $\up_1$ and modified update ${\up_1}'$ only modify tuples for which either $Price \geq 50$ or $Price \geq 60$. For instance, the tuple with ID 11 does not fulfill any of these two conditions. Even through this tuple is modified by both histories, the same modifications are applied and, thus, the final result is the same (see \Cref{fig:updated-example-instance} and \Cref{fig:whatif-example-instance}): the shipping fee of this order was changed to \$8. Our data slicing technique determines selection conditions that filter out such tuples. For instance, for our running example we can apply the condition shown below (checking that either $\up_1$ or ${\up_1}'$ may modify the tuple):
  \begin{align*}
    (Price \geq 50) \lor (Price \geq 60)
  \end{align*}
Initially, we will limit the discussion to data slicing for a single modification $\modi = \up \gets \up'$ where $\up$ and $\up'$ are of the same type (e.g., both are updates). We will show how to construct conditions $\condDSh{\modi}$ and $\condDSm{\modi}$ that we apply to filter irrelevant tuples from the inputs of $\ract{\history}$ and $\ract{\history[\deltaHist]}$. As explained above, for a single modification $\up \gets \up'$  we can assume WLOG that $\up$ is the first update in $\history$, because any update before $\up$ can be ignored for reenactment. Afterwards, we extend the technique for multiple modifications and modifications that insert or delete statements (which also covers modifications that replace a statement with a statement of a different type). In the following, we will use $\qDSh$ to denote $\selection_{\condDSh{\modi}}(\rel)$ and $\qDSm$ to denote $\selection_{\condDSm{\modi}}(\rel)$.

\partitle{Updates}
First, consider a modification $\modi = \up \gets \up'$ where both $\up$ and $\up'$ are updates.
Since only tuples that match the condition of an update operation (the operation's \lstinline!WHERE! clause) can be affected by the operation, a conservative overestimation of $\iDiff{\history(\db)}{\history[\deltaHist](\db)}$ is the set of tuples that are derived from tuples affected by $\up$ in the original history or $\up'$ in the modified history.
Thus, the tuples in $\db$ from which such a tuple is derived have to either match the condition of $u$ ($\condOf{u}$) or the condition of $u'$ ($\condOf{u'}$). This means we can filter the input to the reenactment queries using:
\begin{align}
\condDSh{\modi} = \condDSm{\modi} = \condOf{u} \vee \condOf{u'} \label{eq:update-ds-cond}
\end{align}

\partitle{Deletes}
Let us now consider a single modification $\up \gets \up'$ which replaces a delete $\up = \delete{\cond}$ with a delete $\up' = \delete{\cond'}$. For a tuple $\tup \in \rel$ to contribute to $\iDiff{\ract{\history}(\rel)}{\ract{\history[\deltaHist]}(\rel)}$, it has to be deleted by either $\up$ or $\up'$, but not by both (such tuples do not contribute to any result of $\ract{\history}(\rel)$ or $\ract{\history[\deltaHist]}(\rel)$). Thus, we can filter from $\rel$ all tuples that do not fulfill the condition
\begin{align}
\condDSh{\modi} = \condDSm{\modi} = (\cond \land \neg\,\cond') \vee (\neg\,\cond \land \cond')\label{eq:ds-del-cond}
\end{align}
\iftechreport{Note that for any tuple $\tup_{out}$ to be in the result of $\ract{\history}(\rel)$ ($\ract{\history[\deltaHist]}(\rel)$), it has to be the case that the input tuple $\tup$ in $\rel$ it is derived from has to not fulfill the condition of $\up$ ($\up'$), otherwise $\tup$ would have been deleted. That is, for $\history$, any tuple fulfilling $(\cond \land \neg\,\cond')$ will be filtered out by the delete. Similarly, for $\ahmod$, any tuple fulfilling $(\neg\,\cond \land \cond')$ will be deleted. Thus, we can simplify the data slicing conditions from \Cref{eq:ds-del-cond} by removing this redundant test and get:
  \begin{align*}
    \condDSh{\modi} &= \neg\,\condOf{\up} \land \condOf{\up'}\\
    \condDSm{\modi} &= \condOf{\up} \land \neg\,\condOf{\up'}
  \end{align*}
  Furthermore, for any tuple $t$ ``surviving'' the delete of $\history$ ($\ahmod$) we have that $t$ fulfills the condition $\neg\,\condOf{\up}$ ($\neg\, \condOf{\up'}$). This means the conditions can be further simplified:
  \begin{align*}
    \condDSh{\modi} &= \condOf{\up'}\\
    \condDSm{\modi} &= \condOf{\up}
  \end{align*}
}

\partitle{Inserts with Queries}
Recall that an insert $\aqinsert$ is reenacted using the query $\rel \union \query$. Only tuples that are returned by the query $\query$ need to be considered. Thus, if $\aqinsert$ is the only statement that is modified, then it is sufficient to replace $R \union \query$ in the reenactment query with $\query$. However, for multiple modifications, tuples from the LHS of the union of the reenactment query for a statement $\aqinsert$ may be affected by downstream updates modified by $\deltaHist$. Thus, we cannot simply replace $R \union \query$ with $\query$ if $\aqinsert$ is not the first statement in the history that got modified by $\deltaHist$. To deal with this case, we need a condition that selects tuples which may contribute to the result of $\query$. We can achieve this by pushing the selection conditions of $\query$ down to the relations accessed by $\query$. For that we apply standard selection move-around techniques from query optimization. The final result is a selection condition for each input relation of the query. For instance, for $\ins{\selection_{A=5}(R \join_{A=C} S)}(R)$ over relations $R(A,B)$ and $S(C,D)$, the selection can be pushed to both inputs of the join resulting in condition $A=5$ for $R$ and $C=5$ for $S$.

\partitle{Multiple modifications}
Data slicing can also be applied to \abbrHWs with more than one modification. For a tuple to be in the result of the what-if query, it has to be affected by at least one statement $\up$ such that there exists one modification $\modi \in \deltaHist$ with either $\modi = \up \gets \up'$ or $\modi = \up' \gets \up$ for some statement $\up'$. However, we cannot simply use the disjunction of the data slicing conditions $\condDSh{\modi}$ and $\condDSm{\modi}$ we have developed for single modifications to filter the input. To see why this is the case, consider a modification $\modi = \up \gets \up'$ where $\up$ is the $i^{th}$ update in $\history$. The
input of $\up$ ($\up')$ over which the condition of the update is evaluated is the result of $\hislice{i-1}$ (or $\hisliceOf{\history[\deltaHist]}{i-1}$). To be able to derive a selection condition that can be applied to $\rel$, we have to ``push'' the condition for $\up$ down to determine a condition that returns the set of tuples from $\rel$ that contribute to tuples in $\hislice{i-1}$ fulfilling condition $\condDSh{\modi}$ (or $\condDSm{\modi}$).
For that, we iteratively substitute references to attributes in $\condDSh{\modi}$ (or $\condDSm{\modi}$) with the expressions from the previous statement in $\history$ that defines them. For instance, consider a history $\history = (\up_1 = \update{A \gets 3}{C = 5}, \up_2 = \update{B \gets B + 1}{A < 4})$ and modification $\modi = \up_2 \gets \up_2'$ with $\up_2' = \update{B \gets B + 1}{A < 5}$. To push the condition $A < 4$ of $\up_2$, we substitute $A$ with $\sqlCase{C=5}{3}{A}$ and get $(\sqlCase{C=5}{3}{A}) < 4$.

More formally, consider a modification $\modi = \up_i \gets {\up_i}'$ for a history $\history = (\up_1, \ldots, \up_n)$. Let us first consider how to push $\condDSh{\modi}$ (the case for $\condDSm{\modi}$ is symmetric).
We construct $\pushCond{\condDSh{\modi}}{j}$, the version of $\condDSh{\modi}$ pushed down through $j < i$ updates as shown below. We use $\pushCond{\condDSh{\modi}}{\ast}$ to denote $\pushCond{\condDSh{\modi}}{i-1}$, i.e., pushing the condition through all updates of the history before $\up$. Furthermore, we use an operator $\qpushCond{\cond}{\query}$  to push a condition $\cond$ through a query $\query$. 
\begin{align*}
  \pushCond{\condDSh{\modi}}{0}                                                 & = \condDSh{\modi}                            \\
  \pushCond{\condDSh{\modi}}{j+1}                                               & =
                                    \begin{cases}
                                      \subst{\pushCond{\condDSh{\modi}}{j}}{\vec{A}}{\vec{e}} & \mathbf{if}\, \up_{i-j} = \update{\pset}{\cond} \\
                                      \pushCond{\condDSh{\modi}}{j} \lor \qpushCond{\pushCond{\condDSh{\modi}}{j}}{\query} &\mathbf{if}\, \up_{i-j} = \aqinsert\\
                                      \pushCond{\condDSh{\modi}}{j}             & \mathbf{otherwise}                           \\
                                    \end{cases}
\end{align*}
In the above equation, $\vec{A}$ denotes $(A_1, \ldots, A_n)$ and $\vec{e}$ denotes
\[
  (\sqlCase{\cond}{\pset(A_1)}{A_1},\ldots, \sqlCase{\cond}{\pset(A_n)}{A_n})
\]
Furthermore, $e[\vec{A} \gets \vec{e}]$ denotes the result of substituting each reference to $A_i$ in $e$ with $e_i$ (for all $i \in [1,n]$).

\iftechreport{
The operator $\qpushCond{\cond}{\query}$ mentioned above pushes a selection condition through a query $\query$. So far we have assumed for easy of presentation that a history accesses a single relation $R$. We will stick to this restriction for now and define $\qpushCond{\cond}{\query}$ under this assumption. Afterwards, we will discuss how to generalize data slicing to histories that access multiple relations which is often the case for inserts that use queries.

  \begin{align*}
    \qpushCond{\cond}{R} &= \cond\\
    \qpushCond{\cond}{\selection_{\cond'}(\query)} &= \qpushCond{\cond \land \cond'}{\query}\\
    \qpushCond{\cond}{\projection_{\vec{e}}(\query)} &= \qpushCond{\cond[\vec{A} \gets \vec{e}]}{\query}\\
    \qpushCond{\cond}{\query_1 \union \query_2} &= \qpushCond{\cond}{\query_1} \lor \qpushCond{\cond[\schema{\query_1} \gets \schema{\query_2}]}{\query_2}\\
  \end{align*}

\partitle{Data slicing for histories accessing multiple relations}
To generalize data slicing to histories that access multiple relations, we have to generate a separate slicing condition for every relation accessed by the history. For that we extend our push-down rules for conditions. Note that similar to how we deal with inserting and deleting statements from a history and replacing a statement $\up$ with a statement $\up'$ of a different type, modifications that change what relation is modified by a statement can be rewritten into a deletion of the original statement followed by a insertion of the modified statement. In turn these modifications can be rewritten into modifications that replace a statement with a statement of the same type that modifies the same relation using no-op statements. Thus, from now on we only need to consider modifications that replace a statement $\up$ with a statement $\up'$ where both $\up$ and $\up'$ modify the same relation. We use $\mpushCond{\cond}{\history}{\rel}$ to denote the condition generated for relation $\rel$ by pushing condition $\cond$ through history $\history$. Intuitively, statements that modify a relation $S$ can be ignored when computing the condition for a relation $\rel$ if $\rel \neq S$. For inserts with query, we use $\mqpushCond{\cond}{\query}{\rel}$, explained below, to push $\cond$ through the query $\query$ for relation $\rel$. The relation-specific data slicing conditions for updates, deletes, and inserts are shown below. As before we assume a modification $\up \gets \up'$ and use $\condOf{\up}$ to denote the condition of statement  $\up$ if $\up$ is an update or delete.

\begin{itemize}
\item Update $\aupdate(\rel)$:
  \begin{align*}
    \mcondDSh{\modi}{S} = \condDSm{\modi}{S} =
    \begin{cases}
      \condOf{u} \vee \condOf{u'} &\mathbf{if}\, \rel = S\\
      \T & \mathbf{otherwise}\\
    \end{cases}
  \end{align*}
\item Delete $\adelete(\rel)$:
  \begin{align*}
    \condDSh{\modi} &=
                      \begin{cases}
                        \condOf{u'} &\mathbf{if}\, \rel = S\\
                        \T & \mathbf{otherwise}\\
                      \end{cases}\\
    \condDSm{\modi} &=
                      \begin{cases}
                        \condOf{u} &\mathbf{if}\, \rel = S\\
                        \T & \mathbf{otherwise}\\
                      \end{cases}
  \end{align*}
\end{itemize}

Based on these extended definitions, we then define pushing relation-specific conditions through histories as shown in \Cref{fig:pushing-relation-specific}.

\begin{figure*}[t]
  \centering
\begin{align}
  \label{eq:pushing-queries-multi-rel}
  \mpushCond{\condDSh{\modi}}{0}{\rel}                                                 & = \mcondDSh{\modi}{\rel}                            \\
  \mpushCond{\condDSh{\modi}}{j+1}{rel}                                               & =
                                                                                        \begin{cases}
                                                                                          \subst{\mpushCond{\condDSh{\modi}}{j}{\rel}}{\vec{A}}{\vec{e}} & \mathbf{if}\, \up_{i-j} = \update{\pset}{\cond}(\rel) \\
                                                                                          \revm{\mpushCond{\condDSh{\modi}}{j}{\rel} \lor \mqpushCond{\mpushCond{\condDSh{\modi}}{j}{\rel}}{\query}{S}} &\revm{\mathbf{if}\, \up_{i-j} = \aqinsert}\\
                                                                                          \mpushCond{\condDSh{\modi}}{j}{\rel}             & \mathbf{otherwise}                           \\
                                                                                        \end{cases}
\end{align}
  \caption{Pushing relation-specific data slicing conditions}\label{fig:pushing-relation-specific}
\end{figure*}

Note that in the definition of $\mpushCond{\condDSh{\modi}}{j}{\rel}$ we make use of $\mqpushCond{\mpushCond{\condDSh{\modi}}{j}{\rel}}{\query}{S}$ which we define below. %

  \begin{align*}
    \mqpushCond{\cond}{R}{S} &=
                               \begin{cases}
                                 \cond &\mathbf{if}\, R = S\\
                                 \T &\mathbf{otherwise}
                               \end{cases}\\
    \mqpushCond{\cond}{\selection_{\cond'}(\query)}{\rel} &= \mqpushCond{\cond \land \cond'}{\query}{\rel}\\
    \mqpushCond{\cond}{\projection_{\vec{e}}(\query)}{\rel} &= \mqpushCond{\cond[\vec{A} \gets \vec{e}]}{\query}{\rel}\\
    \mqpushCond{\cond}{\query_1 \union \query_2}{\rel} &= \mqpushCond{\cond}{\query_1}{\rel} \lor \mqpushCond{\cond[\schema{\query_1} \gets \schema{\query_2}]}{\query_2}{\rel}\\
  \end{align*}

}

\begin{exam}\label{ex:ds-example}
Consider our running example history and a modification that replaces $u_3$ (reducing shipping fee by \$2 if the shipping fee is at least \$10 and the order price is at most \$30) with ${\up_3}'$ which applies to orders of $\leq \$40$: ${\up_3}' = \update{F \gets F-2}{P \leq 40 \land F \geq 10}$.
The data slicing condition for $u_3$ and ${\up_3}'$ is $(P \leq 30 \land F \geq 10) \vee (P \leq 40 \land F \geq 10)$ which can be simplified to $(P \leq 40 \land F \geq 10)$. To push this condition through $\up_2$, we have to substitute $F$ (the shipping fee) with the conditional update of the shipping fee corresponding to $\up_2$ and get
$(P \leq 40 \land F'' \geq 10)$ for $F'' = \sqlCase{C=UK \land P \leq 100}{F+5}{F}$. We then have to push this condition through $\up_1$. For that we substitute $F$ again, this time with $F' = \sqlCase{P \geq 50}{0}{F}$. The final data slicing condition for both $\history$ and $\ahmod$ and our modification $\modi = \up_3 \gets {\up_3}'$ is:
\begin{align*}
\pushCond{\condDSh{\modi}}{\ast} &= \pushCond{\condDSm{\modi}}{\ast} = (P \leq 40 \land F'' \geq 10)\\[1mm]
F'' &= \sqlCase{C=UK \land P \leq 100}{F'+5}{F'}\\
  F' &= \sqlCase{P \geq 50}{0}{F}
\end{align*}
Evaluating this condition over the database from \Cref{fig:running-example-instance}, only the tuple with ID 11 has a sufficiently low price $P \leq 40$ and fulfills the condition $F'' \geq 10$ ($F = F' = 5$ and $F'' = F'+5 = 10$). Thus, using this slicing condition we can exclude tuples 12, 13, and 14 from reenactment.
\end{exam}

\partitle{Modifications that insert or delete statements}
Recall that we also allow modifications that insert a new statement at position $i$ ($\minsert{\up}{i}$) or delete the statement at position $i$ ($\mdel{i})$. Note that it is possible to insert new statements into a history without changing its semantics as long as these statements do not modify any data, e.g., a delete $\delete{\F}$ that does not delete any tuples. We refer to such operations as \emph{no-ops}. Using no-ops, we can pad the original history at position $i$ for every insert $\minsert{\up}{i}$. We then can rewrite  $\minsert{\up}{i}$ in $\deltaHist$ into a modification $\up_i \gets \up$ where $\up_i$ is a no-op. A deletion $\mdel{i}$ is rewritten into a modification $\up_i \gets {\up_i}'$ where ${\up_i}'$ is a no-op. Thus, the data slicing method explained above is already sufficient for dealing with inserts $\minsert{\up}{i}$ and deletes $\mdel{i}$.

\begin{theo}[Data Slicing]\label{theo:data-slicing}
  Consider a  $\history$ be a sequence of modifications $\deltaHist = (\modi_1, \ldots, \modi_n)$. Let $\qDSh = \selection_{\bigvee_{i=1}^{n} \pushCond{\condDSh{\modi_i}}{\ast}}(\rel)$
and $\qDSm = \selection_{\bigvee_{i=1}^{n} \pushCond{\condDSm{\modi_i}}{\ast}}(\rel)$.
  Then,
\[
  \iDiff{\ract{\history}(\rel)}{\ract{\history[\deltaHist]}(\rel)}
  =
\iDiff{\ract{\history}(\qDSh(\rel))}{\ract{\history[\deltaHist]}(\qDSm(\rel))}
\]
\end{theo}
\iftechreport{
\newcommand{\tupin}{\tup_{in}}
\newcommand{\tupup}{\tup_{up}}
\newcommand{\dfull}{\Delta}
\newcommand{\dslice}{\Delta_{DS}}

\begin{proof}
  We prove the theorem by induction over the number of modifications ($\card{\deltaHist}$).

  \proofpar{\bf Base Case:}
  We consider a history $\history$ with a single modification $\up \gets \up'$ that replaces the first update of $\history$.
  In the following, we use $\dfull$ to denote
  \[
    \iDiff{\ract{\history}(\rel)}{\ract{\history[\deltaHist]}(\rel)}
  \]
  and $\dslice$ to denote
  \[
    \iDiff{\ract{\history}(\qDSh(\rel))}{\ract{\history[\deltaHist]}(\qDSm(\rel))}
    \]
    Note that histories $\history = (\up, \up_2, \ldots, \up_n)$ and $\history[\deltaHist]$ only differ in their first operation ($\up$ or $\up'$). We proof the claim first for updates ($\up = \aupdate$) and then for deletes ($\up = \adelete$).

\proofpar{$\up = \aupdate$:}
For any tuple $\tup$ to be in $\dfull$, there has to exist a tuple $\tupup$ for which either (a)  $\histslice{2}{n}(\{\tupup\}) = \{\tup\}$ or (b) $\hsliceOf{\ahmod}{2}{n}(\{\tupup\}) = \{\tup\}$, i.e., $\tup$ is the result of applying one the histories excluding the first statement ($\up$ or $\up'$), and for which also either (i) $\tupup \in \up(\rel) \land \tupup \not\in \up'(\rel)$ or (ii) $\tupup \not\in \up(\rel) \land \tupup \in \up'(\rel)$ holds. To see why this has to be true, consider the only two remaining cases: for all tuples fulfilling (a) or (b) either (iii) $\tupup \in \up(\rel) \land \tupup \in \up'(\rel)$ or (iv) $\tupup \not\in \up(\rel) \land \tupup \not\in \up'(\rel)$ holds. In case (iii), the same suffix history $(\up_2, \ldots, \up_n)$ is applied to $\tupup$ in both $\history$ and $\history[\deltaHist]$ which means that $\tup \in \history(\rel)$ and $\tup \in \history[\deltaHist](\rel)$ which contradicts $\tup \in \dfull$. Case (iv) also contradicts the assumption that  $\tup \in \dfull$.
  Let us now only consider case (i) since (ii) is symmetric. Consider any tuple $\tupin \in \rel$ such that $\up(\tupin) = \tupup$ (recall that we use $\up(t) = t'$ as a notational shortcut for $\up(\{t\}) = \{t'\}$). We know that $\up'(\tupin) \neq \tupup$, because $\tupup \not\in \up'(\rel)$. This can only be the case if $\tupin$ fulfills the condition of update $\up$ and/or $\up'$, because if $\tupin$ does not fulfill the condition of either update, then both updates return $\tupin$ unmodified and we get $\up(\tupin) = \up(\tupin) = \tupin$ contradicting $\tupup \not\in \up'(\rel)$.

 Recall that $\condOf{\up}$ and $\condOf{\up'}$  are the conditions of $\up$ and $\up'$, respectively.  So far we have established that for any tuple $\tup$ in the result of either $\history$ or $\history[\deltaHist]$ and the tuple $\tupin \in \rel$ it is derived from by either $\history$ or $\history[\deltaHist]$, we have

 \begin{align}
\tup \in \dfull \Rightarrow \tupin \models \condOf{\up} \lor \condOf{\up'}
 \end{align}

Using the equivalence $\psi_1 \Rightarrow \psi_2 \Leftrightarrow \neg \psi_2 \Rightarrow \neg \psi_1$ we get:

 \begin{align}
   \tupin \not\models \condOf{\up} \lor \condOf{\up'} \Rightarrow \tup \not\in \dfull
    \label{eq:update-slicing-cond}
 \end{align}

Since $\query_{DS}$ only filters out tuples $\tupin$ for which $\tupin \not \models \condOf{\up} \lor \condOf{\up'}$,  \Cref{eq:update-slicing-cond} implies that all tuples filtered by $\query_{DS}$ do not contribute to any tuple in $\dfull$. Thus, we get $\dfull = \dslice$ which concludes the proof for this case.

\proofpar{$\up = \adelete$:}
Now consider the case where $\up$ is a delete statement.

\proofpar{$\dfull \subseteq \dslice$:}
  We prove this direction by contradiction. We have two histories $\history$ and $\history[\deltaHist]$ that only differ in the first statement: $\up = \delete{\cond}$ in $\history$ and $\up' = \delete{\cond'}$ in $\history[\deltaHist]$.
  Consider  a tuple $\tup \in   \iDiff{\ract{\history}(\rel)}{\ract{\history[\deltaHist]}(\rel)}$ and WLOG assume the $\tup \in \ract{\history}(\rel)$ (the other case is symmetric). Then there has to exist $\tupin \in \rel$ such that $\ract{\history}(\{\tupin\}) = \{ \tup \}$, i.e., $\tupin$ is in the provenance of $\tup$. For this to be the case $\tupin \models \neg\,\cond$, i.e.,  $\tupin$ does not fulfill the condition $\cond$ of the delete $\up$ (otherwise it would have been deleted), and $\tupin \models \cond'$ (otherwise $\tup$ would not be in $\iDiff{\ract{\history}(\rel)}{\ract{\history[\deltaHist]}(\rel)}$).
  For sake of contradiction assume that $\tup \not \in \iDiff{\ract{\history}(\qDSh(\rel))}{\ract{\history[\deltaHist]}(\qDSm(\rel))}$. Since the two histories only differ in the first statement, this means that $\tupin$ does not fulfill the selection condition of $\qDSh$. Recall that this selection condition is $\cond'$. Thus, we have $\tupin \models \neg \cond'$ which contradicts $\tupin \models \cond'$.

\proofpar{$\dfull \supseteq \dslice$:}
Consider a tuple $\tup \in \dslice$ and as above let $\tupin \in \rel$ denote the tuple it is derived from. We have to show that $\tup \in \dfull$. Since $\tup \in \dslice$ either $\tup \models \cond \land \tup \not\models \cond'$ or $\tup \not\models \cond \land \tup \models \cond$. Since these two cases are symmetric, WLOG assume that $\tup \models \cond \land \tup \not\models \cond$. Note that the only difference between  $\ract{\history}(\rel)$ and $\ract{\history}(\qDSh(\rel))$ is the selection applied by $\qDSh$. Thus, $\tup \in \ract{\history}(\qDSh(\rel))$. Also $\tup \not\in \ract{\history[\deltaHist]}(\qDSm)$ holds, because $\tupin$ is already filtered out by $\up'$ (it fulfills $\cond'$). It follows that $\tup \in \dfull$.

\proofpar{\bf Inductive Step:}
We again use $\dfull$ to denote $\iDiff{\ract{\history}(\rel)}{\ract{\history[\deltaHist]}(\rel)}$ and $\dslice$ to denote $\iDiff{\ract{\history}(\qDSh(\rel))}{\ract{\history[\deltaHist]}(\qDSm(\rel))}$ in the following. For a tuple $t$, let $t_i$ be denote $\history_i(t)$ and $t_i'$ to denote ($\ahmod_i(t)$). Abusing notation, let $\modi$ in subscripts of $\tup$ denote the position of a modification $\modi$ in $\history$.

\proofpar{$\dfull \subseteq \dslice$:}
Let $\tup \in \dfull$. Given that the tuple $\tup \in \dfull$, we will show that this implies that there exists $\modi = \up \gets \up'$ in $\deltaHist$ such that $\cond_\modi$ is the condition of $\up$ and $\cond_{\modi}'$ is the condition of $\up'$ and for which $\cond_\modi(t_{\modi -1}) \vee \cond_{\modi}'(t_{\modi -1}')$. This claim follows from inductive use of the  argument for single modifications from the proof of \Cref{theo:data-slicing-single-mod}. If a tuple $t \in \rel$ is not affected by any of the modified updates (i.e., the successors $t_i$ of $t$ do not fulfill any of the conditions of these statements), then the final result produced by $\history(\tup)$ and $\ahmod(\tup)$ which contradicts $\tup \in \dfull$.

Assume that we have access to an oracle that given an input tuple $\tup$ determines whether any successor of $\tup$ fulfills the condition from above:

\[
  \exists \modi \in \deltaHist: \cond_\modi(t_{\modi -1}) \vee \cond_{\modi}'(t_{\modi -1}')
\]

Then we could filter input tuples using this oracle without changing the result of the historical what-if query. The only problem is that we cannot simply use a selection with condition, because we have only access to $\tup$, but not its successors $t_i$ (the result of applying the first $i$ updates to $t$). In the remainder of the proof we will show how to filter the input based on a pushed down condition is equivalent to applying the condition to a successor. From this then immediately follows the claim.

Since a reenactment query $\ract{\history}$ is equivalent to the history $\history$, it suffices to show that conditions can be pushed through the relational algebra operators used in $\ract{\history}$. We then apply this repetitively for both $\ract{\history}$ and $\ract{\ahmod}$ to yield $\ract{\history}(\qDSh(\rel))$ and $\ract{\ahmod}(\qDSm(\rel))$.
In the following we will abuse notation and for a query $\query$, denote by $\pushCond{\cond}{i}$ the condition $\cond$ pushed through the top-most $i$ operators.
Consider a tuple $\tup = (c_1, \ldots, c_n)$, query $\query$ consisting of a single relational algebra operator, and condition $\cond$ and let $\tup_{out} = \query(\tup)$.
We need to show that
\begin{align}
  \cond(\query(\{\tup\})) = \pushCond{\cond}{1}(\tup)
  \label{eq:push-down-eq}
\end{align}

Showing this equivalency is possible by a case distinction over the possible relational algebra operators:
\begin{itemize}
	\item \textbf{Projection} Consider a projection $\projection_{\vec{A}}$ with $\vec{A} = e_1 \to A_1, \ldots, e_n \to A_n$. Therefore, $\projection_{\vec{A}}(t) = (e_1(c_1), \ldots, e_n(c_n))$. Applying the definition of $\pushCond{\cond}{1}(\tup)$, we get $\pushCond{\cond}{1}(\tup) = (e_1(c_1), \ldots, e_n(c_n)) = \projection_{\vec{A}}(t)$.
	\item \textbf{Selection} Based on the semantics of selection, $\tup_{out} = \tup$. Based on $\pushCond{\cond}{1} = \cond$, it follows that $\pushCond{\cond}{1}(\tup) = \cond(\tup_{out})$.
	\item \textbf{Union} Considering that a union has two inputs, the tuple $\tup$ may be present in the left, the right, or both inputs to the union. However, no matter which input it stems from, $\tup_{out} = \tup$. Since $\pushCond{\cond}{1} = \cond$, we have $\pushCond{\cond}{1}(\tup) = \cond(\tup_{out})$.
\end{itemize}

Applying \Cref{eq:push-down-eq} iteratively, we can push down all conditions to the input table $\rel$ for both reenactment queries $\ract{\history}$ and $\ract{\ahmod}(\rel)$. This condition can then be applied in a selection over $\rel$. Since this is precisely the condition from $\qDSh$ and $\qDSm$, this implies $\dfull \subseteq \dslice$.

\proofpar{$\dfull \supseteq \dslice$} There are three cases where $\dslice$ could contain tuples not in $\dfull$. The first two cases are symmetric in form, and they are the cases where a tuple $\tup$ is filtered to exclusively $H$ ($\ahmod$ symmetrically). The third case is when a tuple might be inserted by $\qDSh(\rel)$ ($\qDSm(\rel)$).
We can eliminate the third case using the following argument. First observe that $\qDSh(\rel) \subseteq \rel$ as it applies a selection over the reenactment. Considering our updates are monotone this implies that $\history(\qDSh(\rel)) \subseteq \history(\rel)$. Using the same argument, we also have $\ahmod(\qDSm(\rel)) \subseteq \ahmod(\rel)$. Thus, no new tuples can appear in $\dslice$.
  Thus we now only need to consider the first two cases, in which the symmetric difference could possibly produce tuples outside of $\dfull$. For the sake of contradiction, let $\tup$ be a tuple not in $\dfull$ but present in $\dslice$. For a tuple to be in $\dslice$, it needs to match $\cond_\modi$ or $\cond_\modi'$ for at least one $\modi \in \deltaHist$. However, recall that \Cref{eq:update-ds-cond} takes the disjunction over the conditions for $\history$ and $\ahmod$. That is, if a tuple matches at least one, it is included in the slice of the data, making the case that a tuple is filtered to exclusively one history impossible (any tuple modified by just one of the histories is included for reenactment in the other). It then follows from \Cref{theo:data-slicing-single-mod} that $\tup$ must be in $\dfull$ given that it matches the condition $\cond_\modi$ or $\cond_\modi'$ for at least one $\modi \in \deltaHist$. Therefore, we find that $\forall \tup \in \dslice: \tup \in \dfull$, i.e. $\dfull \supseteq \dslice$.

\end{proof}

}

\iftechreport{
\partitle{Discussion}
While for data slicing for historical what-if queries with a single modification, the cost of evaluating the data slicing condition is almost always less then the cost saved by reducing the amount of data to be evaluated by the remainder of the reenactment query. However, for multiple modifications, the cost of conditions for a modification that affects an update later in the input history may approach the cost of the reenactment query itself in the worst-case. This cost depends on several factors: the position of the modified update in the history, the number of attributes referenced by the condition of the update, and how many updates before the modified update have modified attributes referenced by the modified update's condition.
}

\section{Program Slicing}
\label{sec:dep-ana}

In addition to data slicing,
we also optimize the process of answering a historical what-if query $\hwhatif = (\history, \db, \deltaHist)$
by excluding statements from reenactment if their existence has provably no effect on the answer of $\hwhatif$. This is akin to \textit{program slicing}~\cite{cheney07,W81} which is a technique developed by the PL community to determine a slice (a subset of the statements of a program) that is sufficient for computing the values of variables at a given set of locations in the program.
Analog, we define slices of histories wrt. historical what-if queries.
A slice for a historical what-if query $\hwhatif$ consists of subsets of $\history$ and $\history[\deltaHist]$ that can be substituted for the original history and modified history when evaluating the historical what-if query without changing its result. Recall that the result of a historical what-if query is computed as the delta (symmetric difference) between the result of the original and the modified history. That is, only tuples in the delta are relevant for determining slices.
\revdel{We distinguish between \textit{dynamic slices} which take the database $\db$ into account and \textit{static slices} which are independent of the database and are sufficient for any database.}

\begin{defi}[History Slices]
  Let $\hwhatif = (\history, \db, \deltaHist)$ be a historical what-if query over a history $\history = (\up_1, \ldots, \up_n)$.
  Furthermore, let $\idxs = \{ \idx_1, \ldots, \idx_m \}$ be a set of indexes from $[1,n]$ such that $\idx_j < \idx_k$ for $j < k$.
  We call  $(\hslice{\history}{\idxs}, \hslice{\history[\deltaHist]}{\idxs})$  a  \emph{slice} for $\hwhatif$ if
\begin{gather*}
  \begin{aligned}
    &\iDiff{\history(\db)}{\history[\deltaHist](\db)} = \iDiff{\hslice{\history}{\idxs}(\dbver{\idx_1})}{\hslice{\history[\deltaHist]}{\idxs}(\dbver{\idx_1})}
  \end{aligned}
\end{gather*}
\end{defi}

History slices allow us to optimize the evaluation of a historical what-if query by excluding statements from reenactment. Thus, ideally, we would like slices  to be \emph{minimal}, i.e., the result of removing any statement from $\hslice{\history}{\idxs}$ or $\hslice{\history[\deltaHist]}{\idxs}$ is not a slice.
\iftechreport{There may exist more than one minimal slice for a query $\hwhatif$, because the exclusion of one statement may prevent us from excluding another statement.}
\revdel{Note that the condition for static slices is more strict than the condition for dynamic slices, because what subsets of the histories are minimal dynamic slices for a historical what-if query may depend on $\db$. Thus, any static slice is a dynamic slice, but not vice versa and a minimal dynamic slice for a query $\hwhatif$ may be smaller than any static slice for $\hwhatif$.}
A naive method for testing whether $\idxs$ is a slice, is to compute $\iDiff{\history(\db)}{\history[\deltaHist](\db)}$ and compare it against $\iDiff{\hslice{\history}{\idxs}(\dbver{\idx_1})}{\hslice{\history[\deltaHist]}{\idxs}(\dbver{\idx_1})}$. However, this is more expensive then just directly evaluating $\iDiff{\history(\db)}{\history[\deltaHist](\db)}$ which we wanted to optimize. Instead we give up minimality and restrict program slicing to tuple independent statements (\Cref{def:tuple-independence}) which enables us to check that the slice and full histories produce the same result one tuple at a time. Furthermore, we
  design a method that (lossily) compresses the database $\db_C$ and checks this condition (same result for each input tuple) over the compressed database. Since the compression is lossy, a compressed database $\db_C$ represents all databases $\db$ such that compressing $\db$ yields $\db_C$. To ensure that our method produces a valid slice for each such $\db$, we adapt techniques from incomplete databases~\cite{IL84a, lenses15}.
\revdel{Based on this discussion, the reader may assume that dynamic slices are preferable. However, as we have argued above, we have to access the database to compute a minimal dynamic slice while, as we will demonstrate in the following, static slices can be computed based on $\history$ and $\ahmod$ alone. That is, the cost of computing static slices only depends on the length of the history, number of attributes in the schema of the relation updated by $\history$, and the size of expressions in the statements of the history, but is independent of the size of the input database.
Thus, we focus on developing a method that computes static slices independent of $\db$ which based on the above observation implies that the slices determined by our method can not be minimal in general.}

\section{Slicing with Symbolic Execution}
\label{sec:sym-exe}

We adapt concepts from incomplete databases~\cite{IL84a} to reason about the behavior of updates over a set of possible databases represented by a compressed database. This is akin to \textit{symbolic execution}~\cite{cadar13,K76} which is used in software testing to determine inputs that would lead to a particular execution path in the program.
We use \textit{Virtual C-tables}~\cite{pip10,lenses15} (\textit{VC-tables}) as a compact representation of the set of possible worlds represented by a compressed database (to be discussed in \Cref{sec:compr-input-datab}) and demonstrate how to evaluate updates with possible worlds semantics over such representations. That is, the result of a history over a VC-table instance encodes all possible results of the history over every possible world represented by the VC-table. Using a constraint solver, we can then prove existential or universal statements over these possible results. For our purpose, we will check that a candidate slice and the full histories produce the same result for a \abbrHW $\hwhatif$.

\subsection{Incomplete Databases and Virtual C-Tables}
\label{sec:vc-tb}
An incomplete database $\worlds = \{ \world_1, \ldots, \world_n \}$ is a set of deterministic databases called possible worlds. Each $\world_i$ represents one possible state of the database. Queries (and updates) over an incomplete database $\worlds$ are evaluated using possible world semantics where the result of the query (statement) is the set of possible worlds derived by applying the query (statement) to every possible world from $\worlds$: %
$$\query(\worlds) = \{ \query(\world) \mid \world \in \worlds \}$$
For our purpose, it will be sufficient to use an incomplete database consisting of possible worlds containing a single tuple, because we restrict program slicing to tuple independent statements which process every input tuple independent of every other input tuple.
This incomplete database contains one world for any such singleton relation.
We then evaluate updates from the original and modified history and their slices over this incomplete database and search for worlds where the delta is different for the full histories than for the slice.
\iftechreport{However, the number of possible tuples per relation (and, thus, also the number of possible worlds) is exponential in the number of attributes of the relation. For instance, consider a relation with $n$ attributes and a domain with $m$ values. Then there are $m^n$ possible tuples for this relation that we can construct using the values of the domain.}
For efficiency we need a compact representation of an incomplete database. We employ Virtual C-tables~\cite{pip10,lenses15} which extend C-tables~\cite{incomp88} to support scalar operations over values.

A VC-table $\vcrel$ is a relation with tuples whose values are symbolic expressions over a countable set of variables $\varDom$ and where each tuple $\vct$ (we use boldface to indicate tuples with symbolic values) is associated with a condition $\lcond(\vct)$ (the so-called \textit{local condition}). The grammar shown in \Cref{fig:expr-grammar} defines the syntax of valid expressions. A VC-database $\vcdb$ is a set of VC-tables paired with a condition $\gcond$, called a global condition. Let $\dataDomain$ denote a universal domain of values.
A VC-db $\vcdb$ encodes an incomplete database which consists of all possible worlds that can be generated by assigning a value to each variable in $\varDom$, evaluating the symbolic expressions for each tuple in $\vcdb$ and including tuples in the possible world whose local condition $\lcond(\vct)$ evaluates to true. Only assignments for which the global condition $\gcond$ evaluates to true are part of the incomplete database represented by $\vcdb$.
We use $\worldsOf{\vcdb}$ to denote the set of worlds encoded by the VC-database $\vcdb$ (and apply the same notation for VC-tables). For ease of presentation, we will limit the discussion to databases with a single relation and for convenience associate a global condition with this single relation (instead of with a VC-database). However, our method is not subject to this restriction.

\begin{defi}[$\worldsOf{\vcdb}$]\label{def:vcdb-worlds}
  Let $\vcdb$ be a VC-db and let $\allVarAssigns$ be the set of all assignments $\varDom \to \dataDomain$.
  \begin{align*}
\worldsOf{\vcdb} &= \{ \db \mid \exists \varAssign \in \allVarAssigns: \varAssign(\vcdb) = \db \land \varAssign(\gcond)\}
  \end{align*}
Abusing notation, we apply $\varAssign$ to VC-dbs, tuples, and symbolic expressions $\expr$ using the semantics defined below.
  \begin{align*}
\varAssign(\vcdb) &= \{ \varAssign(\vct) \mid \vct \in \vcdb \land \varAssign(\lcond(\vct)) \}
  \end{align*}
  \begin{align*}
        \varAssign(\expr_1 \diamond \expr_2) &= \varAssign(\expr_1) \diamond \varAssign(\expr_2) \,\,\,\text{\textbf{for}}\,\, \diamond \in \{+,-,\cdot,\div,=,\neq,<,\leq,>,\geq,\land,\lor\}\\
        \varAssign(\diamond \expr_1) &= \diamond \varAssign(\expr_1)
    \,\,\,\text{\textbf{for}}\,\, \diamond \in \{\neg, \isnull\}
  \end{align*}\\[-9mm]
  \begin{align*}
    \varAssign( (\expr_1, \cdots, \expr_n)) &= (\varAssign(\expr_1), \cdots, \varAssign(\expr_n))     &\varAssign(\T) &= \T  \\
        \varAssign(\sqlCase{e_1}{e_2}{e_3}) &=
                                          \begin{cases}
                                            \varAssign(e_2) &\mathbf{if}\,\,\varAssign(e_1)\\
                                            \varAssign(e_3) &\mathbf{otherwise}\\
                                          \end{cases}
    &\varAssign(\F) &= \F\\
  \end{align*}

\end{defi}

\begin{exam}
\label{ex:pw-example}
Consider relation \texttt{Order} from \Cref{ex:running-example}. In this example, we just consider the three attributes that are used by updates in the history (\texttt{Country},\texttt{Price},\texttt{ShippingFee}).
A VC-table over this schema is shown on the top left in \Cref{fig:running-vctb}. This VC-table contains a single tuple with three variables $x_{Country}$, $x_{Price}$, and $x_{ShippingFee}$ and a local condition $\T$ (shown on the right of the tuple). Consider the variable assignment $x_{Country} = UK$, $x_{Price} = 10$, and $x_{ShippingFee} = 0$. Applying this assignment, we get the  possible world  $\{(UK, 10, 0)\}$.
\end{exam}

Note that we can encode information about the data distribution of the database of a \abbrHW as part of the global condition of an VC-database. For instance, we can compress the Order relation from \Cref{fig:running-example-instance} into a conjunction of range constraints. The set of tuples fulfilling this condition is a superset of the Order relation.
\begin{align*}
  x_{Country} \in \{UK,US\} \land x_{Price} \geq 20 \land x_{Price} \leq 60 \\
  \land x_{Shipping Fee} \geq 3 \land x_{ShippingFee} \leq 5
\end{align*}

\subsection{Updates on VC-Tables}
\label{sec:up-vc-tb}
Prior work on updating incomplete databases (e.g., \cite{fagin-86-upld,winslett-86-upldcnv,AG85}) does not support VC-tables. For our purpose, we need to be able to evaluate statements over VC-tables with possible world semantics. That is, the possible worlds of  the result of applying a statement  to a VC-table are derived by computing the statement over every possible world of the input.
For an insert $\ainsert$ we just add the concrete tuple $\tup$ with a local condition $\cond(t) = \T$ to the input VC-table $\vcrel$.
For a delete $\delete{\cond}$, for some assignment $\varAssign$, the concrete tuple $\varAssign(\vct)$ derived from a  symbolic tuple $\vct \in \vcrel$ is deleted by the statement if the tuple's local condition evaluates to true $\varAssign(\lcond(\vct)) \models \T$ and the tuple does not fulfill condition $\cond$ ($\varAssign(\vct) \not\models \cond$). We can achieve this behavior by setting the local condition of every tuple $\vct$ to $\lcond(\vct) \land \neg \cond(\vct)$. The symbolic expression $\cond(\vct)$ is computed by substituting any reference to attribute $A$ in  $\cond$ with the symbolic value $\vct.A$.
An update $\aupdate$ can affect a tuple $\vct$ in a VC-table in one of two ways in each possible world ($\varAssign$): (i) either the update's condition evaluates to false and the values of $\varAssign(\vct)$ are not modified or (ii) the update's condition evaluates to true on $\varAssign(\vct)$ and $\pset$ is applied to the values of $\varAssign(\vct)$. We have to provision for both cases.

\iftechreport{
One way to encode this is to return two tuples for every input tuple $\vct$: one tuple with updated local condition that ensures that $\vct$ is only included if $\cond$ evaluates to false on $\vct$ and another tuple $\pset(\vct)$ with a location condition that ensures that $\pset(\vct)$ is only included if $\cond$ evaluates to true. Note that there may exist multiple input tuples $\vct$, $\vct'$, $\vct''$, \ldots which are all projected onto the same output, i.e., there exists $\vct_{out}$ such that $\vct_{out} = \pset(\vct) = \pset(\vct') = \ldots$. Tuple $\vcOf{t_{out}}$ exists in the result of the update as long as $\cond$ holds for at least one of these inputs. That is the local condition of $\vcOf{t_{out}}$ is a disjunction of conditions $\lcond(\vct) \wedge \cond(\vct)$ for any input $\vct$ where $\pset(\vct) = \vcOf{t_{out}}$. Note that we can simplify the resulting instance by evaluating constant subexpressions in symbolic expressions and by removing any tuple $\vct$ for which $\lcond(\vct) \equiv \F$.
Observe that in the worst-case evaluating a sequence of $n$ updates over a VC-table can lead to an instances that is $2^{n-1}$ times larger than the input since we generate two output tuples for each input tuples in the worst case (if no two inputs are projected onto the same symbolic output).
}
We can avoid this exponential blow-up by introducing tuples with fresh variables to represent the updated versions of tuples and by assigning values to these new variables using the global condition.
We show these semantics for statements below. To ensure that there are no name clashes between variables, we generate fresh variables $\{x_{\vct,A_i}\}$ to represent the value of attribute $A_i$ of the tuple produced by applying the statement to tuple $\vct$ from the input VC-table. %
We use $\lcond(\vcrel,\vct)$ to denote the local condition of tuple $\vct$ in relation $\vcrel$ and for convenience  define $\lcond(\vcrel,\vct) = \F$ for any $\vct \not\in \vcrel$. Furthermore, $\cond(\vct)$ for update $\update{\pset}{\cond}$ and tuple $\vct$ denotes the result of substituting references to attributes in $\cond$ with their value in $\vct$.

\begin{defi}[Updates over VC-tables]
  Let $\vcrel$ a VC-table $\schema{\vcrel} = (A_1, \ldots, A_n)$. Update statements over VC-tables are defined as shown below. %
Let  $\pset = (e_1, \ldots, e_n)$. Given a tuple $\vct$, we use  $\vctn$ to denote $(x_{\vct,A_1}, \ldots, x_{\vct,A_n})$.
  \begin{align*}
    \update{\pset}{\cond}(\vcrel)   &= \{ \vctn \mid \vct \in \vcrel \}
    &\lcond(\update{\pset}{\cond}(\vcrel), \vctn )  &= \lcond(\vcrel,\vct)
  \end{align*}
  \begin{align*}
    &\gcond(\update{\pset}{\cond}(\vcrel)) = \gcond(\vcrel) \land \bigwedge_{\vct \in \vcrel} \bigwedge_{i=1}^{n} x_{\vct,A_i} = \sqlCase{\cond(\vct)}{e_i(\vct)}{\vct.A_i}
    \end{align*}
  \begin{align*}
    \delete{\cond}(\vcrel)                & = \{ \vct \mid \vct \in \vcrel \}                                                                                                                                 &
    \lcond(\delete{\cond}(\vcrel), \vct) &= \lcond(\vcrel, \vct) \land \neg \cond(\vct)
  \end{align*}
  \begin{align*}
    \ins{t}(\vcrel)                        & = \vcrel \union \{ t \}                                                                                                                                &
\lcond(\ins{t}(\vcrel),t)  &= \T
    & \lcond(\ins{t}(\vcrel),\vct) &= \lcond(\vcrel,\vct) \tag{for $\vct \neq t$}
  \end{align*}
  \begin{align*}
        \gcond(\ins{t}(\vcrel)) = \gcond(\adelete(\vcrel)) = \gcond(\vcrel)
  \end{align*}
\end{defi}

Using this semantics, the result of a sequence of $n$ statements over a relation with $m$ attributes has the same number of tuples as the input and the number of conjuncts in the global condition is bound by $n \cdot m$. Furthermore,
each conjunct is of size linear in the size of the expressions of the statements ($\cond$ or $\pset$).
\iftechreport{Note that we can reduce the number of variables in an updated VC-table, by reusing variables for attributes that are not affected by an update.} For our use case we execute a sequence of statements over an instance with a single tuple. Thus, it will be convenient to use a different naming schema for variables. We use $x_{A,i}$ to denote the value of attribute $A$ of the version of this single input tuple after the $i^{th}$ update.

\begin{figure*}[t]
  \centering
    \begin{subfigure}[b]{0.25\linewidth}
    \footnotesize
    \begin{tabular}{|c|c|c|l}
        \thead{Country} & \thead{Price} & \thead{ShippingFee} &  \\ \cline{1-3}
            $x_{Country}$ & $x_{Price}$ & $x_{ShippingFee}$ & $\T$ \\ \cline{1-3}
    \end{tabular}
    \[\gcond = \T\]
      \caption{Initial VC-database $\vcdbini$.}\label{fig:initial-vc-table}
    \end{subfigure}
    \begin{subfigure}[b]{0.7\linewidth}
     \centering
    \footnotesize
        \begin{tabular}{|c|c|c|l}
        \thead{Country} & \thead{Price} & \thead{ShippingFee} &  \\ \cline{1-3}
            $x_{Country}$ & $x_{Price}$ & $x_{ShippingFee,2}$ & $\T$ \\ \cline{1-3}
        \end{tabular}
        \begin{align*}
          \gcond &\defas \gcond_1 \land \gcond_2 %
                 \hspace{2cm}\gcond_1 \defas (x_{ShippingFee,1} = \sqlCase{x_{Price} \geq 50}{0}{x_{ShippingFee}})\\
                 & \gcond_2 \defas (x_{ShippingFee,2} = \sqlCase{(x_{Country} = UK \wedge x_{Price} \leq 100)}{x_{ShippingFee,1} + 5}{x_{ShippingFee,1}})
        \end{align*}\\[-3mm]
      \caption{VC-table after evaluating $\history = (u_1, u_2)$.}
   \end{subfigure}\\[-3mm]
  \caption{Running example for evaluating updates over VC-Tables.}
  \label{fig:running-vctb}
\end{figure*}

\begin{exam}\label{ex:up-vctb}
Continuing with \Cref{ex:pw-example}, consider the first two updates from \Cref{fig:Transitive-Transactions-Example} (we abbreviate attribute names as in previous examples): $u_1 = \update{F \gets 0}{P \geq 50}$  %
and $\up_2 = \update{F \gets F + 5}{C = UK \land P \leq 100}$. %
After execution of $u_1$ and $u_2$ over $\vcdbini$ shown in \Cref{fig:initial-vc-table}, we get an instance with a single tuple. Since both updates only modify attribute \texttt{ShippingFee}, all other attributes can reuse the same variable as in the input.  %
The value of attribute \texttt{ShippingFee} is a new variable $x_{ShippingFee,2}$ which is constrained by the global condition that ensures that it is equal to the previous value of this attribute ($x_{ShippinFee,1}$) if the condition of $u_2$ does not hold and otherwise is the result of applying  $\pset_{\up_2}$ to $x_{ShippingFee,1}$. Furthermore, $x_{ShippingFee,1}$ is related to the value of attribute \texttt{ShippingFee} in the input in the same way using a conditional expression based on $u_1$'s condition and update expression (setting shipping fee to $0$ if the price is at least $50$).
\end{exam}

We now prove that our definition of  update semantics for VC-tables complies with possible world semantics. %

\begin{theo}\label{theo:vc-updates-are-possible-worlds-semantics}
  Let $\vcdb$ be a VC-database and $\up$ a statement. We have:
  \begin{align*}
    \worldsOf{\up(\vcdb)} = \up(\worldsOf{\vcdb})
  \end{align*}
\end{theo}
\iftechreport{
\begin{proof}
  WLOG consider an assignment $\varAssign$ to the variables in $\Sigma$  and let $\db_{\varAssign} = \varAssign(\vcdb)$ denote the possible world corresponding to this assignment. Furthermore, observe that in both the VC-database as well as in $\db_{\varAssign}$, applying a statement $\up$ to the input database does only modify the relation $\rel$ ($\vcrel$) affected by $\up$. Thus, it is sufficient for us to reason only about this relation. We will show that $\varAssign(\up(\vcrel)) = \up(\rel_{\varAssign})$.

\proofpar{Insert $\up = \ins{\tup}$:}
Note that $\gcond(\up(\vcrel)) = \gcond(\vcrel)$ and $\varAssign(\gcond(\vcrel)) = \T$ because $\rel_{\varAssign}$ is a possible world of $\vcrel$. Thus, we have $\varAssign(\gcond(\up(\vcrel))) = \T$ and $\varAssign(\up(\vcrel))$ is a possible world of $\up(\vcrel)$.
We have $\up(\vcrel) = \vcrel \union \{ \tup \}$. Note that $\varAssign(\vcrel)$ is defined as applying $\varAssign$ to each tuple $\vct \in \vcrel$. Furthermore, by definition $\lcond(\up(\vcrel),\tup) = \T$.
Thus, $\varAssign(\vcrel \union \{ \tup \}) = \varAssign(\vcrel) \union \{ \tup \} = \rel_{\varAssign} \cup \{ \tup \} = \up(\varAssign(\vcrel))$.

\proofpar{Delete $\up = \adelete$:}
For the same reason as for inserts, $\varAssign(\up(\vcrel))$ is a possible world of $\up(\vcrel)$.
Substituting the definition of deletes, we get $\adelete(\rel_{\varAssign}) = \{ \tup \mid \tup \in \rel_{\varAssign} \land \neg \cond(\tup) \}$. For the VC-table, we get $\adelete(\vcrel) = \{ \vct \mid \vct \in \vcrel \}$ and $\lcond(\adelete(\vcrel), \vct) = \lcond(\vcrel, \vct) \land \neg \cond(\vct)$. WLOG consider a tuple $\tup \in \rel_{\varAssign}$ and a tuple $\vct \in \vcrel$ such that $\varAssign(\vct) = \tup$ and $\varAssign(\lcond(\vcrel, \vct))$. At least one such tuple $\vct$ has to exist, because otherwise $\tup$ would not exist in $\rel_{\varAssign}$. We have to show that $\varAssign(\lcond(\adelete(\vcrel), \vct))$ holds. %
Note that based on the definition of the application of an assignment to an expression, we can push $\varAssign$ through expressions, e.g., $\varAssign(e_1 \land e_2) = \varAssign(e_1) \land \varAssign(e_2)$. Thus,
\begin{align*}
  &\varAssign(\lcond(\adelete(\vcrel), \vct))
  = \varAssign(\lcond(\vcrel, \vct) \land \neg \cond(\vct))
  = \varAssign(\lcond(\vcrel, \vct)) \land \varAssign(\neg \cond(\vct))\\
  = &\T \land \neg \cond(\varAssign(\vct))
  = \neg \cond(\varAssign(\vct))
  = \neg \cond(\tup) = \T
\end{align*}
\proofpar{Update $\up = \aupdate$:}
Note that any tuple $\vct$ in $\up(\vcrel)$ is not in $\vcrel$ and all variables $\vct$ do not occur in any tuple, local condition, or global condition in $\vcrel$. Since these variables do not occur in $\vcrel$, any assignment $\varAssign$ such that $\varAssign(\vcrel) = \rel_{\varAssign}$ can be extended to an assignment $\varAssign'$ over $\up(\vcrel)$ by assigning values to these fresh variables. Furthermore, recall that these fresh variables are only constrained in the global condition of $\gcond(\up(\vcrel))$. We will first show that for each $\varAssign$ such that $\varAssign(\gcond(\vcrel))$ holds, there exists one and only one extension $\varAssign'$ of $\varAssign$ such that $\gcond(\varAssign'(\up(\vcrel)))$ holds. Intuitively this means that for every world in the input there exists exactly one corresponding world in the output $\up(\vcrel)$.

\proofpar{Unique extension of $\varAssign$:}
Note that for each tuple $\vct \in rel$, $\up(\vcrel)$ contains a tuple $\vctn = (x_{\vct,A_1}, \ldots, x_{\vct,A_n})$ where each $x_{\vct,A_i}$ is a fresh variable and $\lcond(\up(\vcrel),\vctn) = \lcond(\up(\vcrel),\vct)$. Recall that

\begin{align}
  \label{}
\gcond(\update{\pset}{\cond}(\vcrel)) = \gcond(\vcrel) \land \bigwedge_{\vct \in \vcrel} \bigwedge_{i=1}^{n} x_{\vct,A_i} = \sqlCase{\cond(\vct)}{e_i(\vct)}{\vct.A_i}
\end{align}

Consider an assignment $\varAssign$ for $\vcrel$ such that $\varAssign(\gcond(\vcrel))$ holds. Since $\gcond(\update{\pset}{\cond}(\vcrel))$ is a conjunction of $\varAssign(\gcond(\vcrel))$ with constraints for each tuple $\vct \in \vcrel$, to prove our claim, it suffices to show that for any such $\vct$ and attribute $A_i$, there exist a unique assignment of $x_{\vct,A_i}$ that satisfies $x_{\vct,A_i} = \sqlCase{\cond(\vct)}{e_i(\vct)}{\vct.A_i}$ given $\varAssign$. As explained above we can push $\varAssign$ through expressions. Thus,

\[
  \sqlCase{\cond(\vct)}{e_i(\vct)}{\varAssign(\vct.A_i)} = \varAssign(\sqlCase{\cond(\vct)}{e_i(\vct)}{\vct.A_i})
\]

Note that $\varAssign(\sqlCase{\cond(\vct)}{e_i(\vct)}{\vct.A_i})$ is a concrete value. It follows that we can only make the equality true by setting $x_{\vct,A_i} = \varAssign(\sqlCase{\cond(\vct)}{e_i(\vct)}{\vct.A_i})$. Since there is a unique assignment for each $\vct.A_i$ for all $\vct \in \vcrel$ and attribute $A_i$ such that the corresponding conjunct in $\gcond(\up(\vcrel))$ evaluates to true, there exists one and only one extension $\varAssign'$ of $\varAssign$ that satisfies $\gcond(\up(\vcrel))$.

\proofpar{$\varAssign'$ correctly models update semantics:}
Consider a possible world $\db_{\varAssign}$ of $\vcrel$ and let $\varAssign'$ be as established above. Consider a tuple $\vct$ and let $\tup = \varAssign(\vct)$ and $\tup_{up} = \up(\varAssign(\tup))$. Note that $\lcond(\vcrel,\vct) = \lcond(\up(\vcrel,\up(\vct)))$. Thus, $\varAssign(\lcond(\vcrel,\vct)) = \varAssign(\lcond(\up(\vcrel,\up(\vct))))$, i.e., $\vct$ exists iff $\up(\vct)$ exists.
We have to consider two cases. Either (i) $\cond(\tup)$ and $\tup_{up} = \pset(\tup)$ or (ii) $\cond(\tup)$ evaluates to false and $\tup_{up} = \tup$.

\proofpar{(i) $\cond(\tup)$ holds:}
WLOG consider attribute $A_i$. In this case $\up(\vct).A_i = \varAssign(\sqlCase{\cond(\vct)}{e_i(\vct)}{\vct.A_i})$ evaluates to $e_i(\varAssign(\vct)) = e_i(\tup)$ which according to the definition of updates is equal to $\tup_{up}.A_i$. Thus, $\tup_{up} = \varAssign'(\up(\vct))$.

\proofpar{(ii) $\cond(\tup)$ does not hold:}
WLOG consider attribute $A_i$. In this case $\up(\vct).A_i = \varAssign(\sqlCase{\cond(\vct)}{e_i(\vct)}{\vct.A_i})$ evaluates to $\varAssign(\vct.A_i) = \tup.A_i$ which according to the definition of updates is equal to $\tup_{up}.A_i$. Thus, $\tup_{up} = \varAssign'(\up(\vct))$.

Since we have shown that for every tuple $\tup_{up}$ we have $\tup_{up} \in \up(\varAssign(\vcrel))$ iff $\tup_{up} \in \varAssign'(\up(\vcrel))$ and $\vcrel_{\varAssign'}$ is the unique world from $\worldsOf{\up(\vcrel)}$ corresponding to $\vcrel_{\varAssign}$, we have shown that $\worldsOf{\up(\vcrel)} = \up(\worldsOf{\vcrel})$.

\end{proof}
}

Note that by induction, \Cref{theo:vc-updates-are-possible-worlds-semantics} implies that evaluating a history $\history$ over a  VC-database also has possible world semantics.

\subsection{Computing Slices with Symbolic Execution}\label{sec:symbolic-exe}

To compute a slice for a historical what-if query $\hwhatif = (\history, \deltaHist, \db)$  %
where $\history$ consists of tuple independent statements only,
we create a VC-database $\vcdbini$ with a single tuple with fresh variables for each relation in the database's schema.
Even though they are tuple independent, we do not consider inserts of the form $\ainsert$ here, because, as we will show in \Cref{sec:optim-reen-hist}, we can split a reenactment query for a history with such inserts into a union of two queries --- one that is the reenactment query for the history restricted to updates and deletes and a second one that only operates on tuples inserted by inserts $\ainsert$ from the history. Since the second query only operates on an instance of size at most $\card{\history}$, it's cost is too low to warrant spending time on slicing it.

\subsubsection{Compressing the Input Database}
\label{sec:compr-input-datab}

Optionally, we compress the input database $\db$ into a set of range constraints that restrict the variables of the single tuple in $\vcdbini$. For that, we decide on a number of groups and for each table select an attribute to group on. We then compute the minimum and maximum values of each non-group-by attribute for each group and generate a conjunction of the range constraints for each attribute. The  disjunction of the constraints generated for the groups, which we denote as $\adbconstr$, is then added to the global condition. Note that every tuple from a table of the database $\db$ corresponds to an assignment of the variables from $\vcdbini$ to the constants of the tuple that fulfills the condition. For attributes with unordered data types, we can just omit the range condition for this attribute.

\begin{exam}[Compressing Databases]\label{ex:compressing-databases}
Consider our running example instance from \Cref{fig:running-example-instance} and let us compress this database into two tuples by grouping on \texttt{Country}. We get the following constraint that we can add to the global condition of $\vcdbini$ to constrain the possible worlds of $\vcdbini$. Here, we omit the constraint for the name attribute and abbreviate attribute names as before.
  \begin{align*}
    \adbconstr  \defas &(x_{C} = UK \land x_{ID} \in \{11,12\} \land
    x_{P} \in [20,50] \land
    x_{F} = 5)\\
    \lor &(x_{C} = US \land x_{ID} \in \{13,14\}
           \land
          x_{P} \in [30,60] \land
          x_{F} \in [3,4])
  \end{align*}
For instance, the first two tuples (group \texttt{UK}) get compressed into one conjunction of range constraints. Since the smallest (greatest) price in this group is $20$ ($50$), the range constraint for $x_{P}$ is $x_{P}  \in [20,50]$.
\end{exam}

\begin{figure*}[t]
  \centering
  \begin{subfigure}{1\linewidth}
    \begin{minipage}{0.3 \textwidth}
      \begin{tabular}{|c|c|c|l}
        \thead{Country} & \thead{Price} & \thead{ShippingFee} &  \\ \cline{1-3}
        $x_{C}$ & $x_{P}$ & $x_{F,2'}$ & $\T$ \\ \cline{1-3}
      \end{tabular}
    \end{minipage}
    \begin{minipage}{0.69 \textwidth}
      \begin{align*}
        \gcond' &= \gcond_{1'} \land \gcond_{2'} \land \adbconstr
                 \hspace{2cm}\gcond_{1'} = (x_{F,1'} = \sqlCase{x_{P} \geq 60}{0}{x_{F}})\\
               & \gcond_{2'} = (x_{F,2'} = \sqlCase{x_{C} = UK \wedge x_{P} \leq 100}{x_{F,1'} + 5}{x_{F,1'}})
      \end{align*}
    \end{minipage}\\[-2mm]
    \caption{VC-database $\ahmod(\vcdbini)$}\label{fig:vc-database-ahmod-vcdbin}
  \end{subfigure}
  \begin{subfigure}{0.35\linewidth}
    \begin{tabular}{|c|c|c|l}
      \thead{Country} & \thead{Price} & \thead{ShippingFee} &  \\ \cline{1-3}
      $x_{C}$ & $x_{P}$ & $x_{F,1''}$ & $\T$ \\ \cline{1-3}
    \end{tabular}
    \begin{align*}
      \gcond'' &= \adbconstr \land x_{F,1''} = \sqlCase{x_{P} \geq 50}{0}{x_{F}}
    \end{align*}\\[-10mm]
    \caption{VC-database $\hislice{\{1\}}(\vcdbini)$}\label{fig:vc-database-hislice-histo}
\end{subfigure}
\hspace{2cm}
\begin{subfigure}{0.35\linewidth}
    \begin{tabular}{|c|c|c|l}
      \thead{Country} & \thead{Price} & \thead{ShippingFee} &  \\ \cline{1-3}
      $x_{C}$ & $x_{P}$ & $x_{F,2'}$ & $\T$ \\ \cline{1-3}
    \end{tabular}
    \begin{align*}
      \gcond''' &= \adbconstr \land
                x_{F,1'''} = \sqlCase{x_{P} \geq 60}{0}{x_{F}}
    \end{align*}\\[-10mm]
    \caption{VC-database $\hisliceOf{\ahmod}{\{1\}}$}\label{fig:vc-database-hisliceof-ahm}
  \end{subfigure}
\vspace{-4mm}
  \caption{VC-database instances for our slicing example (attributes names are abbreviated as: (C)ounty, (P)rice, Shipping(F)ee).}\label{fig:vc-database-instances-for}
\end{figure*}

\subsubsection{Computing Slices}
\label{sec:test-depend-slic}

To determine whether a given set of indices $\idxs$ is a slice for $\hwhatif$, we have to test whether:

\begin{align}
  \label{eq:slice-original-cond}
\iDiff{\history(\db)}{\history[\deltaHist](\db)} = \iDiff{\hslice{\history}{\idxs}(\db)}{\hslice{\history[\deltaHist]}{\idxs}(\db)}
\end{align}

Recall that we restrict program slicing to tuple independent statements (\Cref{def:tuple-independence}). That is, the result produced by such a statement for an input tuple only depends on the values of this tuple and is independent of what other tuples exist in the input. Thus, if both deltas return the same result for every input tuple, then the two deltas are guaranteed to be equal.
Thus,  $\idxs$ is a slice if for all input tuples from $\db$, both deltas return the same result (see \Cref{eq:tuple-level-slice-test} below). Note that this is only a sufficient, but not necessary condition. To see why this is the case, consider two input tuples $t_1$ and $t_2$ and assume that the delta of the results of the full histories returns $s_1$ for $t_1$ and $s_2$ for $t_2$, but the delta of the results of the sliced histories returns $s_2$ for $t_1$ and $s_1$ for $t_2$. The final result is the same, even though the results for the individual input tuples is different.

\begin{align}
  \label{eq:tuple-level-slice-test}
  \begin{split}
    \forall t \in \db:\,\, &\iDiff{\history(\{t\})}{\history[\deltaHist](\{t\})}\\ = &\iDiff{\hslice{\history}{\idxs}(\{t\})}{\hslice{\history[\deltaHist]}{\idxs}(\{t\})}
  \end{split}
\end{align}

For each $t \in \db$, by construction of $\vcdbini$ (the VC-database we use as input for program slicing), there exists a world  $\db_{t} \in \worldsOf{\vcdbini}$ such that $\db_{t} = \{t\}$. Note that since $\vcdbini$ is generated by compressing the input database into a set of range constraints, some worlds may not correspond to a tuple from $\db$. However, our argument only requires that for each $t \in \db$ there exists a world in $\vcdbini$ which implies that if the condition from \Cref{eq:single-inst-slice-test} evaluates to true for every such $\db_{t}$, then \Cref{eq:tuple-level-slice-test} holds. Thus, the formula shown below is a sufficient condition for $\idxs$ to be a slice.

\begin{align}
  \label{eq:single-inst-slice-test}
  \begin{split}
    \forall \db_{t} \in \worldsOf{\vcdbini}:\,\, &\,\,\,\iDiff{\history(\db_{t})}{\history[\deltaHist](\db_{t})}\\ = &\,\,\,\iDiff{\hslice{\history}{\idxs}(\db_{t})}{\hslice{\history[\deltaHist]}{\idxs}(\db_{t})}
  \end{split}
\end{align}

For an input tuple $t$, based on the definition of symmetric difference, $\iDiff{\history(\db_{t})}{\history[\deltaHist](\db_{t})}$ is equal to $\iDiff{\hslice{\history}{\idxs}(\db_{t})}{\hslice{\history[\deltaHist]}{\idxs}(\db_{t})}$ if either (i) $\history(\db_{t}) = \history[\deltaHist](\db_{t})$ and $\hslice{\history}{\idxs}(\db_{t}) = \hslice{\history[\deltaHist]}{\idxs}(\db_{t})$ which means that both deltas return the empty set for $\db_{t}$ or (ii) both deltas return the same set of tuples over $\db_{t}$ which is the case when $\history(\db_{t}) \neq \history[\deltaHist](\db_{t})$ and one of the conditions shown below holds.
\begin{itemize}
\item (a) $\history(\db_{t}) = \hslice{\history}{\idxs}(\db_{t}) \land \history[\deltaHist](\db_{t}) =  \hslice{\history[\deltaHist]}{\idxs}(\db_{t})$
\item (b)
$\history(\db_{t}) = \hslice{\history[\deltaHist]}{\idxs}(\db_{t}) \land \history[\deltaHist](\db_{t}) = \hslice{\history}{\idxs}(\db_{t})$
\end{itemize}
 Thus, \Cref{eq:single-inst-slice-test} is equivalent to:
\begin{align}
  \label{eq:slice-test-cases}
  \begin{split}
    &\forall \db_{t} \in \worldsOf{\vcdbini}:\\
    &\left(\history(\db_{t}) = \history[\deltaHist](\db_{t}) \land \hslice{\history}{\idxs}(\db_{t}) = \hslice{\history[\deltaHist]}{\idxs}(\db_{t})\right)   \\
    \lor &(\history(\db_{t}) \neq \history[\deltaHist](\db_{t}) \land\\
    &\hspace{1cm}(
           \history(\db_{t}) = \hslice{\history}{\idxs}(\db_{t}) \land \history[\deltaHist](\db_{t}) =  \hslice{\history[\deltaHist]}{\idxs}(\db_{t})  \\
  &\hspace{0.7cm}\lor \history(\db_{t}) = \hslice{\history[\deltaHist]}{\idxs}(\db_{t}) \land \history[\deltaHist](\db_{t}) = \hslice{\history}{\idxs}(\db_{t})))
  \end{split}
\end{align}

Based on the semantics of updates over VC-tables, the result of a history over a single tuple instance $\vcdbini$ is an instance with a single tuple whose local condition governs the existence of the tuple in any particular world $\db_{t}$. For a history $\history$ let us denote this tuple as $\singtupH{\history}$. Consider the valuation $\varAssign_{t}$ generating $\db_{t}$. Then for two histories $\history$ and $\history'$, the condition $\history(\db_{t}) = \history'(\db_{t})$ is equivalent to the equation shown below as long as we appropriately rename variables such that the two VC-databases do not share any variables except for the variables from $\vcdbini$. %

\begin{align}
  \label{eq:db-equals-symbolic}
  \begin{split}
      &(\varAssign_{t}(\singtupH{\history}) = \varAssign_{t}(\singtupH{\history'}) \land  \lcond(\varAssign_{t}(\singtupH{\history})) \land \lcond(\varAssign_{t}(\singtupH{\history'})))\\
  \vee &(\neg \lcond(\varAssign_{t}(\singtupH{\history})) \land \neg \lcond(\varAssign_{t}(\singtupH{\history'})))
  \end{split}
\end{align}
Intuitively, this condition means that for the two histories to return the same result over $\db_{t}$, either (i) they both return the same result tuple (equal values and the local conditions of the single result tuples evaluates to true for both histories) or (ii) they both return the empty set (the local conditions of the single result tuples evaluate to false for both histories).

If we substitute this equation into \Cref{eq:slice-test-cases}, then we get a universally quantified first order sentence (a formula without free variables) over the variables from the VC-database $\vcdbini$. We will use $\aslicetest$ to denote the resulting formula (recall that $\adbconstr$ denotes the constraints encoding the compressed database). We can now use a constraint solver to determine whether $\aslicetest$ is true by checking that its negation is unsatisfiable. We use an MILP-solver for this purpose. The translation rules for transforming a logical condition into an MILP program are mostly well-known rules applied in linear programming and many have been used in related work (e.g.,~\cite{MeliouS12}). 
  We are now ready to state the major formal result of this section.

\begin{theo}[Slicing Condition]\label{theo:slicing-condition-co}
Let $\hwhatif = (\history, \db, \deltaHist)$ be a historical what-if query where $\history$ is a history with $n$ statements (updates and deletes). If $\aslicetest$ is true, then $\idxs$ is a slice for $\hwhatif$.
\end{theo}
We first prove that \Cref{eq:tuple-level-slice-test}  implies \Cref{eq:slice-original-cond} for histories consisting of updates and deletes which are both tuple independent. This follows from the definition of tuple independence (\Cref{def:tuple-independence}). \Cref{eq:tuple-level-slice-test} is implied by \Cref{eq:single-inst-slice-test}, because the  worlds of $\vcdbini$ encode a superset of $\db$ by construction and \Cref{theo:vc-updates-are-possible-worlds-semantics} (updates over  VC-databases have possible world semantics). The equivalence of \Cref{eq:slice-test-cases} and \Cref{eq:single-inst-slice-test} follows from the definition of database deltas. Finally, the equivalence of $\history(\db_{t}) = \history'(\db_{t})$ and \Cref{eq:db-equals-symbolic} follows from \Cref{theo:vc-updates-are-possible-worlds-semantics}.
\iftechreport{
\begin{proof}
  We start by proving an auxiliary result that will be used in the main part of the proof: if all statements of a history $\history$ are tuple independent deletes and updates (\Cref{def:tuple-independence}), then for any database $\db$ we have

  \begin{align}
    \label{eq:history-TI}
    \history(\db) &= \bigcup_{\tup \in \db} \history(\{\tup\})
  \end{align}
  We then prove that \Cref{eq:tuple-level-slice-test}  implies \Cref{eq:slice-original-cond} for histories consisting of updates and deletes which are both tuple independent. This follows from the definition of tuple independence (\Cref{def:tuple-independence}). \Cref{eq:tuple-level-slice-test} is implied by \Cref{eq:single-inst-slice-test}, because the  worlds of $\vcdbini$ encode a superset of $\db$ by construction and \Cref{theo:vc-updates-are-possible-worlds-semantics} (updates over  VC-databases have possible world semantics). The equivalence of \Cref{eq:slice-test-cases} and \Cref{eq:single-inst-slice-test} follows from the definition of database deltas. Finally, the equivalence of $\history(\db_{t}) = \history'(\db_{t})$ and \Cref{eq:db-equals-symbolic} follows from \Cref{theo:vc-updates-are-possible-worlds-semantics}.

  \proofpar{\textbf{Union factors through tuple independent histories:}}
    Consider a history $\history = (u_1, \ldots, u_n)$ such that for all $i \in [1,n]$ statement $\up_i$ is tuple independent.  We will prove that this implies \Cref{eq:history-TI}. We proof this claim by induction.

  \proofpar{Base case:}
  $\history = (\up_1)$ for some statement $\up_1$. The claim follows directly from the definition of tuple independence.

  \proofpar{Inductive step:}
  Let $\history = (\up_1, \ldots, \up_n)$ and assume that for any database $\db$ we have $ \hislice{\history(\db)}{[1,n]} = \bigcup_{\tup \in \db} \hislice{\history(\db)}{[1,n]}(\{\tup\})$. We have to show that $\history(\db) = \bigcup_{\tup \in \db} \history(\{\tup\})$. For any tuple $\tup \in \db$ for which $\hislice{\history}{[1,n]}(\tup) \neq \emptyset$  let $s = \hislice{\history}{[1,n]}(\{\tup\})$. WLOG assume that $\db = \{t_1, \ldots, t_k\}$ such for some integer $l$ we have $\hislice{\history}{[1,n]}(\{\tup_i\}) \neq \emptyset$ for $i \leq l$ and $\hislice{\history}{[1,n]}(\{\tup_i\}) = \emptyset$ for $i > l$  (it is always possible to find such an arrangement of the tuples of $\db$). Then,

  \begin{align*}
    &\hislice{\history(\db)}{[1,n]}\\
    = &\bigcup_{\tup \in \db} \hislice{\history(\db)}{[1,n]}(\{\tup\})\\
    = &\bigcup_{\tup \in \db \land \hislice{\history}{[1,n]}(\tup) \neq \emptyset} \hislice{\history(\db)}{[1,n]}(\{\tup\})\\
    = &\{s_1, \ldots, s_l\}
  \end{align*}

  Because $\up_n$ is tuple independent, we know that
  \begin{align*}
    &\up_n(\{s_1, \ldots, s_l\})\\
    = &\bigcup_{s_i \in \{s_1, \ldots, s_l\}} \up_n(s_i)\\
    = & \bigcup_{\tup_i \in \db} \up_n(\hislice{\history}{[1,n]}(\tup_i)) \tag{based on $s = \hislice{\history}{[1,n]}(\tup)$}\\
    = &\bigcup_{\tup_i \in \db} \history(\{\tup_i\})
  \end{align*}

  This concludes the proof of \Cref{eq:history-TI}.

  \proofpar{\textbf{\Cref{eq:tuple-level-slice-test}  implies \Cref{eq:slice-original-cond}}:}
  We will proof this implication by proving the contrapositive:  $\neg (\ref{eq:slice-original-cond}) \Rightarrow \neg (\ref{eq:tuple-level-slice-test})$.

  \begin{align*}
    &\neg (\ref{eq:slice-original-cond}) \\
    = &\neg \iDiff{\history(\db)}{\history[\deltaHist](\db)} = \iDiff{\hslice{\history}{\idxs}(\db)}{\hslice{\history[\deltaHist]}{\idxs}(\db)}\\
    \Leftrightarrow &\iDiff{\history(\db)}{\history[\deltaHist](\db)} \neq \iDiff{\hslice{\history}{\idxs}(\db)}{\hslice{\history[\deltaHist]}{\idxs}(\db)}
  \end{align*}

  For two relations ($\iDiff{\history(\db)}{\history[\deltaHist](\db)}$ and $\iDiff{\hslice{\history}{\idxs}(\db)}{\hslice{\history[\deltaHist]}{\idxs}(\db)}$ in our case) to be different, there has to exist at least one tuple $\tup$ that is in one, but not in the other. WLOG assume that $\tup \in \iDiff{\history(\db)}{\history[\deltaHist](\db)}$:

    \begin{align*}
      \Rightarrow &\exists \tup: \tup \in \iDiff{\history(\db)}{\history[\deltaHist](\db)} \land \tup \not\in \iDiff{\hslice{\history}{\idxs}(\db)}{\hslice{\history[\deltaHist]}{\idxs}(\db)}
    \end{align*}

Since all histories are tuple independent, we have:

\begin{align}
  \label{eq:delta-TI-has-difference}
      \Rightarrow \exists \tup:\,\,\, &\tup \in \diffsym\left(\bigcup_{s \in \db}\history(\{s\}),\bigcup_{s \in \db}\ahmod(\{s\})\right)\\
                    \land &\tup \not\in \diffsym\left(\bigcup_{s \in \db}\hslice{\history}{\idxs}(\{s\}), \bigcup_{s \in \db} \hslice{\ahmod}{\idxs}(\{s\})\right)
    \end{align}

    This can only be the case if there exists at least one tuple $t_{\in} \in \db$ such that the two delta are different. To see why this is the case assume the negation: for all $t_{in} \in \db$ both deltas return the same results which contradicts \Cref{eq:delta-TI-has-difference}.

    \begin{align*}
      \Rightarrow \exists \tup_{in} \in \{t_{in}\}: &\iDiff{\history(\{t_{in}\})}{\history[\deltaHist](\{t_{in}\})}\\
      &\neq \iDiff{\hslice{\history}{\idxs}(\{t_{in}\})}{\hslice{\history[\deltaHist]}{\idxs}(\{t_{in}\})}\\[3mm]
      \Rightarrow \forall \tup \in \{t\}: &\iDiff{\history(\{t\})}{\history[\deltaHist](\{t\})}\\
      &= \iDiff{\hslice{\history}{\idxs}(\{t\})}{\hslice{\history[\deltaHist]}{\idxs}(\{t\})}\\[3mm]
           = &\neg (\ref{eq:slice-original-cond})
    \end{align*}

This concludes the proof that \Cref{eq:tuple-level-slice-test} implies \Cref{eq:slice-original-cond}.

\proofpar{\textbf{\Cref{eq:single-inst-slice-test} implies \Cref{eq:tuple-level-slice-test}:}}
To proof this claim, we first have to show that for each tuple $t \in \db$, it follows that $\db_t = \{t \}$ is in $\worldsOf{\vcdbini}$. Recall that $\vcdbini = \{\vct\}$ where $\vct$ consists of variables only and $\lcond(\vcdbini,\vct) = \T$. Furthermore, $\gcond(\vcdbini) = \adbconstr$ where $\adbconstr$ is a disjunction of conjunctions of range constraints such that each conjunction  ``covers'' all attribute values withing a partition of $\db$. WLOG let $t = (c_1, \ldots, c_m)$ and $\vct = (x_1, \ldots, x_m)$.

\begin{align*}
  &\db_t \in \worldsOf{\vcdbini}\\
  \Leftrightarrow &\exists \varAssign: \varAssign(\vct) = t \land \varAssign(\lcond(\vcdbini,\vct)) \land \varAssign(\gcond(\vcdbini))
\end{align*}

Based on our assumption this requires that $\varAssign(x_i) = c_i$. Since $\lcond(\vcdbini,\vct) = \T$, trivially $\varAssign(\lcond(\vcdbini,\vct)) = \T$. It remains to be shown that $\varAssign(\gcond(\vcdbini)) = \T$. Let $\db_{frag}$ denote the fragment of the partition of $\db$ based on which $\adbconstr$ was created that contains $t$. Recall that the conjunction in $\adbconstr$ produced for $\db_{frag}$ contains one range constraint for each attribute $A_i$, bounding the value of $x_i$ by the smallest and largest value of $A_i$ in $\db_{frag}$. Since $t$'s values are considered when determining the minimum and maximum values, this means that $\varAssign(x_i) = t.A_i$ fulfills each range constraint and, thus, also the conjunction. This implies that $\varAssign(\gcond(\vcdbini))$ and we know that $\db_t \in \worldsOf{\vcdbini}$ which means that \Cref{eq:single-inst-slice-test} implies \Cref{eq:slice-original-cond}.

\proofpar{\textbf{\Cref{eq:slice-test-cases} is equivalent to \Cref{eq:single-inst-slice-test}:}}
\Cref{eq:slice-test-cases} is derived from \Cref{eq:single-inst-slice-test} by substituting the deltas for their definition which preserves equivalence.

\proofpar{\textbf{\Cref{eq:db-equals-symbolic}:}}

Substituting \Cref{eq:db-equals-symbolic} into \Cref{eq:slice-test-cases}, we get the final slicing condition $\aslicetest$. Based on \Cref{theo:vc-updates-are-possible-worlds-semantics}, this substitution does not change the semantics. Thus, if $\aslicetest$ is true, then $\idxs$ is a valid slice.

  \end{proof}

}

\begin{exam}[Testing Slice Candidates]\label{ex:testing-slice-candidates}
Consider our running example database (\Cref{fig:running-example-instance}) and the history $\history = \{u_1,u_2\}$ from \Cref{ex:up-vctb} and let ${u_1}' = \update{Price \geq 60}{ShippingFee \gets 0}$. Let $\vcdbini$ be as shown in \Cref{fig:running-vctb}, but with $\gcond = \gcond_1\land \gcond_2 \land \adbconstr$ where $\adbconstr$ is the database constraint from \Cref{ex:compressing-databases}. Furthermore, consider a \abbrHW $\hwhatif = (\history, \deltaHist)$ for $\deltaHist = ( u_1 \gets u_1' )$ (higher price requirements for waiving shipping fees). To test whether $\idxs = \{1\}$ is a slice, we first have to construct $\slicetest{\hwhatif}{\idxs}{\adbconstr}$ for which we have to evaluate $\history$ $\ahmod$, $\hislice{\{1\}}$, and $\hisliceOf{\ahmod}{\{1\}}$ over $\vcdbini$. The results are shown in \Cref{ex:testing-slice-candidates}. We use $\gcond$ to denote the global condition of $\history(\vcdbini)$, $\gcond'$ for $\ahmod(\vcdbini)$, $\gcond''$ for $\hislice{\{1\}}(\vcdbini)$, and $\gcond'''$ for $\hisliceOf{\ahmod}{\{1\}}(\vcdbini)$.
  Since in the result of both histories and their slices, the local condition of the result tuple is true, we do not have to test whether the local condition is true or false as done in \Cref{eq:db-equals-symbolic} and can instead directly test equality of two history's result tuples to test whether the histories return the same result. Furthermore, observe that all four histories only modify attribute \texttt{ShippingFee}. Thus, it is sufficient to compare tuples on attribute shipping fee to determine whether the result tuples are the same. Applying this simplifications, $\slicetest{\hwhatif}{\idxs}{\adbconstr}$ is equal to:

  \begin{align*}
    \forall x_{I}, x_{P}, x_{F}:\,\,\, &\adbconstr \land \gcond \land \gcond' \land \gcond'' \land \gcond''' \land\\
                                 &\hspace{2mm}(
                                   (x_{F,2} = x_{F,2'} \land x_{F,1''} = x_{F,1'''})\\
                                 &\lor (x_{F,2} \neq x_{F,2'} \land ((x_{F,2} = x_{F,1''} \land x_{F,2'} = x_{F,1'''})\\
                                 &\hspace{2.1cm}\vee (x_{F,2} = x_{F,1'''} \land x_{F,2'} = x_{F,1''})))
      )
  \end{align*}

This formula is not true for all possible input tuples. For instance, if the shipping fee is $x_{F} = 55$ and country is $x_{C} = UK$, then the final shipping fee for $\history$ ($x_{F,2}$) is \$5 and for $\ahmod$ is \$50. Thus, $\history(\varAssign(\vcdbini)) \neq \ahmod(\varAssign(\vcdbini))$. Since the slice candidate $\idxs = \{1\}$ does not apply the second update, we get \$0 (for $\hislice{\idxs}$) and \$45 (for $\hisliceOf{\ahmod}{\idxs}$). Thus, the slice candidate may produce a  result for this database that is different to the one returned by $\hwhatif$ which means that $\idxs$ is not a valid slice.
\end{exam}

\subsubsection{Our Slicing Algorithm}
\label{sec:finding-slice}
Given a set of indexes $\idxs$, we now have a sound method for testing whether $\idxs$ is a slice for a historical what-if query $\hwhatif$. A brute force approach for computing a slice would be to test all possible subsets of indexes to determine the smallest possible slice. Note that even this method is not guaranteed to return a minimal slice, because the test we have devised it not complete. The disadvantage of the brute force approach is that there is an exponential number of candidates (all subsets of the histories). We propose instead a greedy algorithm that considers a linear number of candidates. The algorithm starts with a trivial slice $\idxs_{0} = [1,n]$ where $n$ is the number of updates in the history (recall from \Cref{sec:filter} that we can pad histories such that both $\history$ and $\ahmod$ have $n$ statements). It then iterates for $n$ steps. In each iteration, we remove index $i$ from the current slice $\idxs_{i}$ and test whether $\idxs_{i} - \{i\}$ is still a slice. If yes, then we set $\idxs_{i+1} = \idxs_{i} - \{i\}$. Otherwise, $\idxs_{i+1} = \idxs_{i}$. The final result produced by this algorithm is $\idxs_{n}$ which is guaranteed to be a valid slice.


\section{Optimized Program Slicing for Single Modifications}
\label{sec:optim-progr-slic}

Based on the VC-databases created by this step we then determine a static slice. We introduce a condition called \textit{dependency} that can be checked over the VC-database and determines whether an update's result depends on the modification $\deltaHist$. As we will demonstrate, the set of dependent updates from $\history$ ($\history(\deltaHist)$) is a static slice for the historical what-if query $\hwhatif$.

Observe that a statement can be excluded from reenactment if none of the tuples affected by the statement will be in the difference between $\history$ and $\history[\deltaHist]$. Note that any tuple in the difference has to be affected by at least one of the statements modified by a modification $\modi \in \deltaHist$, because a tuple that is not affected by any statement from $\deltaHist$ will be the same in $\history$ and $\history[\deltaHist]$ and, thus, cannot be in the result of $\hwhatif$.

Conversely, an update has to be included in a static slice if there exists at least one database $\db$ that contains a tuple which is affected by the statement and which is in the result of $\hwhatif$ over $\db$.

\begin{defi}
Let $\history$ be a history and $\history[\deltaHist]$ be a modified history, $t$ be a tuple in the relation, and $u_{old}$ be an unmodified update and $u_{new}$ the corresponding modified update for any modification $m \in \deltaHist$. Let $t_i$ be the tuple at $\history_i(t)$ ($\ahmod_i(t)$). Let $\pos(u)$ be the position of update $u$ in $\history$. We define the following condition for exclusion from a non-minimal slice of the history:
\begin{align*}
\exclusion{H}{\deltaHist}{u_i} =\ &\forall m \in \deltaHist \neg\exists \varAssign \exists t \in \db_{\history[\deltaHist], \pos(u_m)}\\
&(\cond_{u_{orig}}(\varAssign(t_{\pos({u_{m}})})) \land \cond_{u_{i}}(\varAssign(t_{i-1})))\\
&\lor (\cond_{u_{new}}(\varAssign(t_{\pos({u_{m}})})) \land \cond_{u_{i}}(\varAssign(t_{i-1})))
\end{align*}
\end{defi}

We use VC-Tables for symbolic execution of update operations and determining dependency of updates on  modifications by the historical what-if queries. Independent updates can be excluded from reenactment as their output is the same in $\history$ and $\history[\deltaHist]$.

We can determine dependent updates in the history by generating VC-tables for each $u$ and $ u'$ whereas $m = u \gets u'$ and $m \in \deltaHist$. Then, we apply symbolic execution on these VC-tables for the remaining updates in the history. For each examined update ($u_i$), if we can generate a possible word for a tuple that is modified either by both $u_i$ and $u$ or $u_i$ and $u'$, $u_i$ is a dependent update. Since, a possible world shows there is a possibility that a tuple was modified by the historical what-if query and the examined update which must be considered in the answer of the the historical what-if query.

\begin{exam}
In order to detect dependent update in \Cref{fig:running-vctb}, we examine generating a possible world for a tuple in the VC-Table that is modified by both the first ($u_1$) and the second update ($u_2$) in the history in \Cref{fig:Transitive-Transactions-Example}.
 After symbolic execution of the second update where $\cond:= x_{Country}=UK \wedge x_{Price} <=100$, there are four tuples in the VC-Table. The first tuple which has the conditional function
 $x_{Price} >=40 \wedge (x_{Country}=UK \wedge x_{Price} <=100)$ representing it is modified by both updates. The possible world can be generated by evaluating $(x_{Country} \leftarrow UK,x_{Price} \leftarrow 40,x_{ShippingFee} \leftarrow 5)(x_{Price}>=40 \wedge (x_{Country}=UK \wedge x_{Price} <=100))$. As $40>=40 \wedge (UK=UK \wedge 40 <=100):= true$. The possible world $\world''$ after executing the first and second update statements can be $\world''=\lbrace UK,40,5\rbrace$. So, the second update is dependent on the first update as it is possible that  a tuple is modified by both updates.
\end{exam}

\begin{theo}
Consider a historical what-if query $\hwhatif = \ahwhatif$ with a single modification $\deltaHist = \{ m\}$ for $m = \up_1 \gets \up_1'$ over history $\history= u_1,\ldots,u_{n}$.
The set $\idxs = \{ i \mid \neg \exclusion{\history}{\deltaHist}{u_i} \}$ is a slice for $\hwhatif$.
\end{theo}
\iftechreport{
  \begin{proof}

\newcommand{\iex}{\idxs_{excl}}
\newcommand{\iin}{\idxs_{in}}
\newcommand{\iexi}[1]{\idxs_{excl,#1}}
\newcommand{\iini}[1]{\idxs_{in,#1}}
\newcommand{\dslicei}[1]{\Delta_{#1}}

  Consider a history $\history = (u_1, \ldots, u_n)$, set of modifications $\deltaHist = \{m\}$ for $m = u_1 \gets u_1'$, and historical what-if query $\hwhatif = (\history, \db, \deltaHist)$. Let $\idxs_{in} = \{ \upPos(u_i) \mid \neg \exclusion{\history}{\deltaHist}{u_i} \}$ and $\iex = \{  \upPos(u_i) \mid \exclusion{\history}{\deltaHist}{u_i} \}$. We have to prove that $\idxs_{in}$ is a slice for $\hwhatif$. In the following let $\iexi{i}$ denote the first $i$ positions from $\iex$ and $\iini{i} = [1,n] - \iexi{i}$, i.e., from the histories all updates from $\iexi{i}$. We prove the theorem by induction over $i$. In the following we use $\Delta$ to denote $\iDiff{\history(\db)}{\ahmod(\db)}$ and $\dslicei{i}$ to denote $\iDiff{\hslice{\history}{\iini{i}}(\db)}{\hslice{\ahmod}{\iini{i}}(\db)}$.

\proofpar{Base case:}
Consider $\iexi{1} = {j}$ for some $j \in [1,n]$. To prove that $\iini{1}$ is a slice, we have to show that $\Delta = \dslicei{1}$. For sake of contradiction assume that $\Delta \neq \dslicei{1}$. Then there has to exist a tuple $\tup$ such that (i) $\tup \in \Delta \land \tup \not\in \dslicei{1}$ or (ii) $\tup \not\in \Delta \land \tup \in \dslicei{1}$.

\proofpar{$\tup \in \Delta \land \tup \not\in \dslicei{1}$:}
If $\tup \in \Delta$ then either $\tup \in \history(\db) \land \tup \not\in \ahmod(\db)$ or $\tup \not\in \history(\db) \land \tup \in \ahmod(\db)$. Since these two cases are symmetric, WLOG assume that $\tup \in \history(\db) \land \tup \not\in \ahmod(\db)$. It follows that $\exists \tup_{1} \in \db$ such that $\history(\tup_1) = \tup$ and $\neg \exists \tup_{1}' \in \db$ such that $\ahmod(\tup_{1}') = \tup$. Specifically, $\ahmod(\tup_{1}') \neq \tup$. Now let $\tup_i = \hslice{\history}{i}(\tup_1)$ and $\tup_i'= \hslice{\ahmod}{i}(\tup_1)$. Since $j \in \iex$, condition $\exclusion{\history}{\deltaHist}{u_j}$ has to hold which implies $\neg \theta_j(\tup_j)$ and $\neg \theta_j(\tup_{j}')$.

Now let us use $s_i$ to denote $\hslice{\history}{\iini{1} \cap [1,i]}(\tup_1)$ and $s_i'$ to denote $\hslice{\ahmod}{\iini{1} \cap [1,i]}(\tup_1)$.
Since $\hslice{\history}{j-1} = \hslice{\history}{\iini{1} \cap [1,j-1]}$ and $\hslice{\history}{j-1} = \hslice{\history}{\iini{1} \cap [1,j-1]}$, we have $t_{j-1} = s_{j-1}$ and $t_{j-1}' = s_{j-1}'$. Now since $\neg \theta_j(\tup_j)$ and $\neg \theta_j(\tup_{j}')$, $t_j = s_{j-1}$ and $t_j' = s_{j-1}'$. Furthermore, note that $\hslice{\history}{j+1,n} = \hslice{\history}{\iini{1} \cap [j+1,n]}$ and $\hslice{\ahmod}{j+1,n} = \hslice{\ahmod}{\iini{1} \cap [j+1,n]}$. Thus, it follows that $t = \tup_n = s_{n-1}$ (the sliced histories have one less update) and $\tup' = s_{n-1}'$ which contradicts $\tup \not\in \dslicei{1}$.

\proofpar{$\tup \not\in \Delta \land \tup \in \dslicei{1}$:}
Since $\hslice{\history}{\iini{1}}$ (and $\hslice{\ahmod}{\iini{1}}$) are histories, from \Cref{theo:data-slicing} follows that if $s \in \dslicei{1}$ then $s_1 \models \theta_1 \vee \theta_1'$ ($s_i$, $t_i$, $s_i'$ and $t_i'$ are as defined above). Because we know that $\exclusion{\history}{\deltaHist}{u_j}$ we also know that $\neg \theta_j(s_{j-1})$ and $\neg \theta_j(s_{j-1}')$. Applying the same argument as for the opposite direction proven above, this implies that $s \in \Delta$ which contradicts $s \not \in \Delta$.

\proofpar{Induction Step:}
Assume that $\iini{i}$ is a slice for $\hwhatif$, i.e., $\Delta = \dslicei{i}$. We have to show that $\iini{i+1}$ is also a slice, i.e., $\Delta = \dslicei{i+1}$. Note that $\dslicei{i+1}$ differs from $\dslicei{i}$ only in that it excludes an additional update at a position $j$, i.e., $\iini{i+1} = \iini{i} - \{ j \}$ and $\forall l \in \iexi{i}: j > l$. Let $k = \max(\iexi{i})$. The remainder of the argument proceeds analog to the base case. For sake of contradiction assume that there exists a tuple $\tup$ with $\tup \in \Delta \land \tup \not\in \dslicei{i+1}$. Let use again employ the notation $s_i$, $s_i'$, $t_i$ and $t_i'$ as above. Then by applying the argument from the base case iteratively, we can show that $t_{j-1} = s_{j-1}$ and $t_{j-1}'= s_{j-1}'$. Together with $\exclusion{\history}{\deltaHist}{u_j}$ this implies $t_j = s_{j-1}$ and $t_{j}' = s_{j-1}'$ (again using the same argument already applied in the base case) and in turn implies $\tup \in \dslicei{i+1}$. The proof for $\tup \not\in \Delta \land \tup \in \dslicei{i+1}$ is also analog to the base case and, thus, we omit it here.

    $\,$\\[-3.5mm]
  \end{proof}
}

\section{Optimizing Histories with Inserts}
\label{sec:optim-reen-hist}

\newcommand{\honlyup}{\history_{noIns}}
\newcommand{\reenactNoR}[1]{\ract{#1 / R}}

In \Cref{sec:dep-ana} and \Cref{sec:up-vc-tb} we have limited the discussion to histories consisting only of update and delete statements. The reasons for this restriction is that we now introduce an optimization that splits a reenactment query for a history into two parts that can be optimized individually: (i) one part that only simulates update and delete statements over the database at the time of the beginning of the history and (ii) a second part that evaluates the whole history, but only over tuples inserted by insert statements. We use program slicing to optimize (i). The input data size for (ii) is bound by the number of statements in the history and, thus, typically negligible . Note that our symbolic execution technique required by program slicing requires solving a MILP program (an NP-hard problem) whose size is polynomial in the size of the history. Thus, while it may be possible to extend program slicing techniques to deal with inserts, the costs of evaluating (ii) is polynomial in the size of the history and, thus, it is not possible to amortize the cost of program slicing for this part.

We start by stating the idea underlying our optimization, before giving a formal definition of this optimization and proving its correctness.
Recall from \Cref{def:reenactment-queries} that the reenactment query for an insert statement $\ins{\tup}(R)$ is a union between the state of the relation before the insert and a singleton relation containing the inserted tuple. Updates are reenacted using projections and deletions using selection. As an example consider a history $\history$ consisting of a single insert $u_0$ followed by $n$ update statements $u_1$ to $u_n$. \Cref{fig:insert-opt-before} shows the structure of the reenactment query for this history. Using the standard algebraic equivalences shown below which allow us to pulling a union through a projection or selection, we can pull the union up through the projections reenacting the updates of the history. The algebra tree for the resulting query is shown in \Cref{fig:insert-opt-after}. Note that in the rewritten query (i) the right branch only accesses the tuple inserted by the insert statement and (ii) the left input to the union is equal to the reenactment query for a history $\honlyup$ that is the result of deleting the insert statement from $\history$.\\[-6mm]

\begin{align*}
  \projection_A(\query_1  \union \query_2) &\equiv \projection_A(\query_1) \union \projection_{B \to A}(\query_2)\\
  \selection_{\theta}(\query_1 \union \query_2) &\equiv \selection_\theta(\query_1) \union \selection_\theta(\query_2)
\end{align*}

Generalizing this example, the algebraic equivalences shown above are sufficient for rewriting the reenactment query of any history $\history$ into a query $\ract{\honlyup} \union \reenactNoR{\history}$ where  $\reenactNoR{\history}$ is derived from $\ract{\history}$ by replacing the subquery (union) corresponding to the first insert $\up$ in the history with the singleton relation $\{t\}$ containing the tuple inserted by $\up$. Importantly, then we can apply program slicing to optimize $\honlyup$.

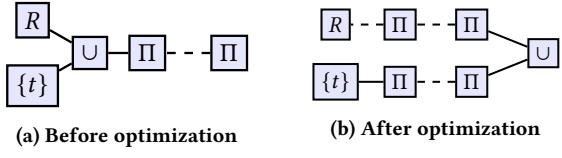
\begin{figure}[t]
  \centering

\begin{subfigure}{0.5\linewidth}
  \centering
  \resizebox{0.8\linewidth}{!}{
  \begin{tikzpicture}
    [
    nod/.style={rectangle,draw=black,fill=blue!10,thick},
    edg/.style={thick},
    ]
    \node[nod] (un) at (2.2,0) {$\projection$};
    \node[nod] (u1) at (1.2,0) {$\projection$};
    \node[nod] (union) at (0.5,0) {$\union$};
    \node[nod] (R) at (-0.2,0.4) {$R$};
    \node[nod] (t) at (-0.2,-0.4) {$\{t\}$};

    \draw[edg,dashed] (un) -- (u1);
    \draw[edg] (u1) -- (union);
    \draw[edg] (union) -- (R);
    \draw[edg] (union) -- (t);
  \end{tikzpicture}
  }
  \caption{Before optimization}\label{fig:insert-opt-before}
\end{subfigure}
\begin{subfigure}{0.45\linewidth}
  \centering
  \resizebox{0.9\linewidth}{!}{
  \begin{tikzpicture}
    [
    nod/.style={rectangle,draw=black,fill=blue!10,thick},
    edg/.style={thick},
    ]
    \node[nod] (union) at (2.7,0) {$\union$};

    \node[nod] (lu1) at (1.7,-0.4) {$\projection$};
    \node[nod] (lun) at (0.7,-0.4) {$\projection$};
    \node[nod] (R) at (-0.2,0.4) {$R$};

    \node[nod] (ru1) at (1.7,0.4) {$\projection$};
    \node[nod] (run) at (0.7,0.4) {$\projection$};
    \node[nod] (t) at (-0.2,-0.4) {$\{t\}$};

    \draw[edg] (union) -- (lu1);
    \draw[edg,dashed] (lu1) -- (lun);
    \draw[edg,dashed] (run) -- (R);

    \draw[edg] (union) -- (ru1);
    \draw[edg,dashed] (ru1) -- (run);
    \draw[edg] (lun) -- (t);
  \end{tikzpicture}
  }
  \caption{After optimization}\label{fig:insert-opt-after}
\end{subfigure}
\vspace{-3mm}
\label{fig:reenact-opt-example}
  \caption{Example structure of an reeactment query with a single insert. The unions can be pulled up to create two separate queries: the left query accesses R and is the same as the reeactment query for the history without inserts while the right one only accesses inserted tuples.}
\end{figure}

\iftechreport{
\begin{figure}[t]
  \centering
  \begin{minipage}{0.4\linewidth}
  \begin{gather*}
    \label{eq:1}
    \frac{\expr \defas \expr_1 < \expr_2}{
      \begin{aligned}
        \var_1 - \var_2 + \bvar \times \upBound &\geq 0\\
        \var_2 - \var_1 + (1- \bvar) \times \upBound &> 0
      \end{aligned}
    }\\
    \frac{ \expr \defas \expr_1 \leq \expr_2}{
      \begin{aligned}
        \var_1 - \var_2 + \bvar \times \upBound &> 0\\
        \var_2 - \var_1 + (1- \bvar) \times \upBound &\geq 0
      \end{aligned}
    }\\
    \frac{\expr := \expr_1 \wedge \expr_2}{
      \begin{aligned}
        \bvar_1 +\bvar_2 -2\bvar -1 &\leq 0 \\
        \bvar_1 +\bvar_2 -2\bvar &\geq 0
      \end{aligned}
    }\\
    \frac{\expr := \expr_1 \vee \expr_2}{
      \begin{aligned}
        \bvar_1 +\bvar_2 -2\bvar &\leq 0 \\
        \bvar_1 +\bvar_2 -\bvar &\geq 0
      \end{aligned}
    }\\
     \frac{\expr \defas \expr_1 + \expr_2}{\var_1 + \var_2 - \var = 0}
  \end{gather*}
  \end{minipage}
  \begin{minipage}{0.49\linewidth}
    \[
      \frac{\expr \defas \sqlCase{\expr_c}{\expr_1}{\expr_2}}{
      \begin{aligned}
        \var_{if} + \var_{else} - \var                             & = 0    \\
        \var_{if} - \var_1                                         & \leq 0 \\
        \var_{if} - \var_1 + \upBound - \upBound \cdot \bvar_{c}   & \geq 0 \\
        \var_{if} - \upBound  \cdot \bvar_{c}                      & \leq 0 \\
        \var_{if} + \bvar_{c}  \cdot \upBound                      & \geq 0 \\
        \var_{else} - \var_2                                       & \leq 0 \\
        \var_{else} - \upBound + \upBound \cdot \bvar_{c}          & \leq 0 \\
        \var_{else} - \var_2 + \upBound \cdot \bvar_{c}            & \geq 0 \\
        \var_{else} + \upBound - \upBound \cdot \bvar_{c}          & \geq 0
      \end{aligned}
    }
  \]
  \begin{gather*}
    \frac{\expr \defas x}{\var_{x}}
     \frac{\expr \defas \neg \expr_1 }{
      \bvar + \bvar_1 =1
    }
  \end{gather*}
  \end{minipage}

  \caption{Compilation rules for translating constraints into an MILP (the remaining comparison operators are omitted since they can be expressed using boolean operations).}
  \label{fig:milp-compilation}
\end{figure}

In these rules we use $\var$, $var_1$, \ldots to denote integer variables and $\bvar$, $\bvar_1$, \ldots to denote boolean variables. Furthermore, $\upBound$ denotes an integer constant that is larger than all integer values used as attribute values.
}

\iftechreport{
\section{MILP Compilation}
\label{sec:sym-lin}

To evaluate the conditions $\aslicetest$ for program slicing, we first translate these conditions into existential form and then translate them into a MILP (mixed integer linear programming) program~\cite{schrijver1998theory}. The resulting program can then be solved using a standard MILP solver, e.g., we use CPLEX~\cite{cplex2009v12} to test the satisfiability of these conditions. %
We now introduce a compilation scheme that translates such logical expressions into linear constraints. 
These rules are applied recursively to a constraint. Each rule generates a set of linear constraints. The MILP generated by these rules for a boolean expression $\expr$ consists of the union of all linear constraints produced by the rules for the subexpressions of $e$. For each subexpression $\expr'$ of $\expr$, the compilation produces a variable $\var'$ ($\bvar'$ if $\expr'$ is boolean) for which any solution to the MILP sets $\var'$ ($\bvar'$) to the value that $\expr'$ evaluates to. An additional constraint $\bvar = 1$ is added to ensure that only solutions that satisfy $\expr$ are produced. %
The translation rules applied here are mostly well-known rules applied in linear programming and many have been used in related work (e.g.,~\cite{MeliouS12}).
}

\section{Related Work}
\label{sec:related-work}

What-if queries determine the effect of a hypothetical change to an input database on the results of query.  What-if queries~\cite{hung17,deutch13,ZG95,bourhis16} are often realized using incremental view maintenance to avoid having to reevaluate the query over the full input including the hypothetical changes. %
The how-to queries of Tiresias~\cite{MeliouS12}  determine how to translate a  change
to a query result into modifications of the input data. Similar to their approach, our system provides support for historical how-to queries, allowing the definition and integrated evaluation of a large set of constrained optimization problems, specifically Mixed Integer Programming problems, on top of a relational database system.
The QFix system~\cite{wang16} is essentially a variation on this where the change to the output has to be achieved by a change to a query (update) workload.
The query slicing technique of QFix is similar to our program slicing optimization. The main difference is that
we apply symbolic execution to a relation with a  single symbolic tuple, i.e.,
the number of constraints we produce is independent of the database instance size.

Several provenance models for relational queries have been studied such as Why-provenance, minimal Why-provenance~\cite{BK01}, and Lineage~\cite{CW00b} and
Provenance semirings by Green et al.~\cite{GK07} generalize these models for positive relational algebra.
Reenactment is a technique for replaying a update operations and  transactional history using queries. The reenactment query for a transaction is equivalent to the transaction for the transactional history under MV-semiring~\cite{AG14,AG17,AG18} semantics to create the same database state with same provenance.

The connection of provenance and program slicing was %
first observed in~\cite{cheney07}.
We present a method that statically analyzes potential provenance dependencies among statements in the history using a method which borrows ideas from symbolic execution~\cite{bucur14,K76,luckow14}, constraint databases~\cite{gomez14,kuper13}, program slicing~\cite{W81}, and expressive provenance models~\cite{AD11d}.
Symbolic execution has been used in different researches such as software testing~\cite{cadar13}.
Cosette~\cite{chu2017} is an automated prover for checking equivalences of SQL queries which converts input queries to constraints over symbolic relations.
Rosette~\cite{torlak2014} is a
language for constraints programming. When executed, the query
function will generate constraints which can then be solved.
Transaction repair approach in~\cite{dashti17} also detects dependency of update operations using closure bound to predicates but they use data objects and a list
of their different versions. In contrast, our approach uses a single symbolic data instance which requires less memory.

\section{Experiments}
\label{sec:experiments}
\begin{figure*}[t]
\begin{minipage}[b]{0.29\linewidth}
  \centering
	\includegraphics[width=1.1\linewidth,trim=0 0 0 0, clip]{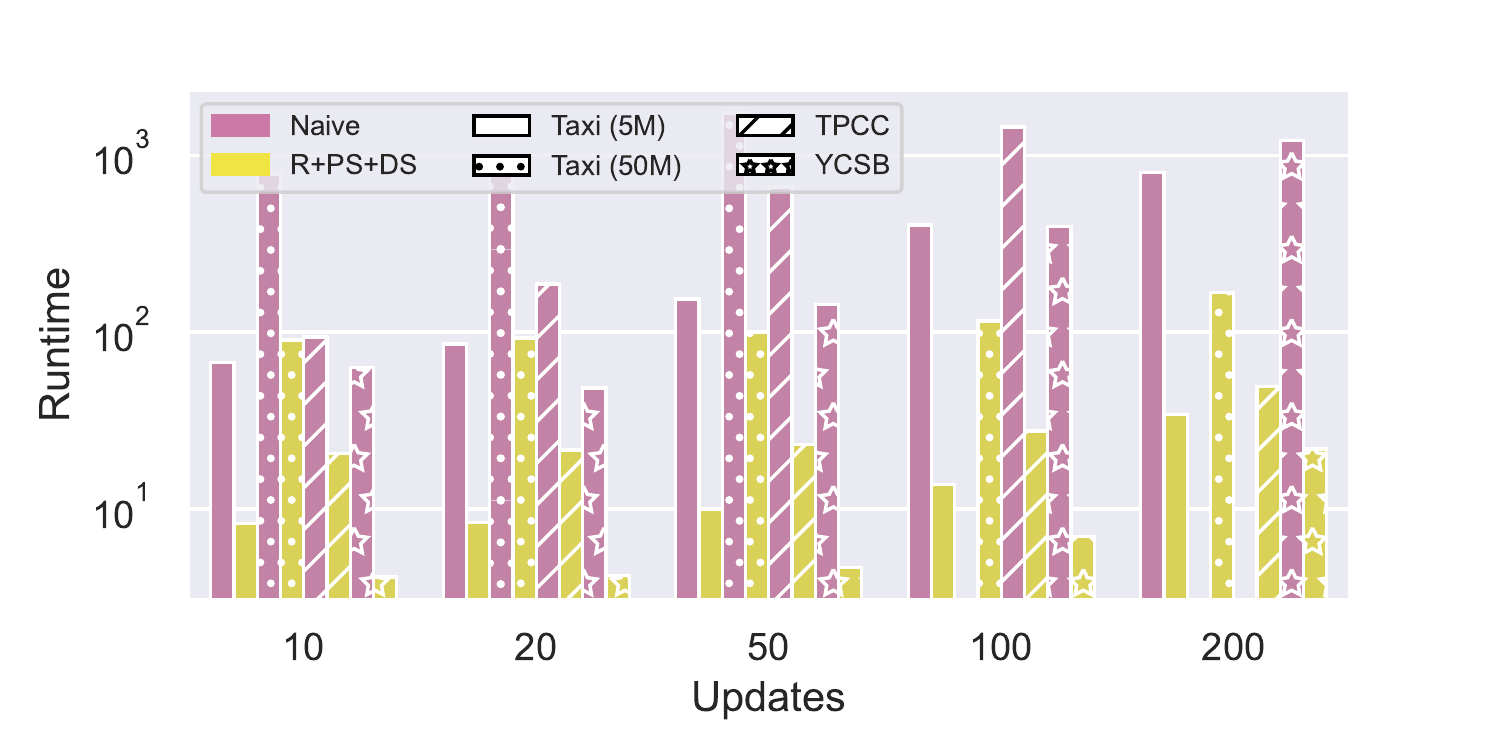}\\
	\vspace{-4mm}
	\caption{Naïve vs. Mahif (sec)}
	\label{fig:Naive vs Mahif}
  \end{minipage}
  \begin{minipage}[b]{0.29\linewidth}
    \centering
    \includegraphics[width=0.95\linewidth,trim=0 0 0 0, clip]{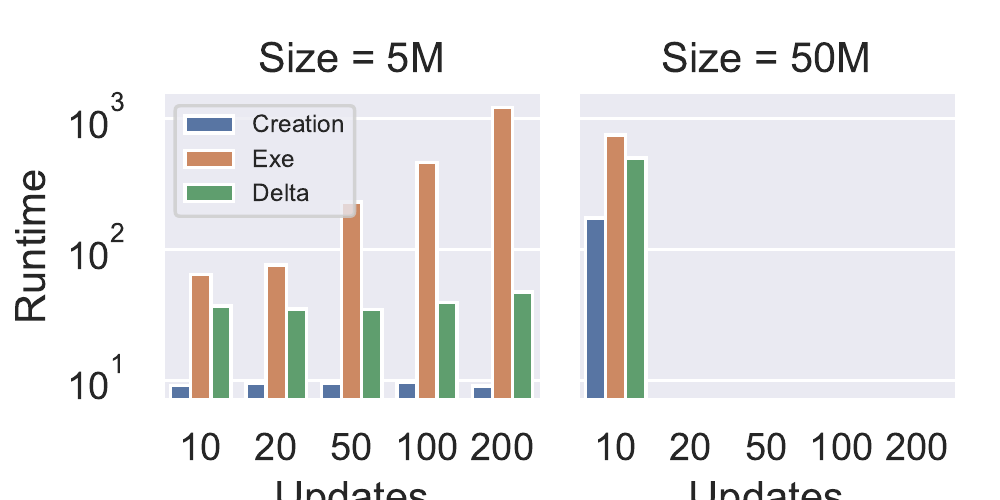}\\
  \vspace{-3mm}
  \caption{Breakdown Naïve}
  \label{fig:Naive Method}
  \end{minipage}
\begin{minipage}[b]{0.32\linewidth}
	\scalebox{0.8}{
	\scriptsize
	\begin{tabular}{|r|rrrr|rrrr|}
      \hline
      & \multicolumn{4}{c|}{\textbf{5M}} & \multicolumn{4}{c|}{\textbf{50M}} \\
      \hline
      \textbf{Updates} & \textbf{PS}     & \textbf{Exe}    & \textbf{R+PS+DS} & \textbf{R}       & \textbf{PS}     & \textbf{Exe}     & \textbf{R+PS+DS} & \textbf{R}        \\
      \hline
		10      & 0.07  & 8.08  & 8.14   & 63.63  & 0.07  & 90.11   & 90.18  & 722.23  \\
		20      & 0.18  & 8.29  & 8.47   & 81.12  & 0.18  & 90.33   & 90.51  & 878.70  \\
		50      & 1.30  & 9.15  & 10.45  & 133.29 & 1.29  & 83.32  & 84.61  & 1414.94 \\
		100     & 8.46  & 18.76 & 27.23  & 218.87  & 8.46  & 108.07 & 116.53 & 2310.84 \\
      200     & 62.13 & 12.36 & 74.49  & 400.71 & 62.22 & 132.07 & 194.29 & 4173.17\\
\hline
	\end{tabular}
	}

	\caption{Breakdown Mahif}
	\label{fig:Mahif Method}
\end{minipage}\\
  \begin{minipage}[b]{0.245\linewidth}
	\centering
	\includegraphics[width=0.8\linewidth,trim=0 0 0 0, clip]{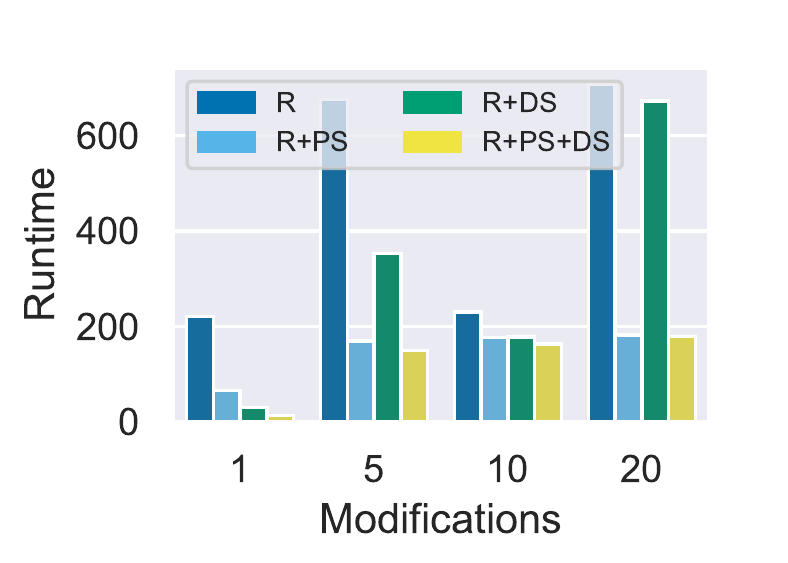}\\
	\vspace{-3mm}
	\caption{Mult. Modifications}
	\label{fig:Multimod}
\end{minipage}
\begin{minipage}[b]{0.265\linewidth}
  \hspace{-0.7cm}
    \includegraphics[width=1.2\linewidth,trim=0 0 0 0, clip]{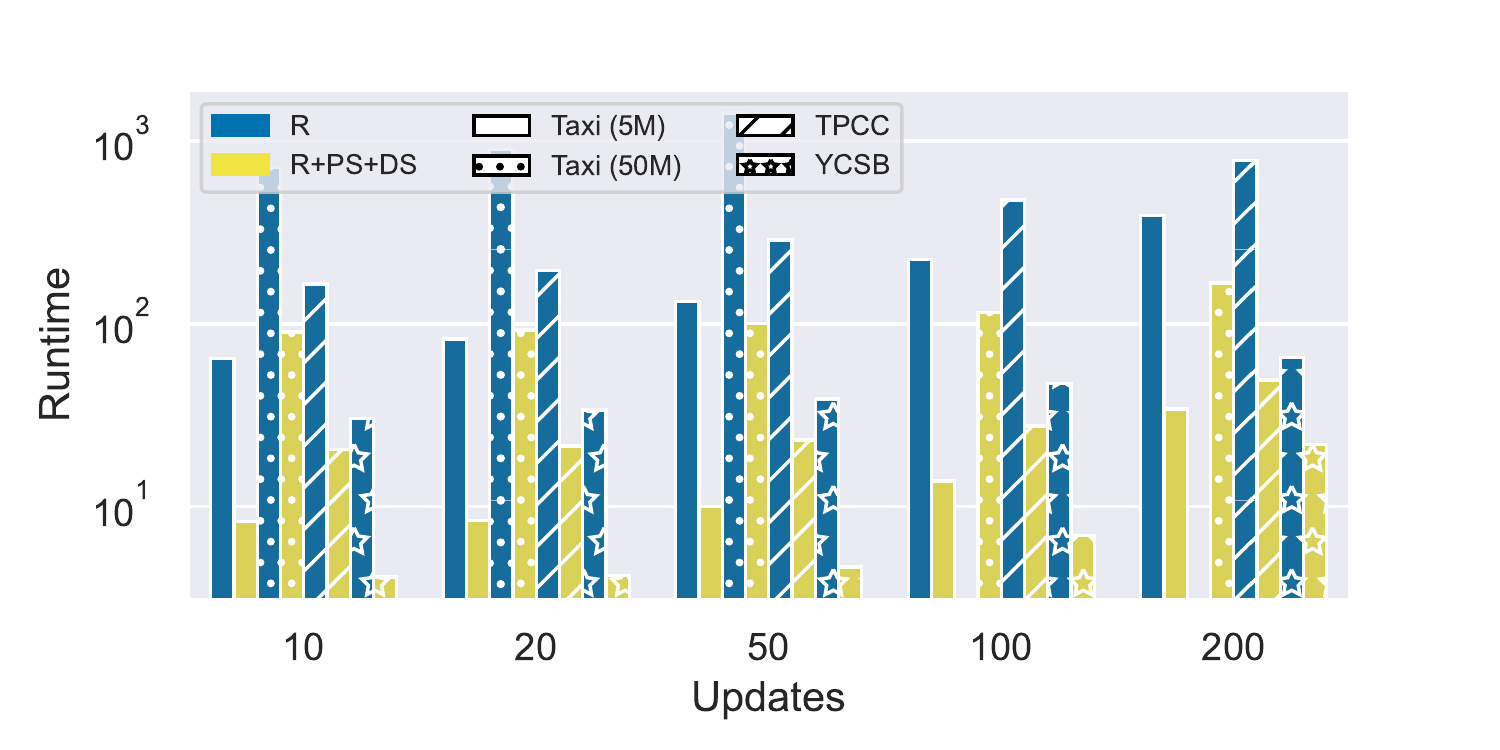}\\
    \vspace{-8mm}
  \caption{Optimization}
  \label{fig:optimization}
  \end{minipage}
  \begin{minipage}[b]{0.245\linewidth}
    \scalebox{1}{\includegraphics[width=1\linewidth,trim=0 0 50pt 0, clip]{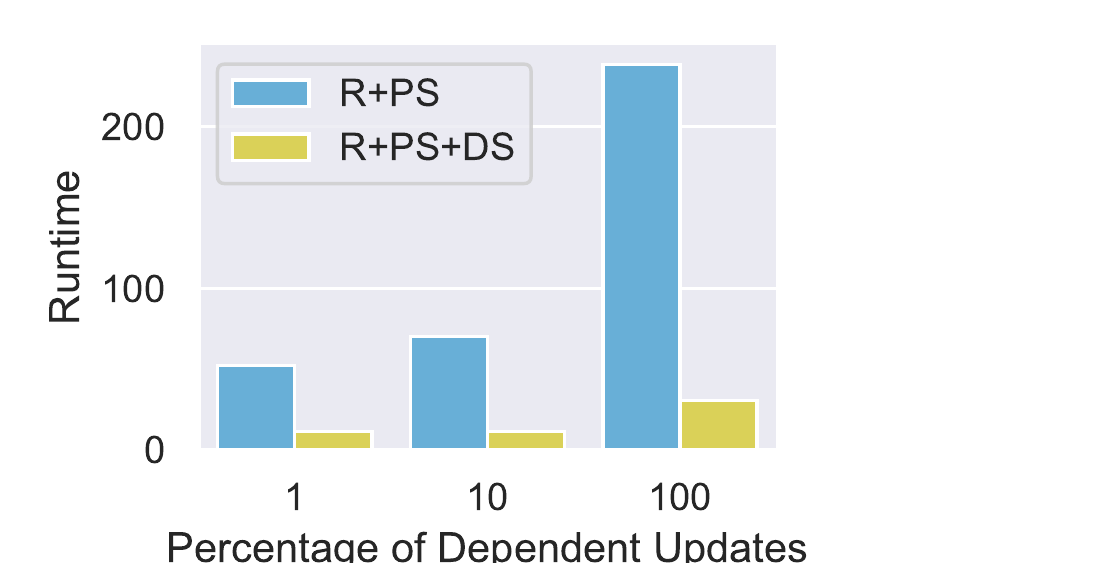}}\\
  \vspace{-8mm}
  \caption{Dependent Updates}
  \label{fig:Dependent Updates}
  \end{minipage}
   \begin{minipage}[b]{0.225\linewidth}
    \includegraphics[width=1\linewidth,trim=0 0 0 0, clip]{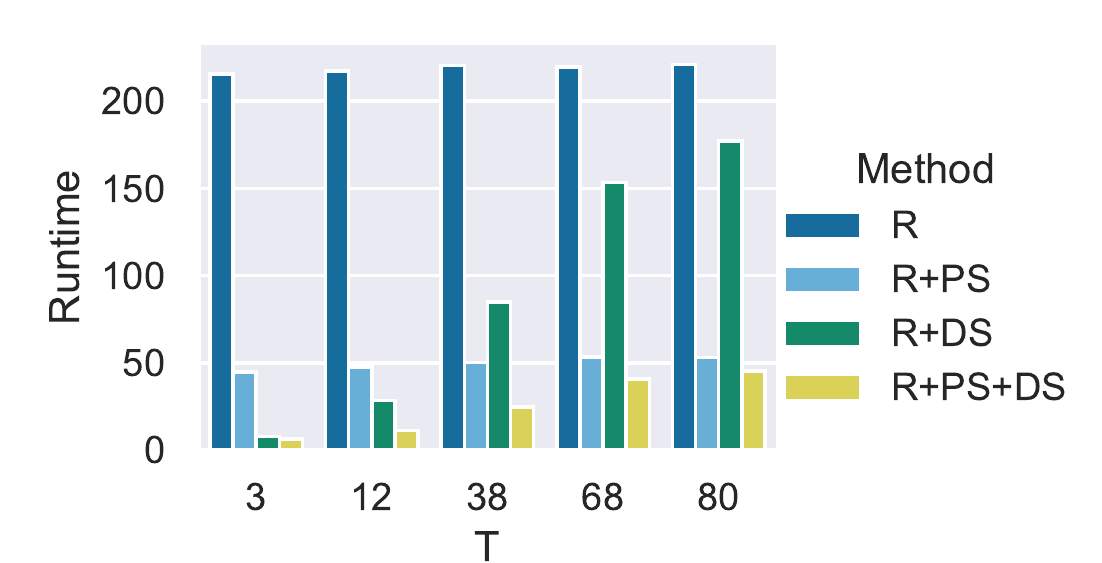}\\
  \vspace{-8mm}
  \caption{Affected Data}
  \label{fig:Affected Data}
  \end{minipage}\\
\end{figure*}

We have conducted experiments to 1) evaluate the performance of our approach and compare it with the naïve approach, 2) examine the effectiveness of the proposed optimizations, and 3) study how our approach scales in database size and %
other important factors.
All experiments were performed on a machine with 2 x AMD Opteron 4238 CPUs (12 cores total), 128 GB RAM, and 4 x 1TB 7.2K
HDs in a hardware RAID 5 configuration. We used PostgreSQL 11.4 as the database backend.
Experiments were repeated with variances in all cases being under 1 second, and the average runtime is reported.

\subsection{Datasets and Workload}
We use a taxi trips dataset from the City of Chicago's open data portal \footnote{https://data.cityofchicago.org/Transportation/Taxi-Trips/wrvz-psew downloaded on 2020-10-13}, \reva{as well as the standard TPC-C \footnote{TPC-C is an On-Line Transaction Processing Benchmark: http://www.tpc.org/tpcc/} and YCSB \cite{CooperSTRS10} benchmarks to examine the performance of our approach}. The \revm{taxi trip} dataset contains data about trips reported to the City of Chicago as a regulatory agency. The original dataset has $\sim100$M rows and 23 attributes.
The dataset contains trip information such as the \texttt{Company} (the taxi company), the \texttt{Taxi ID}, \texttt{Trip Start Timestamp}
(when the trip started), \texttt{Trip Seconds} (duration of the trip in seconds), \texttt{Trip Miles}
(distance of the trip in miles), \texttt{Pickup Community Area}, \texttt{Tips}, \texttt{Tolls}, \texttt{Extras} (tips, tolls and extra charges for the trip), and \texttt{Trip Total} (total cost of the trip).
We used samples from these tables amounting to 10\% ($5M$) and 50\% ($50M$) of the entire taxi dataset as the basis for later experiments. \reva{The TPC-C and YCSB benchmark databases were generated with the Benchbase \cite{DifallahPCC13} application. The TPC-C database was initialized with a scale factor of 100, where the \texttt{stock} relation consisting of 10 million rows was used in workloads. The YCSB database was initialized with a scale factor of 5000, resulting in a singular relation consisting of 5 million rows. The workloads generated by Benchbase for each benchmark were modified to update the proportion of tuples as required by the experiments.}

\subsection{Transactional Workload}

Unless stated otherwise, we use historical what-if queries with a single modification that modifies the first update in a history over a single relation.
We vary the following parameters. $U$ is the number of updates in history (e.g. $U100$ for a 100 updates). Operations in the history that operate over other relations are excluded. $M$ is the number of modifications made to the history.
$D$ is the percentage of updates that are dependent on the update(s) modified by the historical what-if query. We use 10\% as the default value for $D10$.
$T$ is the percentage of tuples in the relation that are affected by each dependent update (the default is 10\%), where $T0$ is less than 1\%. $I$ and $X$ are the percentage of statements in the history that are inserts or deletes respectively.
Non-dependent update statements affect a \reva{fixed} proportion of the data equivalent to the value of $T$, though independent from the tuples modified by dependent updates. %

\subsection{Compared Methods}
We compare the following methods in our experiments.
\textbf{Naïve (N)}: This method creates a copy of the database as of the start time of the history which is modified by the what-if query (\texttt{Creation}), executes $\deltaHist$ over this copy (\texttt{Exe}), and then computes the delta $\iDiff{\history(\db)}{\history[\deltaHist](\db)}$ by running a query over the current database state and the updated copy (\texttt{Delta}).
\textbf{Reenactment Only (R)} creates a reenactment query for $\history$ and for $\deltaHist$.
We use run both reenactment queries over the database, and then compute the delta. \textbf{Reenactment with Data Slicing (R+DS)}: same as the previous method except that we restrict reenactment to the part of data that is determined to be relevant  by our data slicing optimization. \textbf{Reenactment with Program Slicing (R+PS)}: same as the \textit{R} method except that we only reenact the subset of updates in the history determined by our program slicing optimization. \textbf{Reenactment with Program Slicing + Data Slicing (R+PS+DS)}:  we apply both optimization techniques.

\Cref{fig:Naive vs Mahif} shows the naïve method's performance in comparison to \textit{R+PS+DS}. \Cref{fig:optimization} shows the gap between reenactment alone and reenactment with all optimizations enabled. Given the clear efficiency gains in both cases, the naïve method and reenactment alone have been omitted from other experiments to focus on comparing our optimizations to each other. \Cref{fig:Mahif Method} breaks down the cost of \textit{R+PS+DS} into \texttt{PS} and \texttt{Exe}, and together they form the runtime of \textit{R+PS+DS} which should be compared to the cost of \textit{R} (\texttt{Reenact All}).

\begin{figure*}[t]
\begin{minipage}[b]{0.325\linewidth}
    \includegraphics[width=1\linewidth,trim=45pt 0 50pt 0, clip]{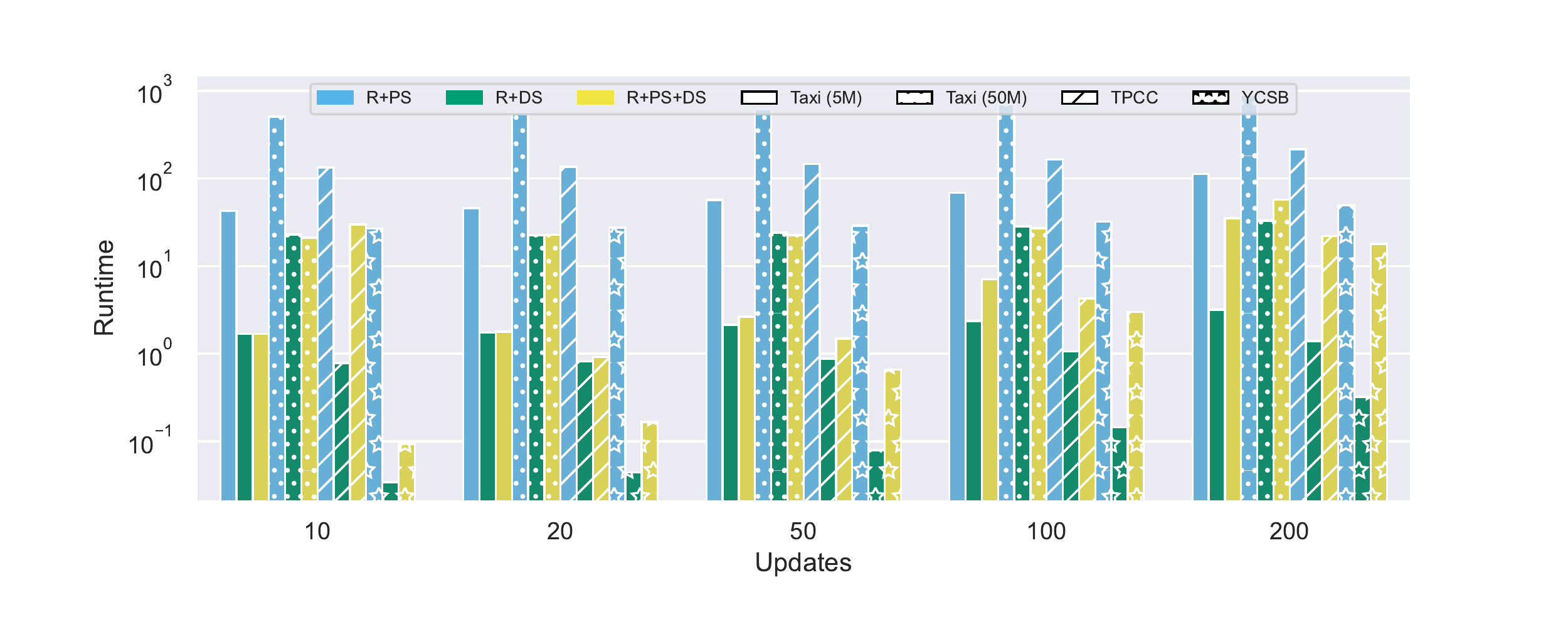}\\
    \vspace{-8mm} 
  \caption{\reva{Datasets} with T0}
  \label{fig:Relation Size}  
  \end{minipage}
 \begin{minipage}[b]{0.325\linewidth}
    \includegraphics[width=1\linewidth,trim=45pt 0 50pt 0, clip]{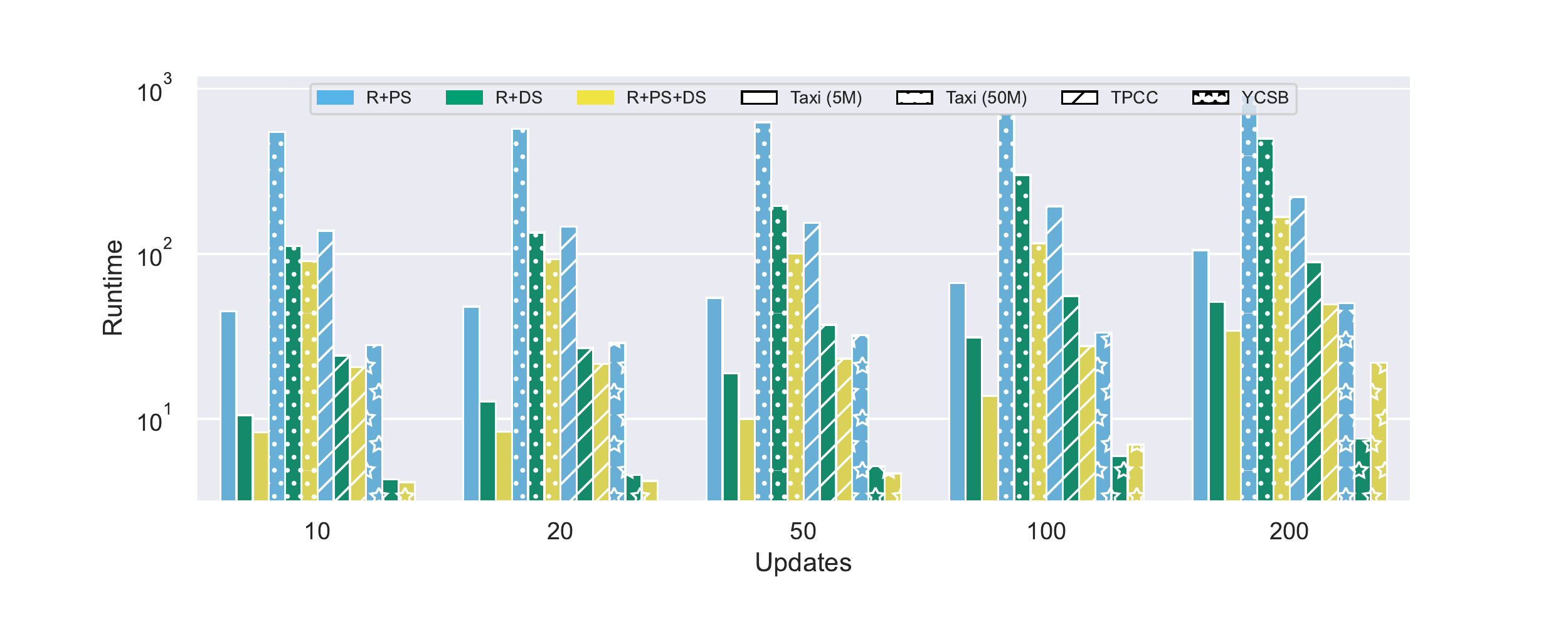}\\
    \vspace{-8mm} 
  \caption{\reva{Datasets} with T10}
  \label{fig:Relation Size1}  
  \end{minipage}
   \begin{minipage}[b]{0.325\linewidth}
    \includegraphics[width=1\linewidth,trim=45pt 0 50pt 0, clip]{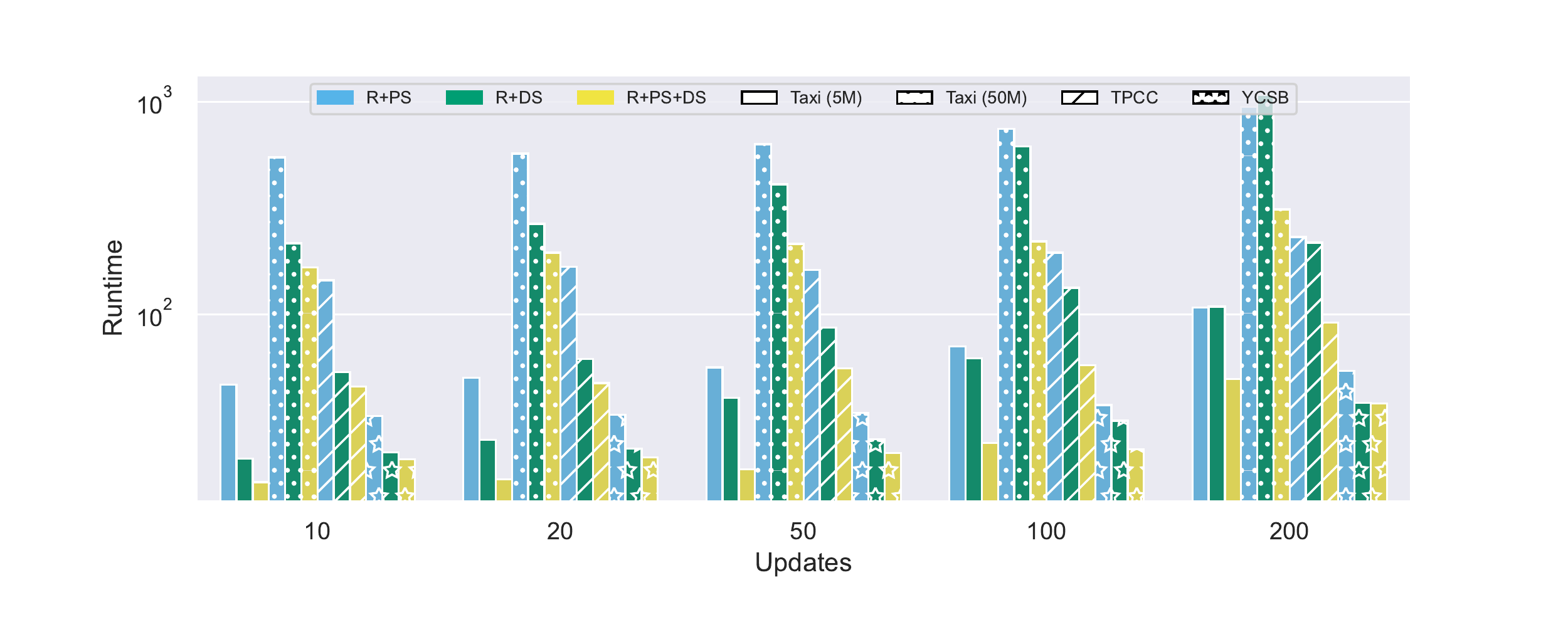}\\
    \vspace{-8mm} 
  \caption{\reva{Datasets} with T25}
  \label{fig:Relation Size2}  
  \end{minipage}
\end{figure*}

\subsection{Optimization Methods}
We evaluate the effectiveness of our proposed optimization methods by varying a set of parameters to observe which workload characteristics benefit each of our proposed optimizations.

\partitle{Varying Datasets (at D10)}
\label{sec:exp-relation-size}
We vary \reva{the datasets} used and number of updates in \Crefrange{fig:Relation Size}{fig:Relation Size2} to illustrate the effectiveness of our approach on increasingly large datasets and selectivities. Overall, we see that our approach scales very well with relation size. As the cost of program slicing is independent of the relation size, larger datasets benefit much more from program slicing. For smaller datasets, program slicing has diminishing returns as the cost of program slicing may be more expensive than it would be to run \textit{R} or \textit{R+DS} over the entire history. At low selectivity (\Cref{fig:Relation Size}), we see that \textit{R+DS} is very competitive with \textit{R+PS+DS} particularly in the smaller dataset (Taxi Trips 5M) as reenactment over the entire history with a small relation and an even smaller proportion of affected data input to the reenactment is cheaper than the cost of solving the MILP problem. \reva{Notably, the YCSB dataset demonstrates that the MILP cost exceeds data slicing, as data slicing is well-served by the physical correlation of key used to update the data.} Given larger proportions of data to be reenacted (\Cref{fig:Relation Size1} and \Cref{fig:Relation Size2}), we see that the combined optimization \textit{R+PS+DS} is consistently an improvement over either of these optimizations individually. The optimal case for our proposed combined optimization \textit{R+PS+DS} is when the size of the input data (the affected data as determined by data slicing) is large enough that calculating and reenacting over a slice is worth the execution cost of MILP.

\partitle{Varying Percentage of Dependent Updates (at T10)}
\label{sec:exp-dep}
\Cref{fig:Dependent Updates} demonstrates the effect the proportion of dependent updates in a given history has on \textit{R+PS}, and how the addition of data slicing (\textit{R+PS+DS}) is an effective way to mitigate these effects. This experiment uses the \revm{$5M$ row taxi trip table, with} the standard defaults of T10 (10\% of tuples affected by modified updates) and U100 (100 updates in the history). As the proportion of dependent updates in the history increase, the data demonstrates that program slicing becomes less effective as more updates have to be included in the slice over the history. At D100 (100\% of updates are dependent), program slicing is not beneficial at all, but also incurs the MILP solver cost as well. However, as can be seen in the figure, data slicing is useful to mitigate this effect, as the input to the reenactment is filtered to include only a fraction of the data, making it more effective than \textit{R+PS} at D100.

\partitle{Varying Affected Data (at U100, D1)}
\label{sec:exp-ds}
The effect of the  percentage of  tuples affected by the historical what-if query is examined  in \Cref{fig:Affected Data}. This experiment is executed for 100 updates on the taxi trips relation with $5M$ rows. For example, $T3$ means 3\% of tuples ($\sim150$K out of 5M) are modified by the historical what-if query. The result demonstrates that varying $T$ does not change the performance of \textit{R+PS}, as the amount of dependent updates remains constant. However, increasing $T$ increases the runtime of \textit{R+DS} and \textit{R+PS+DS} considerably as data slicing becomes less efficient due to the greater amount of input data that needs to be accessed during reenactment. However, at moderate selectivities (T68), we see that \textit{R+PS+DS} provides enough filtering over the history and data to be more performant than either optimization alone. The proportion of tuples affected by the dependent updates is therefore inversely proportional to the effectiveness of data slicing.

\subsection{Mixed workloads}

\begin{figure}[t]
	\begin{minipage}[b]{0.48\linewidth}
		\includegraphics[width=1\linewidth,trim=0 0 0 0, clip]{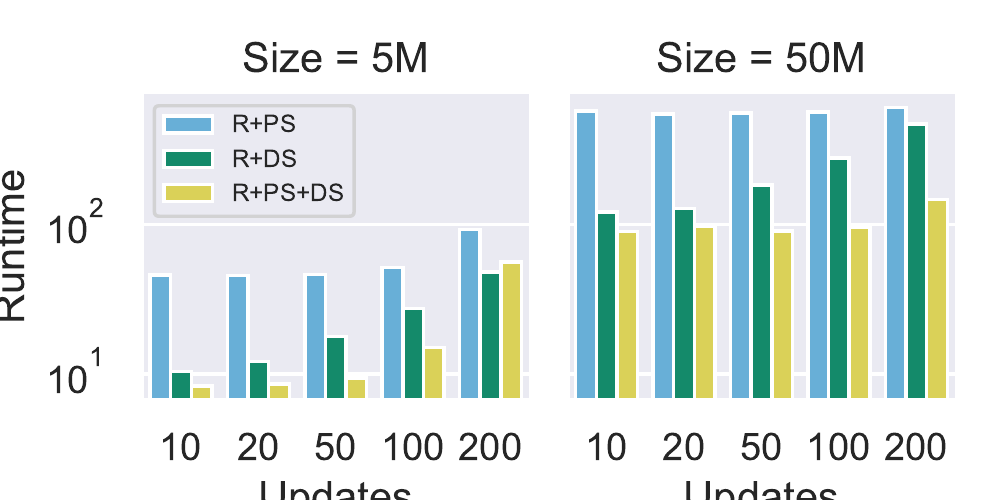}
		\vspace{-6.5mm}
		\caption{Inserts: I10, T10}
		\label{fig:Inserts at I10}
	\end{minipage}
	\begin{minipage}[b]{0.50\linewidth}
		\includegraphics[width=1\linewidth,trim=0 0 0 0, clip]{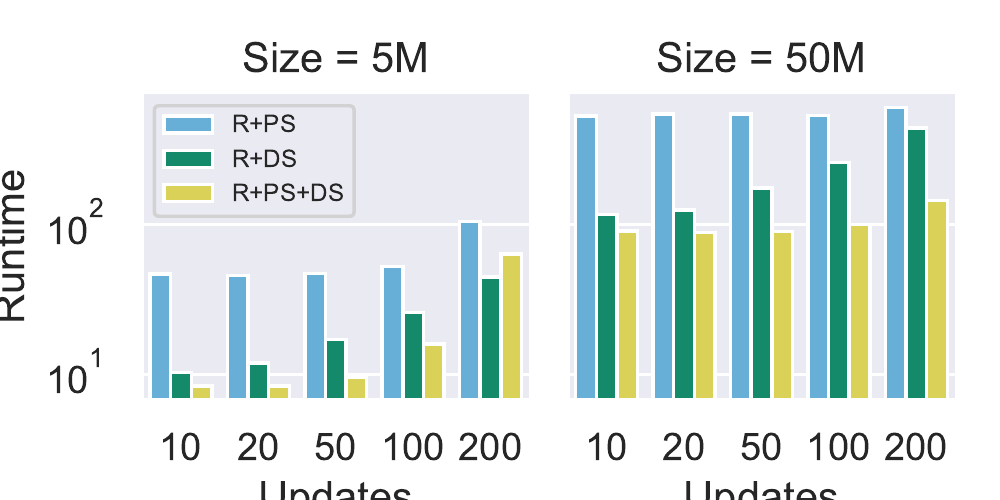}
		\vspace{-6.5mm}
		\caption{Mixed: I10, X10, T10}
		\label{fig:Mixed Updates at IX10}
	\end{minipage}
\end{figure}

\label{sec:mixed-workl-updat}

We now consider workloads that contain deletions, inserts, and updates. Since deletes are handled in a similar fashion, we mainly focus on evaluating the impact of the fraction of inserts on performance. \revm{We use the taxi trip tables for this experiment to demonstrate the scaling factor across otherwise homogeneous tables.}

From  \Cref{fig:Mixed Updates at IX10} we can see that \textit{R+PS+DS} outperforms the other methods introducing workloads of mixed updates, inserts, and deletes. When comparing to similar workloads presented in Figures 21-23, we see that introducing deletes and inserts into our workload in lieu of updates makes the reenactment and its optimizations cheaper with regards to runtime. While deletes require constraints in the MILP solver to program slice, their constraints are fewer in number and cheaper to execute given the lack of CASE expressions. Inserts are much cheaper to process than program slicing an update as we are able to reenact the unsliced prefix of the history on a very small amount of tuples (only the tuples being inserted, typically a very small fraction of a given workload).

\subsection{Varying the number of modifications}\label{sec:vary-numb-modif}
\reva{So far we have evaluated historical what-if queries with a single modification. We now evaluate how multiple modifications affect the performance of Mahif and of the proposed optimizations.}

\reva{\Cref{fig:Multimod} depicts the effect of changing the number of modifications in our historical what-if query to better understand how it affects both program slicing and data slicing. Program slicing becomes much more expensive than its single modification counterpart, given that we have to test each update by comparing the state of its symbolic tuple not only between $\history$ and $\ahmod$, but duplicating these histories while removing the update being tested, in order to not falsely classify an update as independent. This effectively quadruples the individual MILP program size over the single modification case. Data slicing also becomes more expensive as we employ the push down technique described in \Cref{sec:filter}, which in turn creates a selection operator with both a wider condition and a more complicated condition to select. These selection conditions also include some form of reenactment, in order to capture every tuple that would be modified by modifications that occur after the first in the historical what-if query. That is, program slicing and data slicing become more inefficient with multiple modifications due to the overhead necessary to ensure correctness of both optimizations.}

\reva{The data from \Cref{fig:Multimod} shows a decrease in performance from a single modification to the multiple modification case. The nature of the modification (attributes updated, conditions, selectivity, causing dependency elsewhere) significantly impacts the performance of \textit{R+DS}, as evaluating the data slicing conditions pushed down through a long history becomes too expensive. \textit{R+PS+DS} remains an effective optimization \textit{R}, though its cost remains higher than a single modification. In part this is due to the effect of slicing the history, reducing the amount of conditions the data slicing conditions are pushed through, reducing the complexity of its evaluation. It should be noted that as the amount of modifications grows, the program slicing time goes down as these modifications are inherently dependent. That being said, an inflection point is possible where the gains in program slicing execution speedup results in a slowdown from the longer history the data slicing conditions need to be pushed through. In general, multiple modifications are most practical with lower selectivities, in order to provide the filtering that speeds up the execution of the slice.}

\subsection{Summary}
Our experiments show that our approach outperforms the naïve method in most cases despite it not needing any additional storage. The proposed optimization methods are very effective and for large number of updates and relation sizes, they improve performance considerably. However, in cases where the relation size is small or the selectivity is very low, the cost of program slicing will outweigh the execution of reenactment or reenactment with data slicing. Despite the comparatively high cost of reenactment without optimizations, reenactment alone is considerably faster than its naïve equivalent. Our experiments also show that our approach scales well with respect to relation size.

\section{Conclusions}
\label{sec:conclusions}

\revm{We propose historical what-if queries, a new type of what-if queries which allow users to explore the effects of hypothetical changes to the transactional history of a database. Our system Mahif, efficiently answers such queries using reenactment %
and  two novel optimization  techniques (program and data slicing) that exclude irrelevant data and updates from the computation.}
Our experimental evaluation demonstrates the effectiveness of our approach and optimization techniques.
\revc{In future work, we will explore how to augment a user's \abbrHW based on information about unobserved external factors and dependencies between updates, e.g., if a \abbrHW the statement creating a customer from history, then the statements creating the orders of this customer should be removed too. Furthermore, we will explore novel application of our symbolic evaluation technique such as proving equivalence of transactional histories.}

\bibliographystyle{ACM-Reference-Format}
\bibliography{trans}


\begin{thebibliography}{35}


\ifx \showCODEN    \undefined \def \showCODEN     #1{\unskip}     \fi
\ifx \showDOI      \undefined \def \showDOI       #1{#1}\fi
\ifx \showISBNx    \undefined \def \showISBNx     #1{\unskip}     \fi
\ifx \showISBNxiii \undefined \def \showISBNxiii  #1{\unskip}     \fi
\ifx \showISSN     \undefined \def \showISSN      #1{\unskip}     \fi
\ifx \showLCCN     \undefined \def \showLCCN      #1{\unskip}     \fi
\ifx \shownote     \undefined \def \shownote      #1{#1}          \fi
\ifx \showarticletitle \undefined \def \showarticletitle #1{#1}   \fi
\ifx \showURL      \undefined \def \showURL       {\relax}        \fi
\providecommand\bibfield[2]{#2}
\providecommand\bibinfo[2]{#2}
\providecommand\natexlab[1]{#1}
\providecommand\showeprint[2][]{arXiv:#2}

\bibitem[\protect\citeauthoryear{Abiteboul and Grahne}{Abiteboul and
  Grahne}{1985}]%
        {AG85}
\bibfield{author}{\bibinfo{person}{Serge Abiteboul} {and}
  \bibinfo{person}{Gösta Grahne}.} \bibinfo{year}{1985}\natexlab{}.
\newblock \showarticletitle{Update semantics for incomplete databases}. In
  \bibinfo{booktitle}{\emph{VLDB}}. \bibinfo{pages}{1--12}.
\newblock


\bibitem[\protect\citeauthoryear{Amsterdamer, Deutch, and Tannen}{Amsterdamer
  et~al\mbox{.}}{2011}]%
        {AD11d}
\bibfield{author}{\bibinfo{person}{Yael Amsterdamer}, \bibinfo{person}{Daniel
  Deutch}, {and} \bibinfo{person}{Val Tannen}.}
  \bibinfo{year}{2011}\natexlab{}.
\newblock \showarticletitle{{Provenance for Aggregate Queries}}. In
  \bibinfo{booktitle}{\emph{PODS}}. \bibinfo{pages}{153--164}.
\newblock


\bibitem[\protect\citeauthoryear{Arab, Gawlick, Krishnaswamy, Radhakrishnan,
  and Glavic}{Arab et~al\mbox{.}}{2016}]%
        {AG17}
\bibfield{author}{\bibinfo{person}{Bahareh~Sadat Arab}, \bibinfo{person}{Dieter
  Gawlick}, \bibinfo{person}{Vasudha Krishnaswamy}, \bibinfo{person}{Venkatesh
  Radhakrishnan}, {and} \bibinfo{person}{Boris Glavic}.}
  \bibinfo{year}{2016}\natexlab{}.
\newblock \showarticletitle{Reenactment for Read-Committed Snapshot Isolation}.
  In \bibinfo{booktitle}{\emph{CIKM}}. \bibinfo{pages}{841--850}.
\newblock


\bibitem[\protect\citeauthoryear{Arab, Gawlick, Krishnaswamy, Radhakrishnan,
  and Glavic}{Arab et~al\mbox{.}}{2018}]%
        {AG18}
\bibfield{author}{\bibinfo{person}{Bahareh~Sadat Arab}, \bibinfo{person}{Dieter
  Gawlick}, \bibinfo{person}{Vasudha Krishnaswamy}, \bibinfo{person}{Venkatesh
  Radhakrishnan}, {and} \bibinfo{person}{Boris Glavic}.}
  \bibinfo{year}{2018}\natexlab{}.
\newblock \showarticletitle{Using reenactment to retroactively capture
  provenance for transactions}.
\newblock \bibinfo{journal}{\emph{IEEE Transactions on Knowledge and Data
  Engineering}} \bibinfo{volume}{30}, \bibinfo{number}{3}
  (\bibinfo{year}{2018}), \bibinfo{pages}{599--612}.
\newblock


\bibitem[\protect\citeauthoryear{Arab, Gawlick, Radhakrishnan, Guo, and
  Glavic}{Arab et~al\mbox{.}}{2014}]%
        {AG14}
\bibfield{author}{\bibinfo{person}{Bahareh~Sadat Arab}, \bibinfo{person}{Dieter
  Gawlick}, \bibinfo{person}{Venkatesh Radhakrishnan}, \bibinfo{person}{Hao
  Guo}, {and} \bibinfo{person}{Boris Glavic}.} \bibinfo{year}{2014}\natexlab{}.
\newblock \showarticletitle{A Generic Provenance Middleware for Database
  Queries, Updates, and Transactions}. In \bibinfo{booktitle}{\emph{TaPP}}.
\newblock


\bibitem[\protect\citeauthoryear{Balmin, Papadimitriou, and
  Papakonstantinou}{Balmin et~al\mbox{.}}{2000}]%
        {hung17}
\bibfield{author}{\bibinfo{person}{Andrey Balmin}, \bibinfo{person}{Thanos
  Papadimitriou}, {and} \bibinfo{person}{Yannis Papakonstantinou}.}
  \bibinfo{year}{2000}\natexlab{}.
\newblock \showarticletitle{Hypothetical Queries in an {OLAP} Environment}. In
  \bibinfo{booktitle}{\emph{{VLDB}}}. \bibinfo{pages}{220--231}.
\newblock


\bibitem[\protect\citeauthoryear{Bourhis, Deutch, and Moskovitch}{Bourhis
  et~al\mbox{.}}{2016}]%
        {bourhis16}
\bibfield{author}{\bibinfo{person}{Pierre Bourhis}, \bibinfo{person}{Daniel
  Deutch}, {and} \bibinfo{person}{Yuval Moskovitch}.}
  \bibinfo{year}{2016}\natexlab{}.
\newblock \showarticletitle{Analyzing data-centric applications: Why, what-if,
  and how-to}. In \bibinfo{booktitle}{\emph{Data Engineering (ICDE), 2016 IEEE
  32nd International Conference on}}. IEEE, \bibinfo{pages}{779--790}.
\newblock


\bibitem[\protect\citeauthoryear{Bucur, Kinder, and Candea}{Bucur
  et~al\mbox{.}}{2014}]%
        {bucur14}
\bibfield{author}{\bibinfo{person}{Stefan Bucur}, \bibinfo{person}{Johannes
  Kinder}, {and} \bibinfo{person}{George Candea}.}
  \bibinfo{year}{2014}\natexlab{}.
\newblock \showarticletitle{Prototyping symbolic execution engines for
  interpreted languages}.
\newblock \bibinfo{journal}{\emph{SIGARCH Comput Archit News}}
  \bibinfo{volume}{42}, \bibinfo{number}{1} (\bibinfo{year}{2014}),
  \bibinfo{pages}{239--254}.
\newblock


\bibitem[\protect\citeauthoryear{Buneman, Khanna, and Tan}{Buneman
  et~al\mbox{.}}{2001}]%
        {BK01}
\bibfield{author}{\bibinfo{person}{Peter Buneman}, \bibinfo{person}{Sanjeev
  Khanna}, {and} \bibinfo{person}{Wang-Chiew Tan}.}
  \bibinfo{year}{2001}\natexlab{}.
\newblock \showarticletitle{{Why and Where: A Characterization of Data
  Provenance}}. In \bibinfo{booktitle}{\emph{ICDT}}. \bibinfo{pages}{316--330}.
\newblock


\bibitem[\protect\citeauthoryear{Cadar and Sen}{Cadar and Sen}{2013}]%
        {cadar13}
\bibfield{author}{\bibinfo{person}{Cristian Cadar} {and}
  \bibinfo{person}{Koushik Sen}.} \bibinfo{year}{2013}\natexlab{}.
\newblock \showarticletitle{Symbolic execution for software testing: three
  decades later}.
\newblock \bibinfo{journal}{\emph{Commun. ACM}} \bibinfo{volume}{56},
  \bibinfo{number}{2} (\bibinfo{year}{2013}), \bibinfo{pages}{82--90}.
\newblock


\bibitem[\protect\citeauthoryear{Cheney}{Cheney}{2007}]%
        {cheney07}
\bibfield{author}{\bibinfo{person}{James Cheney}.}
  \bibinfo{year}{2007}\natexlab{}.
\newblock \showarticletitle{Program slicing and data provenance.}
\newblock \bibinfo{journal}{\emph{IEEE Data Eng. Bull.}} \bibinfo{volume}{30},
  \bibinfo{number}{4} (\bibinfo{year}{2007}), \bibinfo{pages}{22--28}.
\newblock


\bibitem[\protect\citeauthoryear{Chu, Wang, Weitz, and Cheung}{Chu
  et~al\mbox{.}}{2017}]%
        {chu2017}
\bibfield{author}{\bibinfo{person}{Shumo Chu}, \bibinfo{person}{Chenglong
  Wang}, \bibinfo{person}{Konstantin Weitz}, {and} \bibinfo{person}{Alvin
  Cheung}.} \bibinfo{year}{2017}\natexlab{}.
\newblock \showarticletitle{Cosette: An Automated Prover for SQL.}. In
  \bibinfo{booktitle}{\emph{CIDR}}.
\newblock


\bibitem[\protect\citeauthoryear{Cooper, Silberstein, Tam, Ramakrishnan, and
  Sears}{Cooper et~al\mbox{.}}{2010}]%
        {CooperSTRS10}
\bibfield{author}{\bibinfo{person}{Brian~F. Cooper}, \bibinfo{person}{Adam
  Silberstein}, \bibinfo{person}{Erwin Tam}, \bibinfo{person}{Raghu
  Ramakrishnan}, {and} \bibinfo{person}{Russell Sears}.}
  \bibinfo{year}{2010}\natexlab{}.
\newblock \showarticletitle{Benchmarking cloud serving systems with {YCSB}}. In
  \bibinfo{booktitle}{\emph{Proceedings of the 1st {ACM} Symposium on Cloud
  Computing, SoCC 2010, Indianapolis, Indiana, USA, June 10-11, 2010}},
  \bibfield{editor}{\bibinfo{person}{Joseph~M. Hellerstein},
  \bibinfo{person}{Surajit Chaudhuri}, {and} \bibinfo{person}{Mendel
  Rosenblum}} (Eds.). \bibinfo{publisher}{{ACM}}, \bibinfo{pages}{143--154}.
\newblock
\urldef\tempurl%
\url{https://doi.org/10.1145/1807128.1807152}
\showDOI{\tempurl}


\bibitem[\protect\citeauthoryear{Cplex}{Cplex}{2009}]%
        {cplex2009v12}
\bibfield{author}{\bibinfo{person}{IBM~ILOG Cplex}.}
  \bibinfo{year}{2009}\natexlab{}.
\newblock \showarticletitle{V12. 1: User’s Manual for CPLEX}.
\newblock \bibinfo{journal}{\emph{International Business Machines Corporation}}
  \bibinfo{volume}{46}, \bibinfo{number}{53} (\bibinfo{year}{2009}),
  \bibinfo{pages}{157}.
\newblock


\bibitem[\protect\citeauthoryear{Cui, Widom, and Wiener}{Cui
  et~al\mbox{.}}{2000}]%
        {CW00b}
\bibfield{author}{\bibinfo{person}{Yingwei Cui}, \bibinfo{person}{Jennifer
  Widom}, {and} \bibinfo{person}{Janet~L. Wiener}.}
  \bibinfo{year}{2000}\natexlab{}.
\newblock \showarticletitle{{Tracing the Lineage of View Data in a Warehousing
  Environment}}.
\newblock \bibinfo{journal}{\emph{TODS}} \bibinfo{volume}{25},
  \bibinfo{number}{2} (\bibinfo{year}{2000}), \bibinfo{pages}{179--227}.
\newblock


\bibitem[\protect\citeauthoryear{Dashti, Basil~John, Shaikhha, and Koch}{Dashti
  et~al\mbox{.}}{2017}]%
        {dashti17}
\bibfield{author}{\bibinfo{person}{Mohammad Dashti}, \bibinfo{person}{Sachin
  Basil~John}, \bibinfo{person}{Amir Shaikhha}, {and}
  \bibinfo{person}{Christoph Koch}.} \bibinfo{year}{2017}\natexlab{}.
\newblock \showarticletitle{Transaction Repair for Multi-Version Concurrency
  Control}. In \bibinfo{booktitle}{\emph{Proceedings of the 2017 ACM
  International Conference on Management of Data}}. ACM,
  \bibinfo{pages}{235--250}.
\newblock


\bibitem[\protect\citeauthoryear{Deutch, Ives, Milo, and Tannen}{Deutch
  et~al\mbox{.}}{2013}]%
        {deutch13}
\bibfield{author}{\bibinfo{person}{Daniel Deutch}, \bibinfo{person}{Zachary~G
  Ives}, \bibinfo{person}{Tova Milo}, {and} \bibinfo{person}{Val Tannen}.}
  \bibinfo{year}{2013}\natexlab{}.
\newblock \showarticletitle{Caravan: Provisioning for What-If Analysis.}. In
  \bibinfo{booktitle}{\emph{CIDR}}.
\newblock


\bibitem[\protect\citeauthoryear{Difallah, Pavlo, Curino, and
  Cudr{\'e}-Mauroux}{Difallah et~al\mbox{.}}{2013}]%
        {DifallahPCC13}
\bibfield{author}{\bibinfo{person}{Djellel~Eddine Difallah},
  \bibinfo{person}{Andrew Pavlo}, \bibinfo{person}{Carlo Curino}, {and}
  \bibinfo{person}{Philippe Cudr{\'e}-Mauroux}.}
  \bibinfo{year}{2013}\natexlab{}.
\newblock \showarticletitle{OLTP-Bench: An Extensible Testbed for Benchmarking
  Relational Databases}.
\newblock \bibinfo{journal}{\emph{PVLDB}} \bibinfo{volume}{7},
  \bibinfo{number}{4} (\bibinfo{year}{2013}), \bibinfo{pages}{277--288}.
\newblock
\urldef\tempurl%
\url{http://www.vldb.org/pvldb/vol7/p277-difallah.pdf}
\showURL{%
\tempurl}


\bibitem[\protect\citeauthoryear{Fagin, Kuper, Ullman, and Vardi}{Fagin
  et~al\mbox{.}}{1986}]%
        {fagin-86-upld}
\bibfield{author}{\bibinfo{person}{Ronald Fagin}, \bibinfo{person}{Gabriel~M.
  Kuper}, \bibinfo{person}{Jeffrey~D. Ullman}, {and} \bibinfo{person}{Moshe~Y.
  Vardi}.} \bibinfo{year}{1986}\natexlab{}.
\newblock \showarticletitle{Updating Logical Databases}.
\newblock \bibinfo{journal}{\emph{Adv. Comput. Res.}}  \bibinfo{volume}{3}
  (\bibinfo{year}{1986}), \bibinfo{pages}{1--18}.
\newblock


\bibitem[\protect\citeauthoryear{G{\'o}mez-L{\'o}pez and
  Gasca}{G{\'o}mez-L{\'o}pez and Gasca}{2014}]%
        {gomez14}
\bibfield{author}{\bibinfo{person}{Mar{\'\i}a~Teresa G{\'o}mez-L{\'o}pez} {and}
  \bibinfo{person}{Rafael~M Gasca}.} \bibinfo{year}{2014}\natexlab{}.
\newblock \showarticletitle{Using constraint programming in selection operators
  for constraint databases}.
\newblock \bibinfo{journal}{\emph{Expert Syst Appl}} \bibinfo{volume}{41},
  \bibinfo{number}{15} (\bibinfo{year}{2014}), \bibinfo{pages}{6773--6785}.
\newblock


\bibitem[\protect\citeauthoryear{Green, Karvounarakis, and Tannen}{Green
  et~al\mbox{.}}{2007}]%
        {GK07}
\bibfield{author}{\bibinfo{person}{Todd~J. Green}, \bibinfo{person}{Gregory
  Karvounarakis}, {and} \bibinfo{person}{Val Tannen}.}
  \bibinfo{year}{2007}\natexlab{}.
\newblock \showarticletitle{{Provenance Semirings}}. In
  \bibinfo{booktitle}{\emph{PODS}}. \bibinfo{pages}{31--40}.
\newblock


\bibitem[\protect\citeauthoryear{Imieli{\'n}ski and Lipski}{Imieli{\'n}ski and
  Lipski}{1988}]%
        {incomp88}
\bibfield{author}{\bibinfo{person}{Tomasz Imieli{\'n}ski} {and}
  \bibinfo{person}{Witold Lipski}.} \bibinfo{year}{1988}\natexlab{}.
\newblock \showarticletitle{Incomplete information in relational databases}.
\newblock In \bibinfo{booktitle}{\emph{Readings in Artificial Intelligence and
  Databases}}. \bibinfo{publisher}{Elsevier}, \bibinfo{pages}{342--360}.
\newblock


\bibitem[\protect\citeauthoryear{Imieli{\'n}ski and Lipski~Jr}{Imieli{\'n}ski
  and Lipski~Jr}{1984}]%
        {IL84a}
\bibfield{author}{\bibinfo{person}{Tomasz Imieli{\'n}ski} {and}
  \bibinfo{person}{Witold Lipski~Jr}.} \bibinfo{year}{1984}\natexlab{}.
\newblock \showarticletitle{{Incomplete Information in Relational Databases}}.
\newblock \bibinfo{journal}{\emph{JACM}} \bibinfo{volume}{31},
  \bibinfo{number}{4} (\bibinfo{year}{1984}), \bibinfo{pages}{761--791}.
\newblock


\bibitem[\protect\citeauthoryear{Kennedy and Koch}{Kennedy and Koch}{2010}]%
        {pip10}
\bibfield{author}{\bibinfo{person}{Oliver Kennedy} {and}
  \bibinfo{person}{Christoph Koch}.} \bibinfo{year}{2010}\natexlab{}.
\newblock \showarticletitle{PIP: A database system for great and small
  expectations}. In \bibinfo{booktitle}{\emph{Data Engineering (ICDE), 2010
  IEEE 26th International Conference on}}. IEEE, \bibinfo{pages}{157--168}.
\newblock


\bibitem[\protect\citeauthoryear{King}{King}{1976}]%
        {K76}
\bibfield{author}{\bibinfo{person}{James~C King}.}
  \bibinfo{year}{1976}\natexlab{}.
\newblock \showarticletitle{Symbolic execution and program testing}.
\newblock \bibinfo{journal}{\emph{CACM}} \bibinfo{volume}{19},
  \bibinfo{number}{7} (\bibinfo{year}{1976}), \bibinfo{pages}{385--394}.
\newblock


\bibitem[\protect\citeauthoryear{Kuper, Libkin, and Paredaens}{Kuper
  et~al\mbox{.}}{2013}]%
        {kuper13}
\bibfield{author}{\bibinfo{person}{Gabriel Kuper}, \bibinfo{person}{Leonid
  Libkin}, {and} \bibinfo{person}{Jan Paredaens}.}
  \bibinfo{year}{2013}\natexlab{}.
\newblock \bibinfo{booktitle}{\emph{Constraint databases}}.
\newblock


\bibitem[\protect\citeauthoryear{Luckow, P{\u{a}}s{\u{a}}reanu, Dwyer, Filieri,
  and Visser}{Luckow et~al\mbox{.}}{2014}]%
        {luckow14}
\bibfield{author}{\bibinfo{person}{Kasper Luckow}, \bibinfo{person}{Corina~S
  P{\u{a}}s{\u{a}}reanu}, \bibinfo{person}{Matthew~B Dwyer},
  \bibinfo{person}{Antonio Filieri}, {and} \bibinfo{person}{Willem Visser}.}
  \bibinfo{year}{2014}\natexlab{}.
\newblock \showarticletitle{Exact and approximate probabilistic symbolic
  execution for nondeterministic programs}. In \bibinfo{booktitle}{\emph{ASE}}.
  \bibinfo{pages}{575--586}.
\newblock


\bibitem[\protect\citeauthoryear{Meliou and Suciu}{Meliou and Suciu}{2012}]%
        {MeliouS12}
\bibfield{author}{\bibinfo{person}{Alexandra Meliou} {and} \bibinfo{person}{Dan
  Suciu}.} \bibinfo{year}{2012}\natexlab{}.
\newblock \showarticletitle{Tiresias: {T}he Database Oracle for How-To
  Queries}. In \bibinfo{booktitle}{\emph{SIGMOD}}. \bibinfo{pages}{337--348}.
\newblock


\bibitem[\protect\citeauthoryear{Schrijver}{Schrijver}{1998}]%
        {schrijver1998theory}
\bibfield{author}{\bibinfo{person}{Alexander Schrijver}.}
  \bibinfo{year}{1998}\natexlab{}.
\newblock \bibinfo{booktitle}{\emph{Theory of linear and integer programming}}.
\newblock \bibinfo{publisher}{John Wiley \& Sons}.
\newblock


\bibitem[\protect\citeauthoryear{Torlak and Bodik}{Torlak and Bodik}{2014}]%
        {torlak2014}
\bibfield{author}{\bibinfo{person}{Emina Torlak} {and}
  \bibinfo{person}{Rastislav Bodik}.} \bibinfo{year}{2014}\natexlab{}.
\newblock \showarticletitle{A lightweight symbolic virtual machine for
  solver-aided host languages}. In \bibinfo{booktitle}{\emph{ACM SIGPLAN
  Notices}}, Vol.~\bibinfo{volume}{49}. ACM, \bibinfo{pages}{530--541}.
\newblock


\bibitem[\protect\citeauthoryear{Wang, Meliou, and Wu}{Wang
  et~al\mbox{.}}{2017}]%
        {wang16}
\bibfield{author}{\bibinfo{person}{Xiaolan Wang}, \bibinfo{person}{Alexandra
  Meliou}, {and} \bibinfo{person}{Eugene Wu}.} \bibinfo{year}{2017}\natexlab{}.
\newblock \showarticletitle{Qfix: Diagnosing errors through query histories}.
  In \bibinfo{booktitle}{\emph{SIGMOD}}. \bibinfo{pages}{1369--1384}.
\newblock


\bibitem[\protect\citeauthoryear{Weiser}{Weiser}{1981}]%
        {W81}
\bibfield{author}{\bibinfo{person}{M. Weiser}.}
  \bibinfo{year}{1981}\natexlab{}.
\newblock \showarticletitle{{Program slicing}}.
\newblock \bibinfo{journal}{\emph{ICSE}} (\bibinfo{year}{1981}),
  \bibinfo{pages}{439--449}.
\newblock


\bibitem[\protect\citeauthoryear{Winslett}{Winslett}{1986}]%
        {winslett-86-upldcnv}
\bibfield{author}{\bibinfo{person}{Marianne Winslett}.}
  \bibinfo{year}{1986}\natexlab{}.
\newblock \showarticletitle{Updating Logical Databases Containing Null Values}.
  In \bibinfo{booktitle}{\emph{ICDT}}, Vol.~\bibinfo{volume}{243}.
  \bibinfo{pages}{421--435}.
\newblock


\bibitem[\protect\citeauthoryear{Yang, Meneghetti, Fehling, Liu, and
  Kennedy}{Yang et~al\mbox{.}}{2015}]%
        {lenses15}
\bibfield{author}{\bibinfo{person}{Ying Yang}, \bibinfo{person}{Niccolo
  Meneghetti}, \bibinfo{person}{Ronny Fehling}, \bibinfo{person}{Zhen~Hua Liu},
  {and} \bibinfo{person}{Oliver Kennedy}.} \bibinfo{year}{2015}\natexlab{}.
\newblock \showarticletitle{Lenses: An on-demand approach to etl}.
\newblock \bibinfo{journal}{\emph{Proceedings of the VLDB Endowment}}
  \bibinfo{volume}{8}, \bibinfo{number}{12} (\bibinfo{year}{2015}),
  \bibinfo{pages}{1578--1589}.
\newblock


\bibitem[\protect\citeauthoryear{Zhuge, Garcia-Molina, Hammer, and Widom}{Zhuge
  et~al\mbox{.}}{1995}]%
        {ZG95}
\bibfield{author}{\bibinfo{person}{Yue Zhuge}, \bibinfo{person}{Hector
  Garcia-Molina}, \bibinfo{person}{Joachim Hammer}, {and}
  \bibinfo{person}{Jennifer Widom}.} \bibinfo{year}{1995}\natexlab{}.
\newblock \showarticletitle{View maintenance in a warehousing environment}.
\newblock \bibinfo{journal}{\emph{SIGMOD Record}} \bibinfo{volume}{24},
  \bibinfo{number}{2} (\bibinfo{year}{1995}), \bibinfo{pages}{316--327}.
\newblock


\end{thebibliography}

\end{document}